\documentclass[11pt]{article}
\usepackage[margin=1in]{geometry}
\usepackage{amsmath,amssymb,amsthm,comment}
\usepackage{booktabs,longtable,array,multirow,makecell}
\usepackage{pdflscape}
\usepackage{pgfplots}
\pgfplotsset{compat=1.18}
\usepackage{hyperref}
\usepackage{enumitem}
\usepackage{mathtools,empheq}
\usepackage{xcolor,colortbl}
\usepackage{caption}
\DeclareCaptionFont{tablecaption}{\linespread{1.1}\normalsize}
\usepackage{float}

\DeclareFontShape{OMX}{cmex}{m}{n}{<-> cmex10}{}
\usepackage{bm}
\usepackage{authblk}
\usepackage{eso-pic}
\usepackage{cite}
\usepackage[normalem]{ulem}  

\setlist{nosep,leftmargin=1.6em}

\newcommand{\Z}{\mathbb Z}
\hypersetup{colorlinks=true,linkcolor=blue,citecolor=blue,urlcolor=blue}
\setlist[itemize]{noitemsep,topsep=3pt}
\setlist[enumerate]{noitemsep,topsep=3pt}

\newcommand{\tr}{\operatorname{tr}}

\newcommand{\Pf}{\operatorname{Pf}}

\def\be{\begin{equation}}
\def\ee{\end{equation}}

\numberwithin{equation}{section}

\newcommand{\fund}{\mathrm{fund}}
\newcommand{\adj}{\mathrm{adj}}

\newcommand{\eq}[1]{(\ref{#1})}

\definecolor{neutralgray}{gray}{0.88}

\renewcommand{\baselinestretch}{1.15}

\def\a{\alpha}
\def\b{\beta}

\def\e{\epsilon}

\def\s{\sigma}

\newcommand{\bea}{\begin{eqnarray}}
\newcommand{\eea}{\end{eqnarray}}

\title{Diagonally gauged anomaly-free 6D
supergravities and their vacua}

\author[1]{Xu Guo}
\author[2,3]{Yi Pang}
\author[4,5]{Henning Samtleben}
\author[1]{Ergin Sezgin}

\affil[1]{George P. and Cynthia W. Mitchell Institute for Fundamental
Physics and Astronomy, Texas A\&M University,
College Station, TX 77843-4242, USA}

\affil[2]{Center for Joint Quantum Studies and Department of Physics,
School of Science, Tianjin University, Tianjin 300350, China}

\affil[3]{Peng Huanwu Center for Fundamental Theory,
Hefei, Anhui 230026, China}

\affil[4]{ENSL, CNRS, Laboratoire de physique,
F-69342 Lyon, France}

\affil[5]{Institut Universitaire de France (IUF)}

\date{}

\DeclareFontFamily{U}{msb}{}
\DeclareFontShape{U}{msb}{m}{n}{<->msbm10}{}
\DeclareFontFamily{U}{msa}{}
\DeclareFontShape{U}{msa}{m}{n}{<->msam10}{}
\begin{document}

\maketitle

\vspace{-4em}

\begin{center}
\small
E-mails:
\texttt{xuguo@physics.tamu.edu},
\texttt{pangyi1@tju.edu.cn},
\texttt{henning.samtleben@ens-lyon.fr},
\texttt{sezgin@tamu.edu}
\end{center}

\thispagestyle{empty}

\begin{abstract}

We study anomaly-free six-dimensional $N=(1,0)$ supergravities with $n_T=1$ and diagonal $U(1)_{R+}$ or $Sp(1)_{R+}$ gauging. The diagonal group combines R-symmetry with a commuting action on hypermultiplets. We perform bounded searches for anomaly-free models with one or two additional simple gauge group factors of rank at most 30, impose string-charge lattice and positive kinetic term conditions, and tabulate the resulting spectra. The main scans cover $q_{\max}\leq2$ and $d_{\max}\leq3$, where $q_{\max}$ is the largest absolute hyperino charge and $d_{\max}$ the largest diagonal $Sp(1)$ representation dimension, with supplementary searches at larger values. Every hypermultiplet block is required to be nontrivial under at least one additional simple gauge factor. 
The search has led to a large number of models. Here we give the details of 93 models with $G_+= U(1)_{R+}$ and  $q_{\rm max}\le 2$, and 26 models with $G_+=Sp(1)_{R+}$ and $d_{\rm max}\le 3$, and one or two  simple group factors. Additional 1341 and 494 models with $G_+=U(1)_{R+}$ with $q_{\rm max}=2$ and $Sp(1)_{R+}$ with $d_{\rm max}=3$, respectively, are posted to the GitHub repository. 
We establish the existence of supersymmetric Minkowski vacua for all the anomaly free models we have found with $q_{\rm max}=2$ and $d_{\rm max}=3$. We find that such vacua require $q_{\max}>1$ for $U(1)_{R+}$, and $d_{\max}>2$ for $Sp(1)_{R+}$. We also show that every Minkowski vacuum breaks R-symmetry, with the connected unbroken gauge group contained in the additional gauge factors. At two-derivative order, while our anomaly free models do not admit AdS$_6$ solutions, all those with $q_{\rm max} = 2$ and $d_{\rm max}=3$ have non-supersymmetric and unstable dS$_6$ vacua as well.

\end{abstract}

\newpage
{\hypersetup{linkcolor=black}\tableofcontents}

\newpage
\section{Introduction}
\label{sec:introduction}

R-symmetry gauged $N=(1,0)$ supergravities in six dimensions
\cite{Nishino:1986dc,Nishino:1997ff,Riccioni:2001bg} that are free from
local and global anomalies are of interest because their origin in
string/M/F theory remains unclear. Most searches have considered
$Sp(1)_R$ or $U(1)_R$
\cite{Randjbar-Daemi:1985tdc,Avramis:2005qt,Avramis:2005hc,Becker:2023zyb,Becker:2025xgy,Suzuki:2005vu,Becker:2026etl}.
The R-symmetry rotates the triplet of complex structures on the
hypermultiplet scalar manifold. We take this manifold to be 
\begin{equation}
 {\cal M}_H=\frac{G_H}{H\times Sp(1)_R},
 \label{coset}
\end{equation}
where we take \(G_H\) to be a connected noncompact simple Lie group and \(H\times Sp(1)_R\), up to a finite central quotient, to be its maximal compact subgroup, with \(Sp(1)_R\) identified with the canonical quaternionic \(Sp(1)\) factor. Thus \(G_H/(H\times Sp(1)_R)\) is an irreducible Riemannian symmetric quaternionic-K\"ahler space of noncompact type—a noncompact Wolf space~\cite{Bagger:1983tt,Sezgin:2023hkc}.

One possible diagonal gauge group is $Sp(1)_{R+}$ whose generator is a diagonal sum of $Sp(1)_R$ and a subgroup $Sp(1)_H\subset H$. Another possible diagonal gauge group is $U(1)_{R+}$ whose generator is a diagonal sum of the subgroup $U(1)_R\subset Sp(1)_R$ and a subgroup $U(1)_H\subset H$. Suzuki and Tachikawa~\cite{Suzuki:2005vu} constructed a huge number of anomaly-free models whose gauge groups are products of one of these diagonal groups with $SU(2)$ and $U(1)$ factors, with every hyperino charged under some factor, so that hypermultiplet drones are absent. Here ``drones'' denote gauge-neutral hypermultiplets or $U(1)$ vector multiplets under which no scalar or fermion is charged. 

We study both $U(1)$ and $Sp(1)$ diagonal gaugings, with one tensor multiplet and gauge group
\begin{equation}
G=G_+\times\prod_{i=1}^{n}G_i=\prod_zG_z,
\qquad G_+=U(1)_{R+}\ \hbox{or}\ Sp(1)_{R+},
\qquad z=(+,i).
\label{eq:group}
\end{equation}
Note that $G$ is the total gauge group of the anomaly free model, not to be confused with $G_H$. 
Furthermore we have
\begin{equation}
\prod_iG_i=K\times G_E,\qquad K=\prod_{z'}K_{z'}\subset H.
\end{equation}
where $K\subset H$ and $G_E\subset G$ is external in the sense that it is outside the gauge group $K$ associated with the isometries of $\mathcal{M}_H$. For the $U(1)_{R+}$ gauging, the $U(1)_{R+}$ generator is given by
\begin{equation}
T_{R+}=T_R+T_H,
\label{tr}
\end{equation}
as the sum of the $U(1)_R$ generator $T_R$ and the generator $T_H$ of $U(1)_H\subset H$
that commutes with $K$. For the $Sp(1)_{R+}$ gauging we instead take
\begin{equation}
T_{R+}^r=T_R^r+T_H^r,\quad r=1,2,3,
\label{sp:tr}
\end{equation}
where $T_H^r$ generates $Sp(1)_H\subset H$ and commutes with $K$.

The hyperini transform under $H$. The gravitini and tensorini transform as R-symmetry doublets. For $Sp(1)_{R+}$, its gauginos also carry a diagonal adjoint index and therefore transform in $\mathbf3\otimes\mathbf2$. This distinction is essential in comparing the two anomaly polynomials. 
We extend this search to one or two additional simple gauge factors beyond $SU(2)$, including matter charged under both. Every hypermultiplet block is required to be nontrivial under at least one additional simple gauge factor. This is done in order to reduce the search space to a manageable size, since admitting blocks that are singlets of all $G_i$ would greatly increase the number of ways to distribute $n_H$ over irreducible blocks. We impose anomaly factorization to employ the Green-Schwarz mechanism to cancel the local anomalies, a common unimodular string-charge lattice, global anomaly freedom, and a common domain of positive gauge kinetic terms to avoid ghosts. 

In the search for consistent models, in addition to the requirement that all hyperinos carry a nontrivial representation of the gauge group as mentioned earlier, we also impose bounds on the gauge groups, their rank, and the charge and dimension of the $G_+$ irreps, as stated in \eq{G1}, \eq{G1xG2} and \eq{eq:shared-diagonal-scan}. The search has led to a large number of models. Here we give the details of 93 models with $G_+= U(1)_{R+}$ and  $q_{\rm max}\le 2$, and 26 models with $G_+=Sp(1)_{R+}$ and $d_{\rm max}\le 3$, and one or two simple group factors. A further 1341 and 494 models with $G_+=U(1)_{R+}$ with $q_{\rm max}=2$ and $Sp(1)_{R+}$ with $d_{\rm max}=3$, respectively, are posted to the GitHub repository. 

We establish the existence of supersymmetric Minkowski vacua for all the anomaly free models we have found with $q_{\max}=2$ or $d_{\max}=3$, respectively. Using Paradan's lemma~\cite[Section~2.5, Lemma~2.12]{Paradan:2015}, we prove that Minkowski vacua on connected noncompact Wolf spaces require $q_{\max}>1$ for $U(1)_{R+}$ gauging and $d_{\max}>2$ for $Sp(1)_{R+}$ gauging. Here $q_{\rm max}$ denotes the largest absolute value of the $U(1)_{R+}$ charge, and $d_{\rm max}$ is the maximum dimension of the $Sp(1)_{R+}$ representations carried by the hyperfermions. On the Wolf spaces relevant to our anomaly-free models, these bounds are also sufficient for the moment maps associated with the diagonal gauge group to vanish simultaneously, subject to the stated gauging and matter assumptions. A Minkowski vacuum additionally requires the remaining gauged moment maps to vanish there.
We also prove that every ghost-free Minkowski vacuum breaks R-symmetry, with the connected unbroken gauge group contained in the additional gauge factors. We present the details of four $U(1)_{R+}$ and two $Sp(1)_{R+}$ examples of supersymmetric Minkowski vacua.

At two-derivative order, the nonnegative potential excludes $AdS_6$ vacua. The nonsupersymmetric $dS_6$ solution at the hyperscalar origin, when it exists with ghost-free kinetic terms, is unstable in the dilaton direction.

We note that gaugings of homogeneous quaternionic-K\"ahler manifolds were studied in four-dimensional $N=2$ supergravity, where suitable translational gaugings yield Minkowski vacua with spontaneous supersymmetry breaking, accompanied by breaking of gauged translations~\cite{Tsokur:1996su,Tsokur:1996vk}. Here, compact diagonal R-symmetry gaugings in six dimensions admit Minkowski vacua that preserve all eight supercharges while breaking the diagonal gauge symmetry.

The paper is organized as follows. Section~\ref{sec:anomaly-consistency} develops the anomaly polynomial, its factorization, the unimodularity conditions, the action and the Green-Schwarz cancellation mechanism. Section~\ref{sec:search-strategy} describes the bounded search and summarizes the anomaly-free $G_1\times G_+$ and $G_1\times G_2\times G_+$ models where $G_+$ is either $U(1)_{R+}$ or $Sp(1)_{R+}$. Section~\ref{sec:vacua} studies the scalar potential, the embedding of the diagonal generator, and the resulting maximally symmetric vacua. Section~\ref{sec:charge-bound-wolf-general-unified45} derives the charge bound for the existence of the Minkowski$_6$ vacua and the R-symmetry breaking result, and Section~\ref{sec:conclusions} gives our conclusions. Appendix~\ref{app:conventions} collects our conventions. Appendices~\ref{app:abelian-spectra} and~\ref{sp:app:spectra} list selected $U(1)_{R+}$ and $Sp(1)_{R+}$ spectra, while Appendix~\ref{app:maximal-symmetry} lists further Minkowski vacua and their unbroken groups. 

\section{Anomaly polynomial and consistency conditions}
\label{sec:anomaly-consistency}

In this section, we shall begin with the description of the model and the building blocks needed for the description of the anomalies. We shall then impose the consistency conditions for the absence of the local and global anomalies, and for ghost-free gauge kinetic terms, along the lines discussed in detail in \cite{Becker:2025xgy} in the context of $U(1)_R$ gauged models in $6D$. 

\subsection{The model and building blocks for the anomalies}
\label{sec:anomaly-building-blocks}

The model consists of the $N=(1,0)$ combined supergravity multiplet and a single tensor multiplet $\big(g_{\mu\nu},\psi_{\mu+},B_{\mu\nu},\chi_-,\phi\big)$, $n_V$ vector multiplets $\big(\lambda_+,A_\mu\big)$ associated with the gauge group $G$, and $n_H$ hypermultiplets $\big(\zeta_-,\varphi\big)$. The fermions are symplectic Majorana--Weyl, or equivalently complex Weyl, with their chiralities denoted by $\pm$.

We consider the two choices of diagonal factor in \eq{eq:group}. The hypermultiplet representation blocks are written as
\begin{subequations}
\begin{align}
\rho_H&=\bigoplus_I m_I(R_{I1},\ldots,R_{In};q_I),
&&G_+=U(1)_{R+},\label{eq:spec}\\
\rho_H&=\bigoplus_I m_I(R_{I1},\ldots,R_{In};R_{I+}),
&&G_+=Sp(1)_{R+}.\label{sp:eq:spec}
\end{align}
\end{subequations}
Here $I$ labels the block, $m_I\in\mathbb Z_{\geq0}$ counts its copies, and $w_I$ is the weight of each copy in full-hypermultiplet units. The $R_{Ii}$ are irreducible representations of $G_i$, and $q_I$ or $R_{I+}$ specifies the hyperino representation of the diagonal factor. For $U(1)_{R+}$, we also write $R_{I+}$ for the one-dimensional representation of charge $q_I$. Trivial representations are allowed. We also define
\begin{equation}
d_{Ii}=\dim R_{Ii}\ ,\quad
d_{I+}=\dim R_{I+}\ ,
\quad
w_I=\begin{cases}1,&\text{full hypermultiplet},\\ \frac12,&\text{half-hypermultiplet}.\end{cases}
\end{equation}
The numbers $n_V$ of vector multiplets and $n_H$ of hypermultiplets are given by 
\begin{equation}
n_V=\dim G_++\sum_i\dim G_i,\qquad
n_H=\sum_I m_Iw_Id_{I+}\prod_jd_{Ij}.
\label{sp:eq:multiplet-counts}
\end{equation}

The reality condition is imposed separately on each hypermultiplet
block. Let $R_I$ denote the complete irreducible product representation
specified by that block, including its representation under the
diagonal gauge factor. A full hypermultiplet labelled by $R_I$ has
hyperinos valued in the symplectic completion
$R_I\oplus\overline{R}_I$, of complex dimension $2\dim R_I$. The resulting complete symplectic hyperino representation $\mathcal R_H$ has complex dimension $2n_H$. Imposing the symplectic reality condition halves the
independent fermionic degrees of freedom, giving the equivalent of
$n_H$ complex Weyl fermions.
A half-hypermultiplet instead has hyperinos valued in $R_I$ itself
and is allowed only when $R_I$ is pseudoreal. Thus the irreducible
representations used to label the blocks of $\rho_H$ need not
individually be pseudoreal; the symplectic structure belongs to
their corresponding hyperino representations. Their contributions
to $n_H$ are respectively $\dim R_I$ and $\tfrac12\dim R_I$,
as encoded by the weights $w_I$.

For illustration, consider $E_7\times SU(3)\times U(1)_{R+}$.
Although the $\mathbf{56}$ of $E_7$ is pseudoreal, its product
with the complex $\mathbf{3}$ of $SU(3)$, or with a nonzero
abelian charge, is complex. The examples are summarized in
Table~\ref{tab:block-reality-examples}.
\begin{table}[htbp]
\centering
\renewcommand{\arraystretch}{1.2}
\begin{tabular}{ccc}
\hline
Block $R_I$ & Conjugate $\overline{R}_I$ & Reality type \\
\hline
$(\mathbf{56},\mathbf{3})_0$
  & $(\mathbf{56},\overline{\mathbf{3}})_0$
  & Complex \\
$(\mathbf{56},\mathbf{1})_1$
  & $(\mathbf{56},\mathbf{1})_{-1}$
  & Complex \\
$(\mathbf{133},\mathbf{3})_0$
  & $(\mathbf{133},\overline{\mathbf{3}})_0$
  & Complex \\
$(\mathbf{56},\mathbf{1})_0$
  & Equivalent to itself
  & Pseudoreal \\
\hline
\end{tabular}
\caption{Reality types of irreducible product representations
of $E_7\times SU(3)\times U(1)_{R+}$.
Each of the first three blocks requires the symplectic completion
$R_I\oplus\overline{R}_I$ and contributes a full hypermultiplet.
The fourth also permits a half-hypermultiplet.}
\label{tab:block-reality-examples}
\end{table}
For $U(1)_{R+}$, we have $d_{I+}=1$. A half-hypermultiplet in an irreducible product block requires $q_I=0$ and a pseudoreal nonabelian product representation. For $q_I\ne0$, the $U(1)_{R+}$ factor makes the full product representation complex, even when its nonabelian part $R$ is pseudoreal. Charged matter must therefore be counted as full hypermultiplets, with $w_I=1$ and contribution $\dim R$ to $n_H$ per copy \cite{Suzuki:2005vu,Raghuram:2021wvj}.

For $Sp(1)_{R+}$, an irreducible representation is denoted by its dimension $\mathbf d$ and has spin $(d-1)/2$. Odd-dimensional irreducible representations are real and even-dimensional ones are pseudoreal. A real product representation of the additional factors therefore permits half-hypermultiplets for even $d_{I+}$, a pseudoreal product for odd $d_{I+}$, and a complex product requires full hypermultiplets. The reality types of all additional factors must first be combined; two pseudoreal factors have a real product \cite{Suzuki:2005vu}. Hyperscalars carry the additional $Sp(1)_R$ doublet index.

The single-factor indices are defined by
\begin{subequations}\label{abc}
\begin{align}
\tr_{R_{Ii}}F_i^2&=A^{(i)}_{R_{Ii}}\tr_{\fund_i}F_i^2,\\
\tr_{R_{Ii}}F_i^4&=B^{(i)}_{R_{Ii}}\tr_{\fund_i}F_i^4+C^{(i)}_{R_{Ii}}(\tr_{\fund_i}F_i^2)^2.
\end{align}
\end{subequations}
All gauge generators are anti-Hermitian. For groups without an independent quartic Casimir, we set $B_R=0$ and include the whole quartic trace in $C_R$. For $D_4$, there are two independent primitive quartic invariants, represented by $\tr_{8_v}F^4$ and the Pfaffian $\Pf(F)$. We therefore write
\begin{equation}
D_4:\qquad \tr_RF^4=B_R\tr_{8_v}F^4+C_R(\tr_{8_v}F^2)^2+P_R\Pf(F).
\end{equation}
We normalize $\Pf(F_i)=24x_1x_2x_3x_4$, where $\pm ix_a$ are the vector Cartan eigenvalues, so that $P_{\mathbf8_s}=1/2$ and $P_{\mathbf8_c}=-1/2$.

From \eq{abc}, the diagonal group gives
\begin{subequations}
\begin{align}
U(1)_{R+}:\quad &\tr_{R_{I+}}F_{R+}^2=q_I^2\tr_{\fund_+}F_{R+}^2,\quad
\nonumber\\
&\tr_{R_{I+}}F_{R+}^4=q_I^4(\tr_{\fund_+}F_{R+}^2)^2,
\\
Sp(1)_{R+}:\quad &\tr_{R_{I+}}F_{R+}^2=\frac16d_{I+}(d^2_{I+}-1)\tr_{\fund_+}F_{R+}^2,\quad 
\nonumber\\
&\tr_{R_{I+}}F_{R+}^4=\frac{1}{60}d_{I+}(d_{I+}^2-1)(3d^2_{I+}-7)(\tr_{\fund_+}F_{R+}^2)^2,
\end{align}
\end{subequations}
Explicitly, 
\begin{equation}
\left(A_{R_{I+}},C_{R_{I+}}\right)=
\begin{cases}
(q_I^2,q_I^4)\ ,\ &G_+=U(1)_{R+},\\
\displaystyle\left(\frac16d_{I+}(d_{I+}^2-1),
\frac{1}{60}d_{I+}(d_{I+}^2-1)(3d_{I+}^2-7)\right)\ ,\ &G_+=Sp(1)_{R+}.
\end{cases}
\label{sp:eq:su2-indices}
\end{equation}
For each additional simple gauge factor, the quadratic, quartic and reducible quartic sums are
\begin{align}
A_H^{(i)}&=\sum_Iw_Im_Id_{I+}A^{(i)}_{R_{Ii}}\prod_{j\ne i}d_{Ij},\nonumber\\
B_H^{(i)}&=\sum_Iw_Im_Id_{I+}B^{(i)}_{R_{Ii}}\prod_{j\ne i}d_{Ij},\nonumber\\
C_H^{(i)}&=\sum_Iw_Im_Id_{I+}C^{(i)}_{R_{Ii}}\prod_{j\ne i}d_{Ij}.
\end{align}
The sum $P_H^{(i)}$ uses the same weights as $A_H^{(i)}$. We shall refer to the above entities as the weighted index sums. For the diagonal gauge group $G_+$, the reference representation $\fund_+$ of $G_+$ is the charge-one representation for $U(1)_{R+}$ and the doublet for $Sp(1)_{R+}$. 

For either choice of $G_+$, the weighted sums are
\begin{subequations}\label{sp:eq:S2S4}
\begin{align}
S_2&=\sum_Iw_Im_IA_{R_{I+}}\prod_jd_{Ij},\label{sp:eq:S2}\\
S_4&=\sum_Iw_Im_IC_{R_{I+}}\prod_jd_{Ij}.\label{sp:eq:S4}
\end{align}
\end{subequations}
\begin{equation}
A_H^{(i,+)}=\sum_Iw_Im_IA_{R_{I+}}A^{(i)}_{R_{Ii}}\prod_{j\ne i}d_{Ij}.
\end{equation}
As we shall see below, these arise in the computation of mixed gravitational-$G_+$, pure $G_+$ and gauge-$G_+$ mixed anomalies, respectively. For $U(1)_{R+}$ and an additional simple gauge factor $G_i$ with a cubic invariant, define
\begin{equation}
E_q^{(i)}=\sum_Iw_Im_Iq_IE^{(i)}_{R_{Ii}}\prod_{j\ne i}d_{Ij},
\label{cubic}
\end{equation}
where
\begin{equation}
\tr_{R_{Ii}}F_i^3=E^{(i)}_{R_{Ii}}\tr_{\fund_i}F_i^3,
\qquad E^{(i)}_{\overline{R_{Ii}}}=-E^{(i)}_{R_{Ii}}.
\end{equation}
No $F_{R+}\tr F_i^3$ term occurs for $Sp(1)_{R+}$ because every $Sp(1)$ generator is traceless.

\subsection{The anomaly polynomial and its factorization conditions}
\label{sec:anomaly-factorization}

Using the spin-$1/2$ and spin-$3/2$ anomaly formulae of
\cite{Alvarez-Gaume:1983ihn,Alvarez-Gaume:1984zlq} (see \cite{Bilal:2008qx} for a review), the irreducible gravitational anomaly cancels in either theory when
\begin{equation}
n_H=n_V+244.
\label{eq:grav-anomaly}
\end{equation}
We impose this condition in the full polynomials below, retaining the irreducible gauge terms. %
We write explicitly the contributions to the anomaly polynomial involving the $G_+$ dependence; the weighted hyperino trace contains only the curvature of the additional simple gauge factors:
\begin{equation}
\tr_{\rho_H}F^k:=\sum_Iw_Im_Id_{I+}
\tr_{R_{I1}\otimes\cdots\otimes R_{In}}
\left(\sum_iF_i\right)^k,\qquad k=2,4.    
\end{equation}
Inside this trace $F_i$ acts on the $i$-th tensor factor and as the identity on the others. Recall that $d_{I+}=1$ for the abelian branch; wedge products are understood.

\paragraph{The $U(1)_{R+}$ polynomial.}
The full expression is
\begin{align}
\frac{1}{2\pi}I_8=\frac1{(16\pi^2)^2}\Bigg[&
(\tr R^2)^2+\frac{\tr R^2}{6}\Big((20-n_V+S_2)F_+^2
+\sum_i\tr_{\adj_i}F_i^2-\tr_{\rho_H}F^2\Big)\nonumber\\
&+\frac23\Big(-(n_V+4-S_4)F_+^4-\sum_i\tr_{\adj_i}F_i^4
+\tr_{\rho_H}F^4\nonumber\\
&-6F_+^2\Big(\sum_i A_H^{(i,+)}\tr_{\fund_i}F_i^2
-\sum_i\tr_{\adj_i}F_i^2\Big)
-4iF_+\sum_iE_q^{(i)}\tr_{\fund_i}F_i^3\Big)\Bigg],
\label{ap:3}
\end{align}
where $F_+$ is a real two-form and we have made the replacements 
\be
\qquad F_{R+} \to -iF_+\ ,\qquad \tr_{\fund_+}F_{R+}^2 \to -F_+^2\ .
\label{abelian}
\ee

\paragraph{The $Sp(1)_{R+}$ polynomial.}

The diagonal gauginos transform in $\mathbf3\otimes\mathbf2=\mathbf4\oplus\mathbf2$, and the $G_i$ gauginos in $(\adj_i,\mathbf2)$. Including their symplectic Majorana--Weyl factors, the gauginos, gravitino, tensorino and hyperini give
\begin{align}
\frac{1}{2\pi}I_8=\frac1{(16\pi^2)^2}\Bigg[&
(\tr R^2)^2+\frac{\tr R^2}{6}\Bigg(
\Big(\frac{n_V-12}{2}-S_2\Big)\tr_{\fund_+}F_{R+}^2
+\sum_i\tr_{\adj_i}F_i^2-\tr_{\rho_H}F^2\Bigg)\nonumber\\
&+\frac23\Bigg(
\Big(S_4-\frac{n_V+84}{4}\Big)(\tr_{\fund_+}F_{R+}^2)^2
-\sum_i\tr_{\adj_i}F_i^4+\tr_{\rho_H}F^4\nonumber\\
&+6\tr_{\fund_+}F_{R+}^2
\Big(\sum_iA_H^{(i,+)}\tr_{\fund_i}F_i^2
-\frac12\sum_i\tr_{\adj_i}F_i^2\Big)\Bigg)\Bigg].
\label{sp:ap:3}
\end{align}
The diagonal constants reproduce \cite{Suzuki:2005vu}, with the additional charged-hyperino terms appropriate to diagonal gauging.

Cancellation of the irreducible gauge anomalies requires
\begin{equation}
B_H^{(i)}=B_{\adj_i}^{(i)},\qquad
E_q^{(i)}=0\quad\text{for }G_+=U(1)_{R+},
\label{c12}
\end{equation}
where the first condition applies whenever an independent quartic invariant exists. For $D_4$ one also imposes
\begin{equation}
P_H^{(i)}=P_{\adj_i}^{(i)}=0.
\label{c13}
\end{equation}
The mixed cubic condition has no $Sp(1)_{R+}$ counterpart. 

\paragraph{Factorization conditions.}

The remaining polynomial must factorize as
\begin{equation}
\frac{1}{2\pi}I_8=\frac12\eta_{\alpha\beta}Y^\alpha Y^\beta,
\label{CI8}
\end{equation}
where $\eta_{\a\b}$ is the $SO(1,1)$ invariant metric and the Green--Schwarz four-form is~\cite{Taylor:2011wt}
\begin{equation}
Y^\alpha=\frac1{16\pi^2}\left(\frac12a^\alpha\tr R^2
+2b_{R+}^\alpha\tr_{\fund_+}F_{R+}^2
+\sum_i\frac{2b_i^\alpha}{\lambda_i}\tr_{\fund_i}F_i^2\right)\ ,
\label{defY}
\end{equation}
where $\tr_{\fund_+}F_{R+}^2=-F_+^2$ for $G_+=U(1)_{R+}$. The diagonal normalization is $\lambda_{R+}=1$ for both choices of $G_+$. The normalization factors for $G_i$ are given in Table~\ref{tab:lambda}.
\begin{table}[ht]
\centering\small
\renewcommand{\arraystretch}{1.12}
\begin{tabular}{|c|c|c|c|c|c|c|c|c|c|}
\hline
 & $A_n$ & $B_n$ & $C_n$ & $D_n$ & $E_6$ & $E_7$ & $E_8$ & $F_4$ & $G_2$\\
\hline
$\lambda_z$ & 1 & 2 & 1 & 2 & 6 & 12 & 60 & 6 & 2\\
\hline
\end{tabular}
\caption{Normalization factors $\lambda$.}
\label{tab:lambda}
\end{table}

When these conditions are satisfied, both reducible polynomials take the common form
\begin{align}
\frac{1}{2\pi}I_8=\frac1{(16\pi^2)^2}\Big[&
(\tr R^2)^2+\alpha\,\tr R^2\tr_{\fund_+}F_{R+}^2
+\sum_i\beta_i\,\tr R^2\tr_{\fund_i}F_i^2\nonumber\\
&+\gamma\,(\tr_{\fund_+}F_{R+}^2)^2
+\sum_i\delta_i\,\tr_{\fund_+}F_{R+}^2\tr_{\fund_i}F_i^2
+\sum_i\epsilon_{ii}(\tr_{\fund_i}F_i^2)^2\nonumber\\
&+\sum_{i<j}\epsilon_{ij}\tr_{\fund_i}F_i^2\tr_{\fund_j}F_j^2\Big].
\label{eq:I8-product-reducible}
\end{align}
Direct expansion of \eqref{CI8} yields the following matching equations for either diagonal group:
\begin{align}
a\cdot a&=8\ ,
\nonumber\\
a\cdot b_{R+}&=\alpha,\nonumber\\
2b_{R+}\cdot b_{R+}&=\gamma,\nonumber\\
\frac{a\cdot b_i}{\lambda_i}&=\beta_i,\nonumber\\
\frac{2b_i\cdot b_i}{\lambda_i^2}&=\epsilon_{ii},\nonumber\\
\frac{4b_i\cdot b_{R+}}{\lambda_i}&=\delta_i,\nonumber\\
\frac{4b_i\cdot b_j}{\lambda_i\lambda_j}&=\epsilon_{ij},\qquad i<j.
\label{ae}
\end{align}
The shared $G_i$ coefficients are
\begin{align}
\beta_i&=\frac16\left(A_{\adj_i}^{(i)}-A_H^{(i)}\right),\nonumber\\
\epsilon_{ii}&=\frac23\left(C_H^{(i)}-C_{\adj_i}^{(i)}\right),\nonumber\\
\epsilon_{ij}&=4\sum_Iw_Im_Id_{I+}
A_{R_{Ii}}^{(i)}A_{R_{Ij}}^{(j)}\prod_{k\ne i,j}d_{Ik},\qquad i<j.
\label{eq:beta-product}
\end{align}
The last coefficient is generated only by hypermultiplets charged under both $G_i$ and $G_j$.

The remaining coefficients are
\begin{equation}
\begin{alignedat}[b]{2}
U(1)_{R+}:\quad\alpha&=\frac16(n_V-20-S_2),\quad&
Sp(1)_{R+}:\quad\alpha&=\frac{1}{12}(n_V-12-2S_2),\\
\gamma&=\frac23(S_4-n_V-4),\quad&
\gamma&=\frac16(4S_4-n_V-84),\\
\delta_i&=4\left(A_H^{(i,+)}-A_{\mathrm{adj}_i}^{(i)}\right),\quad&
\delta_i&=4\left(A_H^{(i,+)}-\frac12A_{\mathrm{adj}_i}^{(i)}\right).
\end{alignedat}
\label{eq:alpha-gamma}
\end{equation}

Equations \eqref{ae}, \eqref{eq:beta-product} and \eqref{eq:alpha-gamma} determine the Gram data of the anomaly vectors from the spectral sums defined in Section~\ref{sec:anomaly-building-blocks}. Their embedding in a unimodular string-charge lattice is the subject of the next subsection.

\subsection{Global consistency and lattice constraints}
\label{sec:Sp-solving-anomalies}

For the global-anomaly analysis, we take $G$ to be a direct product of $U(1)$ factors and simply connected simple compact groups~\cite{Becker:2025xgy}. We now impose the requirement that the anomaly vectors belong to a common integral unimodular string-charge lattice. Consistency also requires that the gravitational anomaly coefficient must be a characteristic element of the string-charge
lattice~\cite{Seiberg:2011dr,Monnier:2017oqd,Lee:2020ewl,Monnier:2018},
\begin{equation}
 a\cdot x\equiv x\cdot x\pmod 2
 \qquad \text{for every }x\in\Lambda ,
\label{eq:characteristic-element}
\end{equation}
where $a$ is defined in \eqref{defY} and $\Lambda$ is the string charge lattice. This condition will be manifestly satisfied for all models we consider in this paper. 

\paragraph{Charge lattice and global anomalies.}
We use the light-cone convention
\begin{equation}
 x\cdot y=x^+y^-+x^-y^+,\qquad a=(-2,-2).
 \label{eq:lightcone}
\end{equation}
The vectors
\begin{equation}
 \mathcal V=\begin{cases}
 \{a,b_i,\frac12b_{R+}\},&G_+=U(1)_{R+},\\
 \{a,b_i,b_{R+}\},&G_+=Sp(1)_{R+}
 \end{cases}
 \label{eq:V}
\end{equation}
must lie in one integral unimodular lattice $\Lambda$ of signature $(1,1)$, with $a$ characteristic as in \eqref{eq:characteristic-element}. In the abelian case, the lattice vector is $\bar b_{R+}:=b_{R+}/2$, as required by the minimally normalized abelian characteristic class. In the light-cone convention \eqref{eq:lightcone}, the only two unimodular lattices, up to isometry discussed further below, are represented by
\begin{equation}
 U=\Z(1,0)\oplus\Z(0,1),\qquad
 I_{1,1}=\Z(1,\tfrac12)\oplus\Z(1,-\tfrac12).
 \label{eq:lattices}
\end{equation}
Here $\Z u\oplus\Z v$ means the set of all integer linear combinations $mu+nv$, with $m,n\in\Z$. The basis vectors $u=(1,0)$ and $v=(0,1)$ of $U$ have $u^2=v^2=0$ and $u\cdot v=1$, so their Gram matrix is $\left(\begin{smallmatrix}0&1\\1&0\end{smallmatrix}\right)$. Thus $U$ is the even hyperbolic plane: every vector has even norm $2mn$. For $I_{1,1}$, the basis vectors $e=(1,\tfrac12)$ and $f=(1,-\tfrac12)$ obey $e^2=1$, $f^2=-1$ and $e\cdot f=0$. Their Gram matrix is $\operatorname{diag}(1,-1)$, and integer combinations have norm $m^2-n^2$, which may be odd. Both Gram matrices have determinant $-1$, so both lattices are integral and unimodular with signature $(1,1)$; the half-integer light-cone coordinates of $I_{1,1}$ do not imply fractional lattice pairings.

The vector $a=(-2,-2)$ is characteristic in both realizations: $a=-2u-2v$ in $U$ and $a=-3e+f$ in $I_{1,1}$. Exchanging $x^+$ and $x^-$ leaves $a$ fixed and gives the isometric odd-lattice realization $\Z(\tfrac12,1)\oplus\Z(-\tfrac12,1)$. All vectors in $\mathcal V$ must lie simultaneously in one chosen realization of $U$ or $I_{1,1}$; the choice cannot be made separately for each vector.

For the collection $v_A\in\mathcal V$, the Gram matrix is $\mathcal G_{AB}=v_A\cdot v_B$, with entries fixed by \eqref{ae}. In the abelian case, the entries involving $\bar b_{R+}$ are
\begin{align}
 a\cdot\bar b_{R+}=\frac{\alpha}{2},\qquad
 b_i\cdot\bar b_{R+}=\frac{\lambda_i\delta_i}{8},\qquad
 \bar b_{R+}^{\,2}=\frac{\gamma}{8}.
\end{align}
Integral lattice membership requires integral Gram entries and $a\cdot v_A\equiv v_A^2\pmod2$. The Gram matrix must also have rank at most two, with at most one positive and one negative eigenvalue. For any pair of vectors in a unimodular lattice of signature $(1,1)$, minus their Gram determinant is an integer square: if their integer basis-coordinate matrix is $M$, that determinant is $-(\det M)^2$. These are necessary filters; an explicit common embedding in \eqref{eq:lattices} enforces the full lattice requirement. 

The following divisibility conditions for diagonal $Sp(1)_{R+}$ and the additional exceptional group $G_2$ already follow from integral lattice norms~\cite{Bershadsky:1997,Suzuki:2005vu,Avramis:2005qt,Lee:2020ewl}:
\begin{align}
 4S_4-n_V-84=12b_{R+}^2\quad&\Longrightarrow\quad
 4S_4\equiv n_V\pmod{12},\label{eq:Spglobal}\\
 G_2:\quad b_i^2=\frac{4C_H^{(i)}-10}{3}\quad&\Longrightarrow\quad
 1-4C_H^{(i)}\equiv0\pmod3.\label{eq:Gglobal}
\end{align}
Here $S_4$ is defined in \eqref{sp:eq:S4}, and $C_H^{(i)}$ is the reducible quartic weighted sum of Section~\ref{sec:anomaly-building-blocks}.

\paragraph{Diagonal anomaly-vector constraints.}
Writing $b_{R+}=(r,s)$, the diagonal equations in \eqref{ae}, with $\alpha,\gamma$ given in Section~\ref{sec:anomaly-factorization}, yield
\begin{equation}
 (r,s)=\frac14(-\alpha+\sqrt\Delta,-\alpha-\sqrt\Delta),
 \qquad\Delta=\alpha^2-4\gamma.
 \label{eq:diag}
\end{equation}
For $U(1)_{R+}$, integrality, the characteristic condition \eqref{eq:characteristic-element}, and the Gram determinant condition applied to $a,\frac12b_{R+}$ require
\begin{equation}
 n_V-S_2\equiv8\pmod{12},\qquad S_4-S_2\equiv0\pmod{24},
 \qquad\Delta=4k^2,\quad k\in\Z_{\ge0}.
 \label{eq:Uarith}
\end{equation}
The last condition in \eqref{eq:Uarith} follows because $-\det\operatorname{Gram}(a,b_{R+}/2)=\Delta/4$ must be a perfect square for two vectors in a unimodular lattice of signature $(1,1)$.

For $Sp(1)_{R+}$, the same determinant test is applied to $a,b_{R+}$ and requires
\begin{align}
 \alpha\in\Z,\qquad -\det\operatorname{Gram}(a,b_{R+})
 =\Delta=k^2,\qquad k\in\Z_{\ge0}.
\end{align}
These conditions follow from membership in an integral unimodular lattice. They are necessary tests on the diagonal pair; the complete collection \eqref{eq:V} must still admit a common embedding in one of the lattice realizations \eqref{eq:lattices}. 

\subsection{\texorpdfstring{The $Sp(1)_{R+}$ gauged model}{The Sp(1)R+ gauged model}}
\label{sec:sp1-gauged-model}

The action describing the most general two-derivative couplings of $N=(1,0), 6D$ supergravity was constructed long ago \cite{Nishino:1986dc}. The action is gauge invariant and supersymmetric, since it does not include the Green-Schwarz anomaly counterterm. Its inclusion was considered in \cite{Nishino:1997ff} at the level of equations of motion, and in \cite{Riccioni:2001bg} at the level of pseudo-action involving $n_T$ tensor multiplets, so-called because the self-duality equations for the three-form field strengths do not follow from the action. A proper two-derivative action for $n_T=1$ in the presence of the GS counterterm was obtained in \cite{Bossard:2024ffp}.

We display the two-derivative bosonic action together with the Lorentz Chern--Simons and Green--Schwarz terms needed for anomaly cancellation for the $Sp(1)_{R+}$ gauged model; the remaining higher-derivative completion is not displayed:
\begin{align}
S &= \int_{M_6}\Big[\frac14R\star1-\frac14d\phi\wedge\star d\phi
-\frac12e^{2\phi}H\wedge\star H-\frac12f_zF_z^\ell\wedge\star F_z^\ell\nonumber\\
&\hspace{24mm}-\frac12g_{xy}D\varphi^x\wedge\star D\varphi^y-V\star1\Big]+S_{\rm top},
\label{Lagrangian}\\
S_{\rm top}&=-\int_{M_6}\left(B\wedge X_4^-+\frac12\Omega_3^+\wedge\Omega_3^-\right).
\label{eq:common-topological-action}
\end{align}
We use one common component form for both diagonal gaugings. The diagonal sum has three terms for $Sp(1)_{R+}$ and one for $U(1)_{R+}$; the abelian structure constants vanish. The conversion to the charge-one abelian normalization used elsewhere is given in \eqref{eq:abelian-action-rescaling}.

For each nonabelian gauge factor we take 
\begin{equation}
\tr_{\fund_z}(T_z^\ell T_z^m)=-\frac{\lambda_z}{2}\delta^{\ell m},\qquad
T_R^r=-\frac{i}{2}\sigma^r.
\label{sp:eq:generator-normalization}
\end{equation}
Thus $[T_R^r,T_R^s]=\epsilon^{rs}{}_{t}T_R^t$, and $T_R=-i\sigma^3/2$ for the abelian embedding. The indices $\ell,m,\ldots$ label general gauge adjoints, while $r,s,...$ label $Sp(1)_R$.
Here $\varphi^x$ are coordinates on \eqref{coset}, $g_{xy}$ is the metric on $\mathcal M_H$, and
\begin{align}
H&=dB-\Omega_3^+,\qquad
\Omega_3^\pm=v_z^\pm\omega_{3z}-u^\pm\omega_{3L},\qquad
X_4^\pm=d\Omega_3^\pm,
\label{eq:common-tensor-definitions}\\
D\varphi^x&=d\varphi^x-A_+^r(T_{R+}^r\varphi)^x-A_{z'}^I(T_{z'}^I\varphi)^x,\qquad
f_z=v_z^+e^\phi+v_z^-e^{-\phi}.
\label{fz}
\end{align}
For the diagonal factor $z=+$, we denote the coupling coefficients by $v_{R+}^{\pm}$. The Chern-Simons forms are defined by
\begin{align}
\omega_{3z}&=\delta_{\ell m}\left(A_z^\ell\wedge dA_z^m+\frac13f^m{}_{np}A_z^\ell\wedge A_z^n\wedge A_z^p\right),\quad
\omega_{3L}=\tr\left(\omega\wedge d\omega+\frac23\omega\wedge\omega\wedge\omega\right).
\label{eq:common-cs-forms}
\end{align}
The curvature is
\begin{align}
F_z^\ell&=dA_z^\ell+\frac12f^\ell{}_{mn}A_z^m\wedge A_z^n.
\label{eq:common-curvature}
\end{align}
Turning to the potential, it is given by
\begin{align}
V&=-\frac{1}{4f_+}(C_+^r)_A{}^B(C_+^r)_B{}^A-\sum_{z',I}\frac1{4f_{z'}}(C_{z'}^I)_A{}^B(C_{z'}^I)_B{}^A,
\label{pot}
\end{align}
where $C_+^r$ and $C_{z'}^I$ are the moment maps defined as
\begin{align}
(C_+^r)_A{}^B&=\left[L^{-1}(T_H^r+T_R^r)L\right]_A{}^B,\qquad
(C_{z'}^I)_A{}^B=\left[L^{-1}T_{z'}^I L\right]_A{}^B,
\label{eq:Ci}
\end{align}
where $L$ represents the quaternionic K\"ahler coset with $4n_H$ real hyperscalars, and $T_{z'}^I$ generates $K\subset H$. The indices $a,b=1,\ldots,2n_H$ are reserved for the hyperini and $A, B$ run over $1, 2$, transforming in the ${\bf 2_R}$. The spinor conjugation conventions are collected in Appendix~\ref{app:conventions}. At two-derivative order, the fermion supersymmetry transformations, up to terms cubic in fermions, are
\begin{align}
\delta\psi_\mu^A&=D_\mu\epsilon^A-\frac1{24}e^\phi H_{\nu\rho\sigma}\gamma^{\nu\rho\sigma}\gamma_\mu\epsilon^A,
\nonumber\\
\delta\chi^A&=\frac12\partial_\mu\phi\gamma^\mu\epsilon^A+\frac1{12}e^\phi H_{\mu\nu\rho}\gamma^{\mu\nu\rho}\epsilon^A,
\nonumber\\
\delta\lambda_+^{rA}&=-\frac14F_{+\mu\nu}^r\gamma^{\mu\nu}\epsilon^A+\frac1{2f_+}(C_+^r)^{AB}\epsilon_B,
\nonumber\\
\delta\lambda_{z'}^{I A}&=-\frac14F_{z'\mu\nu}^I\gamma^{\mu\nu}\epsilon^A+\frac1{2f_{z'}}(C_{z'}^I)^{AB}\epsilon_B,
\nonumber\\
\delta\lambda_{z''}^{P A}&=-\frac14F_{z''\mu\nu}^P\gamma^{\mu\nu}\epsilon^A,
\nonumber\\
\delta\zeta^a&=-\gamma^\mu\epsilon_A D_\mu\varphi^xV_x^{aA}.
\label{susy}
\end{align}
Here $\lambda_+^{rA}$, $\lambda_{z'}^{I A}$ and $\lambda_{z''}^{P A}$ are the gauginos of $G_+$, $K$ and $G_E$, respectively; $z''$ labels the factors of $G_E$, whose moment maps vanish identically because $G_E$ acts trivially on the hypermultiplet manifold. $V_x^{aA}$ is the vielbein on ${\cal M}_H$. 

Turning to the relation between $v_z^\a$ and the anomaly vectors $b^\a_z$, we compare the classical and one-loop variations. The field strength $H$ is invariant under the gauge and local Lorentz transformations
\begin{equation}
\delta B=v_z^+\delta_{\ell m}\Lambda_z^\ell dA_z^m-u^+\tr(\Theta d\omega)=\Omega_2^{+(1)},\qquad
\delta\Omega_3^\pm=d\Omega_2^{\pm(1)}.
\label{eq:common-tensor-gauge-variation}
\end{equation}
The topological action varies as
\begin{align}
\delta S_{\rm top}&=-\frac12\int_{M_6}\left(\Omega_2^{+(1)}\wedge X_4^-+X_4^+\wedge\Omega_2^{-(1)}\right).
\label{eq:common-topological-variation}
\end{align}
On the other hand, the 1-loop anomaly follows from $I_8$ given in \eqref{CI8} by descent equations,
\begin{align}
I_8&=dI_7^0,\qquad \delta I_7^0=dI_6^1,\qquad
\delta\Gamma_{\rm 1-loop}=\int_{M_6}I_6^1.
\label{eq:common-anomaly-descent}
\end{align}
The anomaly freedom requires that $\delta\Gamma_{\rm 1-loop}+\delta S_{\rm top}=0$. Introducing the constants $\b_\pm$ by
\begin{equation}
Y_4^\pm=\beta_\pm X_4^\pm,
\label{com}
\end{equation}
the anomaly cancellation takes place for
\begin{equation}
a^\pm=-32\pi^2\beta_\pm u^\pm,\quad
b_i^\pm=-16\pi^2\beta_\pm v_i^\pm,\quad
b_{R+}^\pm=-16\pi^2\beta_\pm v_{R+}^\pm,\quad
\beta_+\beta_-=\frac1{2\pi},
\label{rs}
\end{equation}
where $i$ labels the additional simple gauge factors and $b_{R+}$ is the diagonal $Sp(1)_{R+}$ anomaly vector. We choose $\beta_+=\beta_-=-1/\sqrt{2\pi}$, so $Y_4^\pm=-X_4^\pm/\sqrt{2\pi}$ and $I_8=X_4^+\wedge X_4^-$. All gauge kinetic functions must be positive at a common value of $\phi$.

Turning to the requirement of ghost-free gauge kinetic terms, this will always be possible for all the models we shall find, either for any value of the dilaton, or in a nonempty interval determined by the anomaly coefficients. In the presence of abelian factors in the gauge group $K\subset H$, this requires that there are no spectator abelian group factors under which all matter is neutral~\cite{Park:2011wv}. 

We note that higher-derivative positivity bounds depend on the background and assumptions, so we do not impose $j\cdot a<0$ as a universal consistency condition~\cite{Ozkan:2024euj}. It holds automatically in our branch: $j=(e^{-\phi},e^\phi)/\sqrt2$ and $a=(-2,-2)$ give $j\cdot a=-\sqrt2(e^\phi+e^{-\phi})<0$, adding no restriction to the search.

\subsection{\texorpdfstring{The $U(1)_{R+}$ gauged model}{The U(1)R+ gauged model}}
\label{sec:u1-gauged-model}

We now adapt the preceding formulas to $U(1)_{R+}$ gauging, retaining a single diagonal generator $T_{R+}=T_H+T_R$, with $T_R=-i\sigma^3/2$, and one diagonal vector multiplet.

For charge-one normalization of the R-doublet fermions, make the following replacements in the preceding formulas, with the right-hand sides expressed in the new variables and $\Lambda_+$ the abelian gauge parameter.
\begin{equation}
\begin{aligned}
A_+&\longrightarrow2A_+,\quad F_+\longrightarrow2F_+,\quad
\lambda_+^{A}\longrightarrow2\lambda_+^{A},\quad
\Lambda_+\longrightarrow2\Lambda_+,\\
v_{R+}^{\pm}&\longrightarrow\frac14v_{R+}^{\pm},\quad
f_+\longrightarrow\frac14f_+,\quad
f_+=v_{R+}^+e^\phi+v_{R+}^-e^{-\phi}.
\end{aligned}
\label{eq:abelian-action-rescaling}
\end{equation}
The generators and moment maps retain their definitions in \eqref{eq:Ci}. The supersymmetry parameter, gravitino, tensorino and gaugini of the additional factors are unchanged.

The bosonic action retains the form \eqref{Lagrangian}--\eqref{eq:common-topological-action}, with a single diagonal kinetic term $-\tfrac12f_+F_+\wedge\star F_+$ and
\begin{align}
F_+&=dA_+,\quad
D\varphi^x=d\varphi^x-2A_+(T_{R+}\varphi)^x
-A_{z'}^I(T_{z'}^I\varphi)^x,
\label{eq:abelian-covariant-fields}\\
H&=dB-\Omega_3^+,\quad
\Omega_3^\pm=v_{R+}^\pm A_+\wedge dA_+
+\sum_{z\ne+}v_z^\pm\omega_{3z}-u^\pm\omega_{3L},
\label{eq:abelian-tensor-definitions}\\
V&=-\frac1{f_+}\tr_2(C_+^2)
-\sum_{z',I}\frac1{4f_{z'}}\tr_2\!\left[(C_{z'}^I)^2\right].
\label{eq:abelian-potential}
\end{align}
The tensor gauge transformation and Green--Schwarz terms retain their form under \eqref{eq:abelian-action-rescaling}. In the charge-one abelian normalization, with the anomaly vectors defined in \eqref{defY}, the diagonal anomaly--coupling relation is
\begin{equation}
 b_{R+}^\pm=-8\pi^2\beta_\pm v_{R+}^\pm.
 \label{eq:common-diagonal-b-v}
\end{equation}
We set $\epsilon=\epsilon^1$, $\epsilon^c=\epsilon^2$, and similarly for the gravitino, tensorino and gauginos; the conjugation convention is given in Appendix~\ref{app:conventions}. The complex spinor and its conjugate have opposite unit $U(1)_R$ charges.

The hyperini retain the $Sp(n_H)$ index $a=1,\ldots,2n_H$ used in \eqref{susy}; the symplectic reality condition leaves $n_H$ independent complex Weyl spinors. In this complex notation, the fermion transformations \eqref{susy} become
\begin{equation}
\begin{aligned}
 \delta\psi_\mu&=D_\mu\epsilon-\frac1{24}e^\phi
 H_{\nu\rho\sigma}\gamma^{\nu\rho\sigma}\gamma_\mu\epsilon,\\
 \delta\chi&=\frac12\partial_\mu\phi\gamma^\mu\epsilon
 +\frac1{12}e^\phi H_{\mu\nu\rho}\gamma^{\mu\nu\rho}\epsilon,\\
 \delta\lambda_+&=-\frac14F_{+\mu\nu}\gamma^{\mu\nu}\epsilon
 +\frac1{f_+}\bigl(C_+^{12}\epsilon-C_+^{11}\epsilon^c\bigr),\\
 \delta\lambda_{z'}^I&=-\frac14F_{z'\mu\nu}^I\gamma^{\mu\nu}\epsilon
 +\frac1{2f_{z'}}\bigl[(C_{z'}^I)^{12}\epsilon-(C_{z'}^I)^{11}\epsilon^c\bigr],\\
 \delta\lambda_{z''}^P&=-\frac14F_{z''\mu\nu}^P\gamma^{\mu\nu}\epsilon,\\
 \delta\zeta^{a}&=\gamma^\mu D_\mu\varphi^x
 \bigl(V_x^{a1}\epsilon^c-V_x^{a2}\epsilon\bigr).
\end{aligned}
\label{eq:abelian-gaugino-susy}
\end{equation}
Here $D_\mu\epsilon=D_\mu\epsilon^1$ includes the composite $Sp(1)_R$ connection; its off-diagonal component multiplies $\epsilon^2=\epsilon^c$. The conjugate transformations determine the remaining components.

At the hyperscalar origin $L=1$, the moment map has $(C_+)_A{}^B=-i\sigma^3/2$, so $C_+^{11}=C_+^{22}=0$ and $C_+^{12}=i/2$. The diagonal gaugino transformation therefore reduces to
\begin{equation}
 \left.\delta\lambda_+\right|_{\varphi=0}
 =-\frac14F_{+\mu\nu}\gamma^{\mu\nu}\epsilon
 +\frac{i}{2f_+}\epsilon,
 \qquad
 \left.D_\mu\epsilon\right|_{\varphi=0}
 =(\nabla_\mu-iA_{+\mu})\epsilon.
 \label{eq:abelian-zero-hyper-limit}
\end{equation}

\section{Search strategy and results}
\label{sec:search-strategy}

We implement the local anomaly and global lattice conditions of Section~\ref{sec:anomaly-consistency} for one or two additional simple factors. We use the multiplet counts and full-product reality conditions of Section~\ref{sec:anomaly-building-blocks}, with $n_H$ fixed by \eqref{eq:grav-anomaly} for $n_T=1$. Throughout the search, we exclude hypermultiplet blocks trivial under every additional factor $G_i$.

\paragraph{Diagonal multiplicity variables.}
For the bounded search, we collect the blocks of Section~\ref{sec:anomaly-building-blocks} according to their representation under the diagonal factor. Define
\begin{equation}
 U(1)_{R+}:\quad H_j=\sum_{|q_I|=j}w_Im_I\prod_kd_{Ik},\qquad
 Sp(1)_{R+}:\quad H_d=\sum_{d_{I+}=d}w_Im_I\prod_kd_{Ik}.
 \label{eq:H}
\end{equation}
Regrouping the spectral sums already defined in \eqref{sp:eq:S2S4} gives
\begin{subequations}\label{eq:diagonal-weighted-sums}
\begin{align}
 U(1)_{R+}:&\quad n_H=\sum_jH_j,\quad S_2=\sum_jj^2H_j,\quad S_4=\sum_jj^4H_j,\label{eq:Umom}\\
 Sp(1)_{R+}:&\quad n_H=\sum_ddH_d,\quad S_2=\sum_dA_dH_d,\quad S_4=\sum_dC_dH_d,\label{eq:Smom}
\end{align}
\end{subequations}
Thus $H_j$ is the total weighted multiplicity at absolute charge $j$, while $H_d$ is the weighted multiplicity in the $d$-dimensional diagonal representation, before multiplication by $d$ in $n_H$. Here $A_d=A_{R_{I+}}$ and $C_d=C_{R_{I+}}$ are the indices in \eqref{sp:eq:su2-indices} for $d=d_{I+}$. We denote the largest occupied labels by $q_{\max}=\max_I|q_I|$ and $d_{\max}=\max_I d_{I+}$, respectively, with the maxima taken over blocks with $m_I>0$.
The $H_j$ are integers; $H_d$ is integral for odd $d$ and may be half-integral for even $d$. The complete-representation reality condition is enforced when assigning diagonal labels.

\paragraph{Search bounds.}
We consider the following ranges:
\begin{equation}
G_1 \;:\;\;
\begin{array}{l}
3\le {\rm rank}\,G_1\le30\quad \mbox{or the exceptional group}\ G_2\ ,
\end{array}
\label{G1}
\end{equation}
for one additional simple gauge factor, and
\begin{equation}
G_1\times G_2 \;:\;\;
\begin{array}{l}
5\le {\rm rank}\,G_i \le 30\ \mbox{for}\ i=1,2\ \mbox{or}\ F_4 \ \mbox{for either factor\ }\ .
\end{array}
\label{G1xG2}
\end{equation}
All simple factors in these ranges are included. The lower rank cutoffs are chosen for simplicity, since trial searches below them produced very large numbers of spectra. For two additional factors we scan unordered pairs, including equal
factors and representations charged under both. To keep the search manageable, we restrict the $U(1)_{R+}$ charges and $Sp(1)_{R+}$ representations as follows: 
\begin{subequations}\label{eq:shared-diagonal-scan}
\begin{align}
q_{\max}&\leq2,\quad q_I\in\mathbb Z
\quad\text{for }U(1)_{R+},\label{eq:shared-diagonal-scan a}  \\
d_{\max}&\leq3
\quad\text{for }Sp(1)_{R+}.
\label{eq:shared-diagonal-scan b}
\end{align}
\end{subequations}
The representation search is finite because each nontrivial block obeys
\begin{equation}
 \prod_i d_{Ii}>1,\qquad
 w_Id_{I+}\prod_i d_{Ii}\leq n_H,\qquad
 m_I\leq\frac{n_H}{w_Id_{I+}\prod_i d_{Ii}}.
 \label{sp:eq:finite-Sp-scan}
\end{equation}
Here $d_{I+}=1$ for $U(1)_{R+}$, and $w_I$ is fixed by the full-product
reality conditions of Section~\ref{sec:anomaly-consistency}.
Each sector is labelled by its actual largest occupied charge or representation dimension.
The supplementary searches in Section~\ref{sec:larger-diagonal-representations} relax only the diagonal
bounds in \eqref{eq:shared-diagonal-scan}. For $U(1)_{R+}$, the cases $q_{\max}\leq1$ and $q_{\max}=2$
are disjoint; $q_{\max}=0$ is ordinary $U(1)_R$ gauging. For
$Sp(1)_{R+}$ we similarly separate $d_{\max}\leq2$ from $d_{\max}=3$. Higher-charge counting routines enforce the same conditions without storing every assignment.

\paragraph{Diagonal candidates and matter arithmetic.}
The diagonal vectors are determined by \eqref{eq:diag}; their lattice restrictions were derived in Section~\ref{sec:Sp-solving-anomalies}. For $Sp(1)_{R+}$, those restrictions, together with the identity $2C_d-A_d=12\binom{d+2}{5}$, give
\begin{equation}
 J:=\frac{2S_4-S_2}{12}\in\Z_{\ge0},\qquad
 \alpha\in\Z,\qquad\Delta=(\alpha+4)^2+48-16J=k^2,
 \label{eq:Sarith}
\end{equation}
where $k\in\Z_{\ge0}$. The integrality of $J$ also holds for allowed half-hypermultiplets: the dimension of their representation of the additional factors is even for odd $d$, while $\binom{d+2}{5}$ is even for even $d$.

Arithmetic candidates must also be realizable by nonnegative matter multiplicities. For labels $j=0,1,2$ or $d=1,2,3$, \eqref{eq:Umom}--\eqref{eq:Smom} fix
\begin{subequations}\label{eq:diagonal-multiplicities}
\begin{align}
 U(1)_{R+}:\quad&H_2=\frac{S_4-S_2}{12},\quad H_1=S_2-4H_2,
 \quad H_0=n_H-H_1-H_2,\label{eq:UH}\\
 Sp(1)_{R+}:\quad&H_3=\frac{2S_4-S_2}{12},\quad H_2=S_2-4H_3,
 \quad H_1=n_H-2H_2-3H_3.\label{eq:SH}
\end{align}
\end{subequations}
The abelian search enumerates admissible integral pairs $(r,s)$. Through \eqref{ae} and \eqref{eq:alpha-gamma}, these determine the spectral moments $S_2,S_4$, which are then tested for realization by nonnegative multiplicities $H_j$. The code checks nonnegativity, multiplicity integrality and lattice membership of the diagonal pair $\{a,b_{R+}/2\}$ for $U(1)_{R+}$ or $\{a,b_{R+}\}$ for $Sp(1)_{R+}$. At larger bounds it enumerates the higher-label multiplicities before solving for the remaining ones.

\paragraph{Ordinary and diagonal sectors.}
Using the exact largest occupied charge distinguishes the three abelian sectors:
\begin{align}
 q_{\max}=0:\quad& n_V\equiv8\pmod{12},\quad
 b_{R+}=\left(2,-\frac{n_V+4}{12}\right),\label{eq:qzero}\\
 q_{\max}=1:\quad& H_0\in12\Z_{\ge0},\quad
 b_{R+}=(2,20-H_0/12),\label{eq:qone}\\
 q_{\max}=2:\quad& H_0-3H_2\in12\Z,\quad H_2\in2\Z_{>0},\quad
 (r-2)(s-2)=2H_2.\label{eq:qtwo}
\end{align}

The first is ordinary $U(1)_R$ gauging; the congruence was also obtained in \cite{Becker:2025xgy,Becker:2026etl}. The second has $H_1>0,H_2=0$; the third has $H_2>0$ and may contain both lower charges. The vectors are given up to component interchange, and the square condition in \eqref{eq:Uarith} remains required.

For ordinary $Sp(1)_R$ gauging, $d_{\max}=1$ gives $S_2=S_4=0$ and $n_V=12$ or $96$, in agreement with \cite{Becker:2025xgy}. At $d_{\max}=2$, $2S_4=S_2$ reduces \eqref{eq:Sarith} to
\begin{equation}
 n_V-2S_2\in\{-168,-84,-48,-24,12,96\}.
 \label{eq:six}
\end{equation}
At $d_{\max}=3$, triplets instead give $J=H_3>0$. These arithmetic restrictions are necessary; the remaining tests determine whether a complete spectrum exists.

\paragraph{Constructing and checking spectra.}
Independently, the code enumerates nonabelian matter contents with total dimension $n_H$ satisfying the irreducible quartic conditions in \eqref{c12}, together with \eqref{c13} in the one-factor $D_4$ search.
Using $\beta_i,\epsilon_{ii}$ from \eqref{eq:beta-product}, the equations \eqref{ae} give
\begin{equation}
 b_i=(b_i^+,b_i^-)=\frac{\lambda_i}{4}
 (-\beta_i+\sqrt{\beta_i^2-4\epsilon_{ii}},
  -\beta_i-\sqrt{\beta_i^2-4\epsilon_{ii}}),
 \label{eq:external}
\end{equation}
Both orientations are retained subject to $2b_i\in\Z^2$. For two factors, $4b_1\cdot b_2=\lambda_1\lambda_2\epsilon_{12}$ also fixes their relative orientations; $\epsilon_{12}$ is the mixed-anomaly coefficient between the additional factors.

Before assigning diagonal labels, each nonabelian matter content is paired with a diagonal candidate and tested for common lattice membership and ghost freedom: all gauge-kinetic functions in \eqref{fz} must be positive at a common dilaton value.

For each surviving candidate, we assign absolute charges $j$ or diagonal $Sp(1)_{R+}$ representation dimensions $d$ to the matter transforming under the additional factors, starting with the largest label and preserving the previously determined nonabelian matter content. For $Sp(1)_{R+}$, a block of diagonal dimension $d$ contributes $d$ copies of its additional-factor representation, weighted by $w_I m_I$. The required mixed sums are
\begin{subequations}\label{eq:mixed}
\begin{align}
 U(1)_{R+}:\quad&A_H^{(i,+)}=A_{\adj_i}^{(i)}+\frac{b_i\cdot b_{R+}}{\lambda_i},\label{eq:Umixed}\\
 Sp(1)_{R+}:\quad&A_H^{(i,+)}=\tfrac12A_{\adj_i}^{(i)}+\frac{b_i\cdot b_{R+}}{\lambda_i}.\label{eq:Smixed}
\end{align}
\end{subequations}
At each step, a partial assignment is discarded if the remaining matter cannot supply the mixed-anomaly sum still required by \eqref{eq:mixed}. For two factors, the mixed condition for the second additional factor is checked on completion. Abelian assignments must also admit charge signs satisfying the mixed cubic condition $E_q^{(i)}=0$ in \eqref{c12} for every additional factor with a cubic invariant; the weighted sum is defined in \eqref{cubic}. Nonzero abelian charges require full hypermultiplets.

For $U(1)_{R+}$ with one or two additional simple factors, each absolute-charge assignment of the nonabelian matter, together with its anomaly vectors, is counted as one spectrum if at least one assignment of charge signs satisfies the mixed cubic-anomaly conditions \eqref{c12}. Different admissible sign assignments are not counted as additional spectra. This counting convention does not assert that different admissible sign assignments define physically equivalent theories. Tables~\ref{tab:u1-one-combined} and~\ref{tab:u1-two-combined} display one compatible signed representative for each selected spectrum, together with $N_{\rm sign}$ as defined in Appendix~\ref{app:abelian-spectra}.

For a fixed additional gauge group, a class groups spectra with the same anomaly-vector pair $(b_1,b_{R+})$ or triple $(b_1,b_2,b_{R+})$, up to simultaneous interchange of the two light-cone coordinates and permutation of identical additional factors. Different lattice embeddings do not create additional classes.

\subsection{Results with one additional simple factor}
\label{sec:one-factor-results}
\paragraph{\texorpdfstring{$U(1)_{R+}$ gauging.}{U(1)R+ gauging.}}
Under \eq{G1} and \eq{eq:shared-diagonal-scan a} we have found 665 charge-allocation spectra grouped into 242 classes for $q_{\max}\leq1$, and 18,732 spectra grouped into 1,896 classes for $q_{\max}=2$, as shown in Table~\ref{t2}.

The $q_{\max}\leq1$ block of Table~\ref{tab:u1-one-combined} in Appendix~\ref{app:abelian-spectra} lists 15 spectra, omitting $A_3$, $B_3$ and $C_3$. The $q_{\max}=2$ block lists 21 spectra in 17 classes, also omitting $A_5$.

\paragraph{\texorpdfstring{$Sp(1)_{R+}$ gauging.}{Sp(1)R+ gauging.}}
The gravitational and irreducible nonabelian anomaly equations exclude models with one additional classical simple gauge factor of rank $10$--$30$. For the remaining groups, including $A_3$, we exhaust all allowed diagonal singlet, doublet and triplet allocations and solve for the anomaly vectors without bounding their components. We require all gauge kinetic terms to be positive at a common value of the dilaton.

Under \eqref{G1} and \eq{eq:shared-diagonal-scan b}, we find 127 spectra grouped into 60 anomaly-vector classes for $d_{\max}\leq2$, and 479 spectra grouped into 316 classes for $d_{\max}=3$, as shown in Table~\ref{sp:tab:sp1-count-one}. Gauge-group automorphisms and full-product conjugation of full complex blocks are identified. Different lattice embeddings do not create additional spectra, whereas distinct spectra sharing the same anomaly vectors are counted separately. None of these spectra with one additional simple gauge factor have $d_{\max}=1$.

Omitted entries have no spectra in the corresponding range of
\eqref{G1} and \eq{eq:shared-diagonal-scan b}. The selected spectra in Appendix~\ref{sp:app:spectra} omit models with one additional rank-three simple gauge factor for brevity;
their complete spectra are retained with the stored catalogues~\cite{DGcode}.

\begin{table}[H]
\captionsetup{justification=raggedright,singlelinecheck=false}
\centering
\begin{minipage}[t]{0.52\textwidth}
\vspace{0pt}
\color{black}\hypersetup{linkcolor=blue,citecolor=blue,urlcolor=blue}
\centering\small
\setlength{\tabcolsep}{4pt}
\begin{tabular}{lrrrr}
\toprule
 & \multicolumn{2}{c}{$q_{\max}\leq1$} & \multicolumn{2}{c}{$q_{\max}=2$}\\
\cmidrule(lr){2-3}\cmidrule(lr){4-5}
$G_1$ & classes & spectra & classes & spectra\\
\midrule
$A_{3}$ & 200 & 619 & 1{,}742 & 18{,}550\\
$A_{5}$ & 9 & 9 & 54 & 71\\
$A_{6}$ & 2 & 3 & 4 & 7\\
$A_{7}$ & 0 & 0 & 1 & 1\\
$B_{3}$ & 9 & 10 & 25 & 27\\
$B_{6}$ & 1 & 1 & 0 & 0\\
$C_{3}$ & 19 & 21 & 58 & 63\\
$C_{4}$ & 1 & 1 & 2 & 2\\
$C_{5}$ & 0 & 0 & 1 & 1\\
$D_{4}$ & 0 & 0 & 1 & 1\\
$G_{2}$ & 1 & 1 & 8 & 9\\
\midrule
\textbf{Total} & \textbf{242} & \textbf{665} & \textbf{1{,}896} & \textbf{18{,}732}\\
\bottomrule
\end{tabular}
\caption{Numbers of $G_1\times U(1)_{R+}$ spectra under \eqref{G1} and \eq{eq:shared-diagonal-scan a}. Groups in the stated search range that are not listed yield no spectra. Appendix~\ref{app:abelian-spectra} presents the full spectra for those excluding rank 3 groups, and in the case of $q_{\rm max}=2$ the group $A_5$ as well. }
\label{t2}
\end{minipage}\hfill
\begin{minipage}[t]{0.44\textwidth}
\vspace{0pt}
\centering\small
\setlength{\tabcolsep}{4pt}
\begin{tabular}{lrrrr}
\toprule
 & \multicolumn{2}{c}{$d_{\max}\leq2$} & \multicolumn{2}{c}{$d_{\max}=3$}\\
\cmidrule(lr){2-3}\cmidrule(lr){4-5}
$G_1$ & classes & spectra & classes & spectra\\
\midrule
$A_3$ & 47 & 113 & 269 & 425\\
$A_5$ & 0 & 0 & 5 & 7\\
$B_3$ & 6 & 6 & 16 & 20\\
$C_3$ & 3 & 4 & 22 & 23\\
$C_4$ & 4 & 4 & 3 & 3\\
$C_6$ & 0 & 0 & 1 & 1\\
\midrule
\textbf{Total} & \textbf{60} & \textbf{127} & \textbf{316} & \textbf{479}\\
\bottomrule
\end{tabular}
\caption{Anomaly-vector classes and spectra for $G_1\times Sp(1)_{R+}$ under \eqref{G1} and \eq{eq:shared-diagonal-scan b}. Unlisted groups give no spectra. Appendix~\ref{sp:app:spectra} lists the spectra with rank-three groups omitted.}
\label{sp:tab:sp1-count-one}\label{sp:tab:sp1-count-one-low}\label{sp:tab:sp1-count-one-high}
\end{minipage}
\end{table}

\subsection{Results with two additional simple factors}
\label{sec:two-factor-results}

\paragraph{\texorpdfstring{$U(1)_{R+}$ gauging.}{U(1)R+ gauging.}}
Under \eq{G1xG2} and \eq{eq:shared-diagonal-scan a}, we find 126 spectra in 66 anomaly-vector classes for $q_{\max}\leq1$, and 1,222 spectra in 524 classes for $q_{\max}=2$, across 31 and 46 gauge-group pairs, respectively (Table~\ref{t3}).

\begin{table}[H]
\color{black}\hypersetup{linkcolor=blue,citecolor=blue,urlcolor=blue}
\centering
\renewcommand{\baselinestretch}{1}\small
\setlength{\tabcolsep}{4pt}
\begin{minipage}[t]{0.49\textwidth}
\centering
\begin{tabular}[t]{@{}lrrrr@{}}
\toprule
 & \multicolumn{2}{c}{$q_{\max}\leq1$} & \multicolumn{2}{c}{$q_{\max}=2$}\\
\cmidrule(lr){2-3}\cmidrule(lr){4-5}
$G_1\times G_2$ & classes & spectra & classes & spectra\\
\midrule
$A_{5}\times A_{5}$ & 5 & 14 & 93 & 426\\
$A_{5}\times A_{6}$ & 10 & 22 & 32 & 66\\
$A_{5}\times A_{7}$ & 1 & 1 & 23 & 31\\
$A_{5}\times A_{9}$ & 2 & 2 & 8 & 8\\
$A_{5}\times A_{10}$ & 0 & 0 & 5 & 5\\
$A_{5}\times A_{11}$ & 0 & 0 & 9 & 14\\
$A_{5}\times A_{12}$ & 0 & 0 & 3 & 3\\
$A_{5}\times A_{13}$ & 0 & 0 & 2 & 2\\
$A_{5}\times A_{16}$ & 0 & 0 & 1 & 1\\
$A_{5}\times B_{5}$ & 3 & 4 & 15 & 24\\
$A_{5}\times B_{6}$ & 3 & 10 & 29 & 61\\
$A_{5}\times B_{7}$ & 0 & 0 & 4 & 4\\
$A_{5}\times B_{8}$ & 0 & 0 & 1 & 1\\
$A_{5}\times C_{5}$ & 0 & 0 & 7 & 7\\
$A_{5}\times C_{8}$ & 0 & 0 & 1 & 1\\
$A_{5}\times D_{5}$ & 8 & 10 & 90 & 105\\
$A_{5}\times D_{6}$ & 7 & 16 & 68 & 180\\
$A_{5}\times D_{7}$ & 2 & 2 & 30 & 33\\
$A_{5}\times D_{9}$ & 2 & 2 & 9 & 10\\
$A_{5}\times D_{10}$ & 1 & 1 & 4 & 4\\
$A_{5}\times D_{12}$ & 1 & 1 & 2 & 2\\
$A_{5}\times E_{6}$ & 0 & 0 & 7 & 12\\
$A_{5}\times E_{7}$ & 0 & 0 & 12 & 15\\
$A_{5}\times F_{4}$ & 0 & 0 & 12 & 20\\
$A_{6}\times A_{10}$ & 0 & 0 & 2 & 2\\
$A_{6}\times B_{5}$ & 1 & 4 & 1 & 1\\
$A_{6}\times D_{5}$ & 1 & 1 & 4 & 5\\
$A_{6}\times D_{7}$ & 1 & 1 & 1 & 1\\
$A_{6}\times D_{9}$ & 1 & 1 & 6 & 8\\
\bottomrule
\end{tabular}
\end{minipage}
\hfill
\begin{minipage}[t]{0.49\textwidth}
\centering
\begin{tabular}[t]{@{}lrrrr@{}}
\toprule
 & \multicolumn{2}{c}{$q_{\max}\leq1$} & \multicolumn{2}{c}{$q_{\max}=2$}\\
\cmidrule(lr){2-3}\cmidrule(lr){4-5}
$G_1\times G_2$ & classes & spectra & classes & spectra\\
\midrule
$A_{6}\times E_{7}$ & 0 & 0 & 1 & 1\\
$A_{6}\times F_{4}$ & 0 & 0 & 1 & 2\\
$A_{7}\times A_{9}$ & 0 & 0 & 1 & 1\\
$A_{7}\times B_{8}$ & 1 & 1 & 0 & 0\\
$A_{7}\times C_{5}$ & 1 & 1 & 1 & 1\\
$A_{7}\times D_{5}$ & 0 & 0 & 1 & 1\\
$A_{7}\times D_{6}$ & 0 & 0 & 1 & 1\\
$A_{7}\times D_{7}$ & 0 & 0 & 1 & 1\\
$A_{7}\times E_{6}$ & 1 & 1 & 0 & 0\\
$A_{7}\times F_{4}$ & 1 & 1 & 0 & 0\\
$A_{8}\times A_{8}$ & 0 & 0 & 1 & 3\\
$A_{8}\times A_{17}$ & 1 & 1 & 0 & 0\\
$A_{9}\times D_{6}$ & 1 & 3 & 5 & 13\\
$A_{10}\times D_{5}$ & 2 & 2 & 9 & 9\\
$A_{13}\times D_{9}$ & 0 & 0 & 1 & 1\\
$A_{17}\times E_{8}$ & 1 & 1 & 0 & 0\\
$B_{5}\times E_{7}$ & 0 & 0 & 1 & 1\\
$B_{6}\times D_{5}$ & 0 & 0 & 2 & 2\\
$B_{7}\times D_{5}$ & 0 & 0 & 1 & 1\\
$B_{8}\times C_{9}$ & 1 & 1 & 0 & 0\\
$C_{9}\times F_{4}$ & 1 & 1 & 0 & 0\\
$C_{10}\times E_{7}$ & 1 & 1 & 0 & 0\\
$D_{5}\times E_{6}$ & 1 & 1 & 0 & 0\\
$D_{6}\times D_{7}$ & 2 & 17 & 14 & 124\\
$D_{6}\times E_{6}$ & 0 & 0 & 1 & 7\\
$D_{7}\times E_{7}$ & 0 & 0 & 1 & 1\\
$D_{18}\times E_{7}$ & 1 & 1 & 0 & 0\\
$E_{6}\times E_{7}$ & 1 & 1 & 0 & 0\\
\midrule
\textbf{Total} & \textbf{66} & \textbf{126} & \textbf{524} & \textbf{1{,}222}\\
\bottomrule
\end{tabular}
\end{minipage}
\caption{Anomaly-vector classes and spectra for $G_1\times G_2\times U(1)_{R+}$ under \eq{G1xG2} and \eq{eq:shared-diagonal-scan a}. Unlisted group pairs give no spectra. The $q_{\max}\leq1$ results include seven neutral spectra.}
\label{t3}
\end{table}

Appendix~\ref{app:abelian-spectra} retains only factors of rank $6$--$30$ and $F_4$, omitting $D_6\times D_7$ for brevity: Table~\ref{tab:u1-two-combined} lists 15 spectra in the $q_{\max}\leq1$ block and 42 in the $q_{\max}=2$ block.
Seven of these spectra are neutral (shaded gray). Interestingly, two of them, namely the models with $A_8 \times A_{17}\times U(1)_R$ and $D_{18}\times E_7\times U(1)_R$, are new. $E_6 \times E_7$ with $\frac12(\bf 1, \bf 912)$ is the model of \cite{Randjbar-Daemi:1985tdc}; $C_9 \times F_4$ with $\frac12(\bf 18,\bf 52)$ is the model of \cite{Avramis:2005hc}; $B_8\times C_9$ and $C_{10}\times E_7$ are models 11 and 9 of~\cite{Becker:2025xgy}; $A_{17}\times E_8$ with $(\bf 816,\bf 1)$ is model III of \cite{Becker:2026etl}, with the same anomaly vectors after $x^+ \leftrightarrow x^-$. 

\paragraph{\texorpdfstring{$Sp(1)_{R+}$ gauging.}{Sp(1)R+ gauging.}}
The complete search gives 48 spectra: 20 grouped into 19 anomaly-vector classes for $d_{\max}\leq2$, and 28 grouped into 23 classes for $d_{\max}=3$.

All displayed counts are exact; omitted pairs have no spectra in the
corresponding range of \eqref{G1xG2} and \eq{eq:shared-diagonal-scan b}; seven spectra contain matter
charged under both additional simple gauge factors.

Appendix~\ref{sp:app:spectra} contains two spectral tables, each with separate
$d_{\max}\leq2$ and $d_{\max}=3$ blocks.
The table for one additional simple gauge factor omits rank-three factors; the table for two additional simple gauge factors omits spectra with a rank-five factor while retaining $F_4$.

\begin{table}[H]
\centering
\renewcommand{\baselinestretch}{1}\small
\setlength{\tabcolsep}{4pt}
\begin{minipage}[t]{0.49\textwidth}
\centering
\begin{tabular}[t]{@{}lrrrr@{}}
\toprule
 & \multicolumn{2}{c}{$d_{\max}\leq2$} & \multicolumn{2}{c}{$d_{\max}=3$}\\
\cmidrule(lr){2-3}\cmidrule(lr){4-5}
$G_1\times G_2$ & classes & spectra & classes & spectra\\
\midrule
$A_{5}\times A_{5}$ & 0 & 0 & 1 & 1\\
$A_{5}\times A_{6}$ & 1 & 1 & 1 & 1\\
$A_{5}\times B_{5}$ & 1 & 1 & 1 & 1\\
$A_{5}\times B_{6}$ & 0 & 0 & 1 & 1\\
$A_{5}\times B_{7}$ & 0 & 0 & 4 & 6\\
$A_{5}\times D_{5}$ & 0 & 0 & 1 & 1\\
$A_{5}\times D_{6}$ & 0 & 0 & 5 & 5\\
$A_{5}\times D_{7}$ & 0 & 0 & 2 & 2\\
$A_{5}\times E_{7}$ & 1 & 1 & 0 & 0\\
$A_{5}\times F_{4}$ & 0 & 0 & 2 & 3\\
$A_{6}\times C_{5}$ & 0 & 0 & 1 & 1\\
$A_{7}\times B_{5}$ & 0 & 0 & 1 & 1\\
$A_{7}\times D_{5}$ & 1 & 1 & 0 & 0\\
$A_{10}\times D_{7}$ & 1 & 1 & 0 & 0\\
$B_{5}\times D_{7}$ & 1 & 1 & 0 & 0\\
$B_{5}\times E_{6}$ & 1 & 1 & 0 & 0\\
\bottomrule
\end{tabular}
\end{minipage}
\hfill
\begin{minipage}[t]{0.49\textwidth}
\centering
\begin{tabular}[t]{@{}lrrrr@{}}
\toprule
 & \multicolumn{2}{c}{$d_{\max}\leq2$} & \multicolumn{2}{c}{$d_{\max}=3$}\\
\cmidrule(lr){2-3}\cmidrule(lr){4-5}
$G_1\times G_2$ & classes & spectra & classes & spectra\\
\midrule
$B_{6}\times E_{7}$ & 1 & 1 & 0 & 0\\
$B_{6}\times F_{4}$ & 0 & 0 & 1 & 1\\
$B_{7}\times C_{6}$ & 1 & 1 & 0 & 0\\
$B_{8}\times D_{5}$ & 1 & 1 & 0 & 0\\
$B_{9}\times B_{13}$ & 1 & 1 & 0 & 0\\
$C_{5}\times D_{5}$ & 1 & 1 & 0 & 0\\
$C_{5}\times D_{6}$ & 0 & 0 & 1 & 3\\
$C_{5}\times E_{6}$ & 1 & 1 & 0 & 0\\
$C_{7}\times F_{4}$ & 0 & 0 & 1 & 1\\
$C_{8}\times D_{8}$ & 1 & 1 & 0 & 0\\
$D_{5}\times D_{6}$ & 1 & 1 & 0 & 0\\
$D_{5}\times E_{6}$ & 1 & 1 & 0 & 0\\
$D_{8}\times E_{6}$ & 1 & 2 & 0 & 0\\
$D_{9}\times F_{4}$ & 1 & 1 & 0 & 0\\
$E_{6}\times E_{7}$ & 1 & 1 & 0 & 0\\
\midrule
\textbf{Total} & \textbf{19} & \textbf{20} & \textbf{23} & \textbf{28}\\
\bottomrule
\end{tabular}
\end{minipage}
\caption{Full-search counts of anomaly-vector classes and spectra for $G_1\times G_2\times Sp(1)_{R+}$ under \eqref{G1xG2} and \eq{eq:shared-diagonal-scan b}; unlisted group pairs yield no spectra.}
\label{sp:tab:sp1-count-two}\label{sp:tab:sp1-count-two-low}\label{sp:tab:sp1-count-two-high}
\end{table}

Figure~\ref{fig:spectra-rank} summarizes these spectra by the rank of the additional simple factors, combining the $q_{\max}=0,1,2$ sectors for $U(1)_{R+}$ and the $d_{\max}=1,2,3$ sectors for $Sp(1)_{R+}$. 

\begin{figure}[H]
\centering
\renewcommand{\baselinestretch}{1}\small
\captionsetup{font=small,labelfont=bf}
\definecolor{DGrankUone}{HTML}{0072B2}
\definecolor{DGrankSpone}{HTML}{D55E00}
\newcommand{\DGrankseries}[2]{%
  \addplot[only marks,#1,forget plot]
    coordinates {#2};%
}
\newcommand{\DGranklegend}{%
  \addlegendimage{only marks,color=DGrankUone,mark=*,mark size=2.8pt,
    mark options={fill=DGrankUone},line width=1pt}
  \addlegendentry{$U(1)_{R+}$}
  \addlegendimage{only marks,color=DGrankSpone,mark=o,mark size=4.3pt,
    mark options={fill=none},line width=1.2pt}
  \addlegendentry{$Sp(1)_{R+}$}
}
\pgfplotsset{DGrankplot/.style={/pgfplots/.cd,
  width=0.97\linewidth,height=6.3cm,
  axis lines=left,axis line style={black!50},
  ymin=0.65,ylabel={Number of spectra},
  tick label style={font=\small},label style={font=\small},
  scaled y ticks=false,minor y tick num=0,minor tick length=0pt,
  grid=major,major grid style={black!10},
  legend columns=1,legend cell align=left,
  legend style={at={(0.97,0.97)},anchor=north east,
    font=\small,fill=white,draw=none,inner sep=3pt,row sep=3pt},
  clip=false},
  DGrankUseries/.style={color=DGrankUone,mark=*,mark size=2.8pt,
    mark options={fill=DGrankUone},line width=1pt},
  DGrankSpseries/.style={color=DGrankSpone,mark=o,mark size=4.3pt,
    mark options={fill=none},line width=1.2pt}}
{\small\bfseries (a) One additional simple factor}\par
\vspace{1mm}
\begin{tikzpicture}
\begin{semilogyaxis}[DGrankplot,
  xmin=2.7,xmax=8.3,xtick={3,4,5,6,7,8},
  ymax=40000,ytick={1,10,100,1000,10000},
  yticklabels={1,10,100,{1,000},{10,000}},
  xlabel={Rank of $G_1$}]
\DGrankseries{DGrankSpseries}{
(3,591)
(4,7)
(5,7)
(6,1)
}
\DGrankseries{DGrankUseries}{
(3,19290)
(4,4)
(5,81)
(6,11)
(7,1)
}
\node[anchor=west,text=DGrankUone,font=\small]
  at (axis cs:3.12,23000) {19,290};
\DGranklegend
\end{semilogyaxis}
\end{tikzpicture}\par
\vspace{5mm}
{\small\bfseries (b) Two additional simple factors}\par
\vspace{1mm}
\begin{tikzpicture}
\begin{semilogyaxis}[DGrankplot,
  xmin=9.5,xmax=26.5,xtick={10,12,14,16,18,20,22,24,26},
  ymax=1000,ytick={1,10,100,1000},
  yticklabels={1,10,100,{1,000}},
  xlabel={Total rank: $\operatorname{rank}(G_1)+\operatorname{rank}(G_2)$}]
\DGrankseries{DGrankSpseries}{
(10,6)
(11,17)
(12,12)
(13,5)
(14,2)
(16,1)
(17,1)
(22,1)
}
\DGrankseries{DGrankUseries}{
(10,592)
(11,382)
(12,98)
(13,150)
(14,24)
(15,47)
(16,20)
(17,8)
(18,2)
(21,1)
(22,1)
(25,3)
}
\DGranklegend
\end{semilogyaxis}
\end{tikzpicture}
\caption{Distribution of anomaly-free spectra by rank for
(a) one and (b) two additional simple factors, within the group ranges
\eqref{G1} and \eqref{G1xG2} and diagonal bounds \eqref{eq:shared-diagonal-scan}.
Filled blue circles combine the $U(1)_{R+}$ sectors with $q_{\max}=0,1,2$;
larger open orange circles combine the $Sp(1)_{R+}$ sectors with $d_{\max}=1,2,3$.
Each point sums spectra over the searched groups at that rank;
compatible charge-sign assignments are not counted separately.
Rank excludes the diagonal factor and is summed over the two additional
factors in (b). Panel (a) omits the rank-2 group $G_2$ and includes the full
$A_3$ catalogue; panel (b) starts at total rank 10, omitting rank 9.
Vertical scales are logarithmic; zero counts are omitted.}
\label{fig:spectra-rank}
\end{figure}

\subsection{\texorpdfstring{Larger $G_+$ representations}{Larger G+ representations}}
\label{sec:larger-diagonal-representations}

Keeping the additional-group range \eqref{G1} and all consistency tests
unchanged, we extend the searches with one additional simple factor
beyond the diagonal bounds in \eqref{eq:shared-diagonal-scan}.
Table~\ref{tab:extended-diagonal-counts} gives the total number of anomaly-vector classes
and spectra at $q_{\max}=3,4$ for $U(1)_{R+}$ and $d_{\max}=4,5$
for $Sp(1)_{R+}$, using the counting convention defined above. 
Table~\ref{tab:qmax4-representatives} shows one example per group at its smallest $q_{\max}\geq3$: eight at $q_{\max}=3$, $D_4$ at $4$, and $B_6$ at $7$. For $B_6$, the searches at $q_{\max}=5,6$ were exhaustive and yielded no solutions; the search at $q_{\max}=7$ stopped after the first solution.

\begin{table}[H]
\centering
\captionsetup{hypcap=false}
\begin{minipage}[t]{0.59\linewidth}
\centering\small
\setlength{\tabcolsep}{4pt}
\renewcommand{\arraystretch}{1.16}
\textbf{$U(1)_{R+}$}\par\smallskip
\begin{tabular}{@{}lrrrr@{}}
\toprule
 & \multicolumn{2}{c}{$q_{\max}=3$} & \multicolumn{2}{c}{$q_{\max}=4$}\\
\cmidrule(lr){2-3}\cmidrule(lr){4-5}
$G_1$ & classes & spectra & classes & spectra\\
\midrule
$A_{3}$ & 10{,}243 & 778{,}242 & 33{,}476 & 20{,}283{,}577\\
$A_{5}$ & 401 & 613 & 1{,}528 & 4{,}888\\
$A_{6}$ & 20 & 34 & 51 & 92\\
$B_{3}$ & 157 & 199 & 712 & 997\\
$C_{3}$ & 337 & 433 & 1{,}321 & 2{,}824\\
$C_{4}$ & 8 & 8 & 16 & 16\\
$C_{5}$ & 5 & 5 & 5 & 5\\
$D_{4}$ & 0 & 0 & 1 & 1\\
$G_{2}$ & 28 & 46 & 98 & 264\\
\midrule
\textbf{Total} & \textbf{11{,}199} & \textbf{779{,}580} & \textbf{37{,}208} & \textbf{20{,}292{,}664}\\
\bottomrule
\end{tabular}
\end{minipage}
\hfill
\begin{minipage}[t]{0.39\linewidth}
\centering\small
\setlength{\tabcolsep}{4pt}
\renewcommand{\arraystretch}{1.16}
\textbf{$Sp(1)_{R+}$}\par\smallskip
\begin{tabular}{@{}lrrrr@{}}
\toprule
 & \multicolumn{2}{c}{$d_{\max}=4$} & \multicolumn{2}{c}{$d_{\max}=5$}\\
\cmidrule(lr){2-3}\cmidrule(lr){4-5}
$G_1$ & classes & spectra & classes & spectra\\
\midrule
$A_{3}$ & 535 & 854 & 320 & 372\\
$A_{5}$ & 4 & 5 & 2 & 2\\
$B_{3}$ & 63 & 64 & 25 & 27\\
$C_{3}$ & 8 & 8 & 18 & 18\\
$C_{4}$ & 3 & 3 & 5 & 5\\
$C_{6}$ & 1 & 1 & 0 & 0\\
\midrule
\textbf{Total} & \textbf{614} & \textbf{935} & \textbf{370} & \textbf{424}\\
\bottomrule
\end{tabular}
\end{minipage}
\caption{Anomaly-vector classes and spectra for $G_1\times G_+$ at larger diagonal charges or representation dimensions. Counts are taken separately at each exact maximum; a class may occur in more than one sector. For $U(1)_{R+}$, compatible charge-sign assignments are not counted separately. Unlisted groups give zero.}
\label{tab:extended-diagonal-counts}\label{tab:qmax4-summary}\label{sp:tab:sp1-d45-counts}
\end{table}

\section{\texorpdfstring{The potential and the $6D$ vacua}{The potential and the 6D vacua}}
\label{sec:vacua}

\subsection{\texorpdfstring{Conditions for maximally symmetric $6D$ vacua}{Conditions for maximally symmetric 6D vacua}}
\label{sec:vacuum-conditions}

We consider the two gauge groups in \eqref{eq:group} with the potential
\eqref{pot}, in the domain where all gauge kinetic coefficients $f_z(\phi)$ are positive.
For $U(1)_{R+}$ we use the charge-one normalization obtained by
\eqref{eq:abelian-action-rescaling}. With vanishing form fluxes and constant scalars, a Minkowski vacuum therefore requires
\begin{equation}
C_+^r=0,\quad C_{z'}^{\ell}=0\quad\hbox{for every generator of }K.
\label{eq:CC0}
\end{equation}

These conditions also annul the scalar variations of the potential.
Here $r=3$ for $U(1)_{R+}$, when we write $C_+=C_+^3$,
and $r=1,2,3$ for $Sp(1)_{R+}$. Each map is an
$\mathfrak{sp}(1)_R$-valued matrix with three real components.

The dilaton remains free within the common domain $f_z(\phi)>0$;
the fermion transformations preserve all eight supersymmetries. All the anomaly free models we have found admit no $AdS_6$ vacua at two-derivative order. On the other hand, dS$_6$ vacua are allowed as follows.

At the scalar origin, $C_{z'}=0$ and the diagonal maps equal their
R-symmetry generators. In the respective gauge-field normalizations,
$V(0,\phi)=1/(2f_+)$ for $U(1)_{R+}$ and $3/(8f_+)$ for $Sp(1)_{R+}$.
If $v_{R+}^\pm>0$ and all other kinetic coefficients are positive at
the stationary point, the nonsupersymmetric de Sitter solution obeys
\begin{equation}
 \phi_0=\frac12\log\frac{v_{R+}^-}{v_{R+}^+},\quad
 V_0=\begin{cases}
 (4\sqrt{v_{R+}^+v_{R+}^-})^{-1},&G_+=U(1)_{R+},\\
 3(16\sqrt{v_{R+}^+v_{R+}^-})^{-1},&G_+=Sp(1)_{R+},
 \end{cases}
 \quad \partial_\phi^2V(0,\phi_0)=-V_0.
 \label{eq:origin-dS-stationary}
\end{equation}
The de Sitter solution specified by \eqref{eq:origin-dS-stationary} satisfies the two-derivative equations of motion and is unstable in the dilaton direction. Curvature-squared corrections can modify the solution and its stability; these effects are not analysed here. Its hyperscalar stability at two-derivative order is discussed in Section~\ref{sec:charge-bound-wolf-general-unified45}.

\subsection{Diagonal embeddings on the unitary Wolf space}
\label{sec:unitary-embeddings}

The scalar manifold must realize the complete symplectic hyperino representation \(\mathcal R_H\) associated with the spectrum \(\rho_H\) defined in Section~\ref{sec:anomaly-building-blocks}. 
We use the scalar manifold
\begin{align}
\mathcal M_H=
\frac{SU(n_H,2)}{S(U(n_H)\times U(2))}
\label{eq:cosetSU}
\end{align}
for spectra for which the symplectic hyperino representation admits a gauge-invariant decomposition into conjugate representation blocks
\begin{align}
\mathcal R_H=V_H\oplus\overline V_H,
\qquad
\dim_{\mathbb C}V_H=n_H,
\end{align}
with the invariant antisymmetric form pairing \(V_H\) nondegenerately with \(\overline V_H\) and vanishing when both arguments lie in $V_H$, or both in $\overline V_H$. The space \(V_H\) is the \(n_H\)-dimensional complex representation on which the \(U(n_H)\) isotropy factor acts. Full hypermultiplets and pairs of identical half-hypermultiplets admit such a decomposition; an unpaired pseudoreal half-hypermultiplet does not.

Use $r=3$ alone for $U(1)_{R+}$ and $r=1,2,3$ for $Sp(1)_{R+}$.
The Hermitian matrices acting on $V_H$ are
\begin{equation}
 Q_H^3\big|_I=q_I\mathbf1_I\quad\text{for }U(1)_{R+},\quad
 Q_H^r\big|_{(R,\mathbf d)}=2(\mathbf1_R\otimes J^r_{(d)})\quad\text{for }Sp(1)_{R+}.
 \label{eq:hyperino-generator-blocks}
\end{equation}
where $J^r_{(d)}$ are the three $d\times d$ Hermitian spin-$j$
matrices, with $j=(d-1)/2$. The action of $T_H^r$ on $V_H$ is $-iQ_H^r/2$; on an $Sp(1)_{R+}$ block, this equals $-i\mathbf1_R\otimes J^r_{(d)}$, fixing the factor of two.

In an orthonormal weight basis,
\begin{align}
 J^3\lvert j,m\rangle&=m\lvert j,m\rangle,
 &m&=-j,-j+1,\ldots,j,
 \label{eq:Sp-spin-weights}\\
 J^\pm\lvert j,m\rangle&=\sqrt{(j\mp m)(j\pm m+1)}\,
 \lvert j,m\pm1\rangle,
 &J^\pm&=J^1\pm iJ^2.
 \label{eq:Sp-spin-ladder}
\end{align}
Since $\operatorname{tr}J^r=0$, we have $\operatorname{tr}Q_H^r=0$
for all three $Sp(1)_{R+}$ generators.

In the abelian case we suppress the superscript $3$ and write $Q_H$;
for full-hypermultiplet blocks,
$\operatorname{tr}Q_H=\sum_I m_Iq_I\prod_i d_{Ii}$, while compatible
neutral half-hypermultiplet blocks contribute zero.

Parametrize the scalar manifold by an $n_H\times2$ complex matrix
$Z^{aA}$ in the domain $\mathbf1_2-Z^\dagger Z>0$.
Here $a=1,\ldots,n_H$ labels $V_H$\footnote{In Section~\ref{sec:sp1-gauged-model}, $a=1,\ldots,2n_H$ instead labels the full hyperino representation ${\cal R}_H$.}, and $A$ runs over $1, 2$, transforming in the ${\bf 2_R}$.

A coset representative is
\begin{equation}
 L(Z)=\begin{pmatrix}A&ZB\\BZ^\dagger&B\end{pmatrix},
 \quad A=(\mathbf1_{n_H}-ZZ^\dagger)^{-1/2},
 \quad B=(\mathbf1_2-Z^\dagger Z)^{-1/2}.
 \label{eq:L-representative-extended}
\end{equation}
Consider an isotropy element $g=\operatorname{diag}(U,V)$, where
$U\in U(n_H)$, $V\in U(2)$ and $\det U\det V=1$.
Write $V=bS$, with $b^2=\det V$ and $S\in SU(2)_R$.
The identity $gL(Z)g^{-1}=L(UZV^{-1})$ shows that the compensator is
$h=g$ and gives
\begin{equation}
 Z'=UZV^{-1}=(b^{-1}U)ZS^{-1}.
 \label{eq:unitary-isotropy-scalar-action}
\end{equation}
The $V_H$ index is therefore rotated by $b^{-1}U$. The hyperino
constituent $\zeta_H\in V_H$ is an $SU(2)_R$ singlet and transforms as
\begin{equation}
\zeta_H'=b^{-1}U\zeta_H.
\label{eq:unitary-isotropy-hyperino-action}
\end{equation}
Each $T_H^r$ commutes with $SU(2)_R$, so it has the form
\begin{equation}
 T_H^r=-\frac{i}{2}\begin{pmatrix}M^r&0\\0&c^r\mathbf1_2\end{pmatrix},
 \quad M^r=(M^r)^\dagger.
 \label{eq:TH-block-ansatz}
\end{equation}
For a fixed $r$, the element $e^{2\epsilon T_H^r}$ acts on the hyperino
by $e^{-i\epsilon(M^r-c^r\mathbf1_{n_H})}$, which should reproduce the desired $e^{-i\epsilon Q_H^r}$ transformation on the hyperino. We thus derive
\begin{equation}
M^r=Q_H^r+c^r\mathbf1_{n_H}.
\label{eq:TH-hyperino-matching}
\end{equation}
Imposing the traceless condition on $T_H^r$ determines
\begin{equation}
0=\operatorname{tr}M^r+2c^r
=\operatorname{tr}Q_H^r+(n_H+2)c^r,
\quad c^r=-\frac{\operatorname{tr}Q_H^r}{n_H+2}.
\label{eq:TH-trace-condition}
\end{equation}
Hence
\begin{equation}
T_H^r=-\frac{i}{2}\left[
\begin{pmatrix}Q_H^r&0\\0&\mathbf0_2\end{pmatrix}
-\frac{\operatorname{tr}Q_H^r}{n_H+2}\mathbf1_{n_H+2}\right],
\label{eq:TH-embedding}
\end{equation}
Using $T_R^r=-i\sigma_r/2$ in the lower block, the diagonal sums
$T_{R+}^r=T_R^r+T_H^r$ become
\begin{empheq}[box=\fbox]{equation}
 \begin{aligned}
 T_{R+}&=-\frac{i}{2}\left[
 \begin{pmatrix}Q_H&0\\0&\sigma_3\end{pmatrix}
 -\frac{\operatorname{tr}Q_H}{n_H+2}\mathbf1_{n_H+2}\right]
 &&\text{for }U(1)_{R+},\\[2mm]
 T_{R+}^r&=-\frac{i}{2}
 \begin{pmatrix}Q_H^r&0\\0&\sigma_r\end{pmatrix}
 &&\text{for }Sp(1)_{R+}.
 \end{aligned}
 \label{eq:TRplus-embedding}
\end{empheq}
Here $c^r=0$ for $Sp(1)_{R+}$ because $\operatorname{tr}Q_H^r=0$.

For $g=e^{2\epsilon^rT_{R+}^r}=\operatorname{diag}(U,V)$, the two blocks are

\begin{equation}
U=e^{-i\epsilon^r(Q_H^r+c^r\mathbf1_{n_H})},\quad
V=e^{-i\epsilon^r(\sigma_r+c^r\mathbf1_2)}.
\label{eq:diagonal-UV-blocks}
\end{equation}
Thus $V=bS$ gives, continuously from the identity,
\begin{equation}
 b=e^{-i\epsilon^rc^r},\quad S=e^{-i\epsilon^r\sigma_r},\quad
 b^{-1}U=e^{-i\epsilon^rQ_H^r}.
 \label{eq:diagonal-compensator-factors}
\end{equation}
In particular, $b=1$ for $Sp(1)_{R+}$. The scalar transformation is therefore
\begin{align}
Z'&=UZV^{-1}=e^{-i\epsilon^rQ_H^r}Ze^{i\epsilon^r\sigma_r},
\label{eq:scalar-transformation-expansion}
\end{align}

On the $I$-th block, write $a_I=(u,p)$ in the $Sp(1)$ case, where
$u$ labels the representation of $K$ and any copies, and
$p=1,\ldots,d_{I+}$. The block actions are 
\begin{subequations}\label{eq:diagonal-block-actions}
\begin{align}
(Q_H)^{a_I}{}_{b_I}&=q_I\delta^{a_I}{}_{b_I}
&&\text{for }U(1)_{R+},\\
(Q_H^r)^{(u,p)}{}_{(v,k)}&=2\delta^u{}_v(J^r_{(d_{I+})})^p{}_k
&&\text{for }Sp(1)_{R+}.
\end{align}
\end{subequations}
The corresponding infinitesimal transformations are 
\begin{subequations}\label{eq:Z-finite-transformation}
\begin{align}
\delta Z^{a_I A}
&=-i\epsilon\bigl(q_IZ^{a_I A}-Z^{a_I B}(\sigma_3)_B{}^A\bigr)
&&\text{for }U(1)_{R+},\\
\delta Z^{upA}
&=-i\epsilon^r\bigl(2(J^r_{(d_{I+})})^p{}_kZ^{ukA}
-Z^{upB}(\sigma_r)_B{}^A\bigr)
&&\text{for }Sp(1)_{R+}.
\end{align}
\end{subequations}
Here $I$ is fixed and not summed. The two scalar columns in the $U(1)_{R+}$
case carry charges $q_I-1$ and $q_I+1$.

Under the same compensator, $\zeta_H$ transforms by
$b^{-1}U=e^{-i\epsilon^rQ_H^r}$, while its Dirac adjoint
$\bar\zeta_H$ transforms from the right by $e^{i\epsilon^rQ_H^r}$.
The $\overline V_H$ constituent is fixed by the symplectic Majorana--Weyl
reality condition and is not an additional independent multiplet.

The gravitino, tensorino and gauginos carry an $SU(2)_R$ doublet index.
The compensator $S=e^{-i\epsilon^r\sigma_r}$ acts as $\Psi'_A=S_A{}^B\Psi_B$ on lower indices and $\Psi'{}^A=(S^{-1})_B{}^A\Psi^B$ on upper indices. Expanding to first order gives
\begin{align}
\delta\zeta&=-i\epsilon^rQ_H^r\zeta,
&\delta\bar\zeta&=i\epsilon^r\bar\zeta Q_H^r,
\label{eq:fermion-hyperino}\\
\delta\psi_\mu^A&=+i\epsilon^r(\sigma_r)_B{}^A\psi_\mu^B,
&\delta\chi^A&=+i\epsilon^r(\sigma_r)_B{}^A\chi^B,
\label{eq:fermion-R-doublets}\\
\delta\lambda_z^{I A}&=+i\epsilon^r(\sigma_r)_B{}^A\lambda_z^{I B}.
&&
\label{eq:fermion-gaugino}
\end{align}
The last line applies to gauginos of the other gauge factors, whose
adjoint index $I$ is unchanged because $T_H^r$ commutes with those
factors. For the diagonal gauginos,  
\begin{subequations}\label{eq:diagonal-Sp-gaugino-action}
\begin{align}
 \delta\lambda_+^A
 &=+i\epsilon(\sigma_3)_B{}^A\lambda_+^B
 &&\text{for }U(1)_{R+},\\
 \delta\lambda_+^{rA}
 &=2\epsilon^s\epsilon^{st}{}_{r}\lambda_+^{tA}
 +i\epsilon^s(\sigma_s)_B{}^A\lambda_+^{rB}
 &&\text{for }Sp(1)_{R+}.
\end{align}
\end{subequations}
The latter transforms in $\mathbf3\otimes\mathbf2_R$.

On this space, the $\mathfrak{sp}(1)_R$ projection in \eqref{eq:Ci}
is the traceless part of the lower block of $L^{-1}T_{R+}^rL$:
\begin{equation}
 C_+^r=-\frac{i}{2}\bigl[B(\sigma_r-Z^\dagger Q_H^rZ)B\bigr]_0,
 \quad [M]_0=M-\frac12\operatorname{tr}(M)\mathbf1_2.
 \label{eq:SU2R-projection-QH-general_v2}
\end{equation}
For a generator $T_I$ of $K$ acting on $V_H$,
\begin{equation}
 C_I=-\bigl[BZ^\dagger T_IZB\bigr]_0.
 \label{eq:merged-Sp-unitary-maps}
\end{equation}
The $G_E$ moment maps vanish identically because $G_E$ acts trivially
on the hypermultiplet manifold.
For $Sp(1)_{R+}$, all three maps $C_+^r$ must vanish simultaneously
at the same scalar point $Z$; for $U(1)_{R+}$, there is only $C_+=C_+^3$.

\subsection{Examples of Minkowski vacua}
\label{sec:explicit-Minkowski}

We give six examples: four with $U(1)_{R+}$ and two with $Sp(1)_{R+}$.
All six spectra contain only full hypermultiplets and admit the unitary
scalar manifold \eqref{eq:cosetSU}. Furthermore, all solutions necessarily break the $U(1)_{R+}$ or $Sp(1)_{R+}$ symmetry, as we shall show in Section~\ref{sec:r-symmetry-breaking}.

Throughout the six examples below, $\rho'_H$ denotes the remaining massless hypermultiplet spectrum after symmetry breaking, and $n'_H$ its hypermultiplet count.
Interestingly, five of these spectra have been encountered in the search for anomaly free 6D models without R-symmetry gauging~\cite{Avramis:2005hc, Becker:2025xgy, Hamada:2025vga}. The one which has not appeared before has an $SU(3)$ symmetry group which has been left out of these searches.

\section*{$U(1)_{R+}$ gauging.}

Using the moment maps of Section~\ref{sec:unitary-embeddings} with the ansatz
\begin{equation}
Z=zv^T,\quad v=\begin{pmatrix}1\\0\end{pmatrix},
\quad z^\dagger z<1,
\label{eq:redZ}
\end{equation}
all the moment-map equations reduce to
\begin{equation}
z^\dagger(\mathbf1_{n_H}+Q_H)z=2,
\quad z^\dagger T_{z'}^Iz=0\quad\hbox{for every }z',I.
\label{eq:conZZ}
\end{equation}
Let $q_z$ denote the average hyperino charge along $z$, weighted by
the squared amplitudes in its charge blocks. Then
\begin{equation}
q_z=\frac{z^\dagger Q_Hz}{z^\dagger z},\quad
z^\dagger z=\frac{2}{1+q_z}<1
\quad\Longrightarrow\quad q_z>1.
\label{eq:cb}
\end{equation}
For $z$ supported in a single charge block, $q_z=q_I$.
For $v=(0,1)^T$, the first equation instead becomes
$z^\dagger(\mathbf1_{n_H}-Q_H)z=2$; the coordinate bound gives
$1-q_z=2/(z^\dagger z)>2$, hence $q_z<-1$.

One sufficient way to solve the nonabelian equations is to choose
$N\subset K$ such that the adjoint of $K$ has no $N$-singlet, and take
$z$ to be $N$-invariant. Then $z^\dagger T_{z'}^{\ell}z$ is an invariant
covector of the adjoint and must vanish. This subgroup construction does
not establish maximality of $N$. The full unbroken group must instead
be obtained by solving
\begin{equation}
 \rho(k)e^{-i\epsilon(Q_H-\mathbf1)}z=z,
 \quad (k,e^{-i\epsilon})\in K\times U(1)_{R+},
 \label{eq:unbroken-gauge-group}
\end{equation}
where $\rho(k)$ denotes the action of $k\in K$ on $V_H$. Pure continuous $U(1)_{R+}$ is broken at the solutions \eqref{eq:cb}.

The four abelian examples are from the $q_{\max}=2$ blocks of
Tables~\ref{tab:u1-one-combined} and~\ref{tab:u1-two-combined}.
For a block in a real representation of $K$, with hyperino charge $+2$, choose a real unit vector $e$.
Since $Q_He=2e$, \eqref{eq:conZZ} fixes the representative
\begin{equation}
 z=\sqrt{\frac23}\,e,\quad Z=z(1,0),
 \quad z\big|_{\rm other\ blocks}=0.
 \label{eq:real-vector-example-vev}
\end{equation}
Then $e^TTe=0$ for every real antisymmetric generator of $K$, so all
moment maps vanish. A real nonabelian variation cannot cancel the
imaginary $R+$ variation at this point.

In the $SU(7)$ example below, the active $\mathbf{28}$ is complex; its $SU(7)$ moment maps instead vanish by the $SO(7)$-invariance argument given there.

For each model we choose a subgroup $N$ and exhibit
its $n_s$ singlets, their charges and the resulting vacuum equations. Here
$\mathbf1$ denotes a unit $N$-singlet.
The displayed parametrizations use the rank-one ansatz and need not
exhaust all hypermultiplet directions. In each example the broken
nonabelian generators contain no $N$-singlet. The broken $U(1)_{R+}$
vector consumes one singlet hypermultiplet, leaving $n_s-1$ massless
singlet hypermultiplets locally. The tensor dilaton is separate. The
connected groups are understood up to finite central identifications
fixed by the global gauge group; no maximality over other vacua is assumed.

\subsubsection*{\texorpdfstring{\fbox{$SU(7)\times U(1)_{R+}$ model}}{SU(7) x U(1)R+ model}}

Up to simultaneous reversal of all hyperino charges, this model has the spectrum\footnote{The $SU(6)$ vacuum with $n_s=8$ listed on the symplectic Wolf space in Table~\ref{tab:Sp-maximal-combined} also exists on the unitary Wolf space used here.}
\begin{equation}
\begin{aligned}
\rho_H={}5\cdot\mathbf7_{-2}\oplus\mathbf7_{-1}
\oplus3\cdot\mathbf{21}_{-1}\oplus2\cdot\mathbf{21}_0
\oplus\mathbf{28}_{+2}
\oplus\mathbf{35}_{-2}\oplus\mathbf{35}_{+1}
\oplus\mathbf{48}_0,\quad n_H=293,
\end{aligned}
\label{eq:A6-example-spectrum}
\end{equation}
with $b_1=(2,3)$ and $b_{R+}=(30,9)$. We choose $N=SO(7)\subset SU(7)$, under which
\begin{equation}
 \mathbf{48}\longrightarrow\mathbf{21}\oplus\mathbf{27},\quad
 \mathbf{28}\longrightarrow\mathbf{27}\oplus\mathbf1.
 \label{eq:A6-example-branching}
\end{equation}
The $\mathbf7$, $\mathbf{21}$ and $\mathbf{35}$ remain irreducible.
The only singlet is in $\mathbf{28}_{+2}$. Realizing this representation as symmetric matrices with the trace inner product, its general singlet is
\begin{equation}
 z=z_1\mathbf1,\quad
 z\big|_{\mathbf{28}_{+2}}=\frac{z_1}{\sqrt7}\,\mathbf1_7,
 \quad Z=z(1,0).
 \label{eq:A6-example-singlets}
\end{equation}
We use the first column with $q_z=+2$, so the scalar has charge $+1$.
The moment-map equations and coordinate bound become
\begin{equation}
 3|z_1|^2=2,\quad |z_1|^2<1.
 \label{eq:A6-example-constraint}
\end{equation}
Since the $SU(7)$ adjoint in \eqref{eq:A6-example-branching} contains
no $SO(7)$-singlet, equivariance forces all $SU(7)$ moment maps to
vanish on this $SO(7)$-invariant configuration.

A representative is
\begin{equation}
 z\big|_{\mathbf{28}_{+2}}=\sqrt{\frac{2}{21}}\,\mathbf1_7,
 \quad z\big|_{\rm other\ blocks}=0,
 \label{eq:A6-example-vev}
\end{equation}
for which the matrix before the traceless projection in \eqref{eq:SU2R-projection-QH-general_v2} is $-\mathbf1_2$. Its connected unbroken group is $SO(7)$. The model is ghost-free for every finite $\phi$.

The remaining massless hypermultiplet spectrum is
\begin{equation}
\begin{aligned}
\rho'_H={}6\cdot\mathbf7
\oplus6\cdot\mathbf{21}\oplus\mathbf{27}
\oplus2\cdot\mathbf{35}
,\quad n'_H=265,
\end{aligned}
\end{equation}

\subsubsection*{\texorpdfstring{\fbox{$G_2\times U(1)_{R+}$ model}}{G2 x U(1)R+ model}}

This model has the spectrum
\begin{equation}
\begin{aligned}
 \rho_H={}&2\cdot\mathbf7_{+2}\oplus2\cdot\mathbf7_{+1}
 \oplus\mathbf{14}_{+1}\oplus9\cdot\mathbf{14}_0
 \oplus\mathbf{27}_{+1}\oplus\mathbf{64}_{+1},
 \quad n_H=259,
\end{aligned}
\label{eq:G2-example-spectrum}
\end{equation}
with $b_1=(13/2,7)$ and $b_{R+}=(9,6)$. We choose
$N=SU(3)\subset G_2$, under which
\begin{align}
\mathbf{adj}_{G_2}=\mathbf{14}&\longrightarrow\mathbf8\oplus\mathbf3\oplus\overline{\mathbf3},\quad
\mathbf7\longrightarrow\mathbf1\oplus\mathbf3\oplus\overline{\mathbf3},
\nonumber\\
\mathbf{27}&\longrightarrow\mathbf1\oplus\mathbf3\oplus\overline{\mathbf3}
\oplus\mathbf6\oplus\overline{\mathbf6}\oplus\mathbf8,
\nonumber\\
\mathbf{64}&\longrightarrow\mathbf3\oplus\overline{\mathbf3}
\oplus\mathbf6\oplus\overline{\mathbf6}\oplus2\cdot\mathbf8
\oplus\mathbf{15}\oplus\overline{\mathbf{15}}.
\label{eq:G2-example-branching}
\end{align}
The last branching follows from
$\mathbf7\otimes\mathbf{14}=\mathbf7\oplus\mathbf{27}\oplus\mathbf{64}$.
The general singlet is
\begin{equation}
z=z_1\mathbf1\oplus z_2\mathbf1\oplus z_3\mathbf1
\oplus z_4\mathbf1\oplus z_5\mathbf1,
\quad Z=z(1,0),
\label{eq:G2-example-singlets}
\end{equation}
where the terms come, in order, from $2\cdot\mathbf7_{+2}$,
$2\cdot\mathbf7_{+1}$ and $\mathbf{27}_{+1}$. Their hyperino charges are
$(2,2,1,1,1)$ and scalar charges explained below \eq{eq:Z-finite-transformation} are $(1,1,0,0,0)$.
The remaining vacuum equations reduce to
\begin{equation}
3\bigl(|z_1|^2+|z_2|^2\bigr)
+2\bigl(|z_3|^2+|z_4|^2+|z_5|^2\bigr)=2,
\quad \sum_{j=1}^5|z_j|^2<1.
\label{eq:G2-example-constraint}
\end{equation}
The strict bound requires $0<|z_1|^2+|z_2|^2\leq2/3$.
Taking $z_1=\sqrt{2/3}$ and the other coefficients zero gives
\eqref{eq:real-vector-example-vev} in a $\mathbf7_{+2}$ block.
The model is ghost-free for every finite $\phi$. Moreover, the remaining massless hypermultiplet spectrum is
\begin{equation}
\begin{aligned}
\rho'_H={}4\cdot\mathbf1\oplus30\cdot\mathbf3\oplus4\cdot\mathbf6\oplus13\cdot\mathbf8\oplus2\cdot\mathbf{15},\quad n'_H=252.
\end{aligned}
\end{equation}
\subsubsection*{\texorpdfstring{\fbox{$SU(7)\times\mathrm{Spin}(18)\times U(1)_{R+}$ model}}{SU(7) x Spin(18) x U(1)R+ model}}
This model has the spectrum
\begin{equation}
\begin{aligned}
 \rho_H={}&(\mathbf1,\mathbf{18})_{+2}\oplus2(\mathbf1,\mathbf{18})_0
 \oplus6(\mathbf7,\mathbf1)_0\oplus8(\mathbf7,\mathbf1)_{-1}
 \oplus2(\mathbf7,\mathbf1)_{-2}\\
 &\oplus(\mathbf7,\mathbf{18})_{+1}
 \oplus2(\mathbf{21},\mathbf1)_0\oplus2(\mathbf{21},\mathbf1)_{-1}
 \oplus(\mathbf{35},\mathbf1)_{+1}\oplus(\mathbf{35},\mathbf1)_{-1},
 \quad n_H=446,
\end{aligned}
\label{eq:A6D9-example-spectrum}
\end{equation}
with $b_1=(1,4)$, $b_2=(1,-2)$ and $b_{R+}=(10,10)$.
We choose $N=SU(7)\times\mathrm{Spin}(17)$. The nontrivial branchings are
\begin{align}
(\mathbf1,\mathbf{18})&\longrightarrow(\mathbf1,\mathbf{17})\oplus(\mathbf1,\mathbf1),\quad
(\mathbf7,\mathbf{18})\longrightarrow(\mathbf7,\mathbf{17})\oplus(\mathbf7,\mathbf1).
 \label{eq:A6D9-example-branching}
\end{align}
The $(\mathbf7,\mathbf1)$, $(\mathbf{21},\mathbf1)$ and
$(\mathbf{35},\mathbf1)$ are unchanged. The general singlet is
\begin{equation}
 z=z_1\mathbf1\oplus z_2\mathbf1\oplus z_3\mathbf1,
 \quad Z=z(1,0),
 \label{eq:A6D9-example-singlets}
\end{equation}
from the three  copies of $(\mathbf1,\mathbf{18})$, with hyperino charges
$(2,0,0)$ and scalar charges $(1,-1,-1)$. The remaining vacuum equations become
\begin{equation}
 3|z_1|^2+|z_2|^2+|z_3|^2=2,
 \quad |z_1|^2+|z_2|^2+|z_3|^2<1.
 \label{eq:A6D9-example-constraint}
\end{equation}
Thus $1/2<|z_1|^2\leq2/3$. The choice $z_1=\sqrt{2/3}$,
$z_2=z_3=0$ is \eqref{eq:real-vector-example-vev} in
$(\mathbf1,\mathbf{18})_{+2}$. Its connected unbroken group is
$SU(7)\times\mathrm{Spin}(17)$. The common positive kinetic domain is $e^{2\phi}>2$. 

The remaining massless hypermultiplet spectrum is
\begin{equation}
\begin{aligned}
 \rho'_H={}&2(\mathbf1,\mathbf{1})
 \oplus2(\mathbf1,\mathbf17)\oplus17(\mathbf7,\mathbf1)
 \oplus(\mathbf7,\mathbf{17})
 \oplus4(\mathbf{21},\mathbf1)\oplus2(\mathbf{35},\mathbf1),
 \quad n'_H=428,
\end{aligned}
\end{equation}
\subsubsection*{\texorpdfstring{\fbox{$\mathrm{Spin}(12)\times E_6\times U(1)_{R+}$ model}}{Spin(12) x E6 x U(1)R+ model}}
This model has the spectrum
\begin{equation}
\begin{aligned}
\rho_H={}&3(\mathbf1,\mathbf{27})_{+1}
\oplus(\mathbf{12},\mathbf1)_{+2}\oplus3(\mathbf{12},\mathbf1)_0
\oplus3(\mathbf{32},\mathbf1)_0\\
&\oplus(\mathbf{32}',\mathbf1)_{+1}
\oplus2(\mathbf{66},\mathbf1)_{+1},\quad n_H=389,
\end{aligned}
\label{eq:D6E6-example-spectrum}
\end{equation}
with $b_1=(2,3)$, $b_2=(1,-3/2)$ and $b_{R+}=(8,6)$.
We choose $N=\mathrm{Spin}(11)\times E_6$. The branchings are
\begin{align}
 \mathbf{adj}_{\mathrm{Spin}(12)}&\longrightarrow(\mathbf{55},\mathbf1)\oplus(\mathbf{11},\mathbf1),
 \nonumber\\
 (\mathbf{12},\mathbf1)&\longrightarrow(\mathbf{11},\mathbf1)\oplus(\mathbf1,\mathbf1),\quad
 (\mathbf{32},\mathbf1),\ (\mathbf{32}',\mathbf1)\longrightarrow(\mathbf{32},\mathbf1),
 \nonumber\\
 (\mathbf{66},\mathbf1)&\longrightarrow(\mathbf{55},\mathbf1)\oplus(\mathbf{11},\mathbf1),
 \quad (\mathbf1,\mathbf{27})\longrightarrow(\mathbf1,\mathbf{27}).
 \label{eq:D6E6-example-branching}
\end{align}
The general singlet is
\begin{equation}
 z=z_1\mathbf1\oplus z_2\mathbf1\oplus z_3\mathbf1\oplus z_4\mathbf1,
 \quad Z=z(1,0),
 \label{eq:D6E6-example-singlets}
\end{equation}
from the four copies of $(\mathbf{12},\mathbf1)$, with hyperino charges
$(2,0,0,0)$ and scalar charges $(1,-1,-1,-1)$. The remaining equations are
\begin{equation}
 3|z_1|^2+|z_2|^2+|z_3|^2+|z_4|^2=2,
 \quad \sum_{j=1}^4|z_j|^2<1.
 \label{eq:D6E6-example-constraint}
\end{equation}
Again $1/2<|z_1|^2\leq2/3$, and $z_1=\sqrt{2/3}$ with the other
coefficients zero gives \eqref{eq:real-vector-example-vev}.
The connected unbroken group is $\mathrm{Spin}(11)\times E_6$. The common kinetic domain is
$e^{2\phi}>3/2$.

The remaining massless hypermultiplet spectrum is 
\begin{equation}
\begin{aligned}
\rho'_H={}&3(\mathbf1,\mathbf1)
\oplus3(\mathbf1,\mathbf27)\oplus5(\mathbf{11},\mathbf1)
\oplus4(\mathbf{32},\mathbf1)\oplus2(\mathbf{55},\mathbf1),\quad n'_H=377,
\end{aligned}
\end{equation}

For the symplectic Wolf space, Appendix~\ref{app:maximal-symmetry} gives the unbroken groups of the largest dimension found and their singlet counts for all representatives in the $q_{\max}=2$ blocks of Tables~\ref{tab:u1-one-combined} and~\ref{tab:u1-two-combined}.

\section*{$Sp(1)_{R+}$ gauging}

On the unitary Wolf space, choose $R\otimes\mathbf d\subset V_H$ with
$d\geq3$ and a unit vector $e\in R$, $e^\dagger e=1$.
Here $R$ may combine $K$-representation blocks carrying the same $\mathbf d$.
In the weight order $j,j-1,\ldots,-j$, with $j=(d-1)/2$, take
\begin{equation}
 Z\big|_{R\otimes\mathbf d}
 =e\otimes\begin{pmatrix}\operatorname{diag}(x,y)\\\mathbf0_{(d-2)\times2}\end{pmatrix},
 \quad Z\big|_{\rm other\ blocks}=0,\quad 0<x,y<1.
 \label{eq:merged-Sp-two-amplitudes}
\end{equation}
The nonzero entries $x$ and $y$ occupy the first two weight rows. Substituting into
\eqref{eq:SU2R-projection-QH-general_v2}, the three conditions
$C_+^r=0$ reduce to
\begin{equation}
\sqrt{d-1}\,xy=1,\quad
\frac{1-(d-1)x^2}{1-x^2}
=\frac{-1-(d-3)y^2}{1-y^2}.
\label{eq:merged-Sp-amplitude-equations}
\end{equation}
The solution with $0<x^2,y^2<1$ is
\begin{equation}
x^2=\frac{1+(d-2)/\sqrt d}{d-1},\quad
y^2=\frac{\sqrt d}{\sqrt d+d-2}.
\label{eq:merged-Sp-profile}
\end{equation}
Both values lie strictly between zero and one for $d>2$, so this
is a finite zero of all three diagonal maps. Substituting this solution
into \eqref{eq:merged-Sp-unitary-maps} gives the remaining maps for
this ansatz:
\begin{equation}
C_I=-\frac{e^\dagger T_Ie}{d-2}\,\sigma_3.
\label{eq:merged-Sp-external-residual}
\end{equation}
They vanish when $e^\dagger T_Ie=0$ for every generator of $K$.
In particular, this holds for a real representation $R$ and a real
vector $e$ in an orthonormal basis, where the generators are real
antisymmetric.  

\subsubsection*{\texorpdfstring{\fbox{$SU(6)\times Sp(1)_{R+}$ model}}{SU(6) x Sp(1)R+ model}}
The unitary-compatible $A_5$ model in
Table~\ref{sp:tab:sp1-one-combined} has the spectrum
\begin{equation}
 \begin{aligned}
 \rho_H={}&11(\mathbf6,\mathbf1)\oplus3(\mathbf6,\mathbf3)
 \oplus2(\mathbf{20},\mathbf1)\oplus2(\mathbf{20},\mathbf2)
 \oplus(\mathbf{21},\mathbf2),\quad n_H=282,
 \end{aligned}
 \label{eq:merged-Sp-SU6-example}
\end{equation}
with $b_1=(2,3)$ and $b_{R+}=(6,4)$. The four half-hypers in
$(\mathbf{20},\mathbf1)$ have been paired into two full hypers.
We choose $N=SU(5)$, under which
\begin{align}
 \mathbf6&\longrightarrow\mathbf5\oplus\mathbf1,\quad
 \mathbf{20}\longrightarrow\mathbf{10}\oplus\overline{\mathbf{10}},\quad
 \mathbf{21}\longrightarrow\mathbf{15}\oplus\mathbf5\oplus\mathbf1,\quad
 \mathbf{35}\longrightarrow\mathbf{24}\oplus\mathbf5\oplus\mathbf{\bar{5}}\oplus\mathbf1.
 \label{eq:merged-Sp-SU6-branching}
\end{align}
The $N$-singlets have diagonal hyperino representations
$\mathbf1(\times 11)\oplus\mathbf3(\times 3)\oplus\mathbf2$, giving $n_s=22$.
Each complex full hyper supplies both a representation and its
conjugate subject to the symplectic reality condition. For two of the triplet hypers choose
conjugate representation blocks in $V_H$, giving active
$(\mathbf6,\mathbf3)$ and $(\overline{\mathbf6},\mathbf3)$ blocks.
Let $e$ be the real unit vector fixed by $SU(5)$ in $\mathbf6$.
Take
\begin{equation}
\begin{aligned}
Z\big|_{(\mathbf6,\mathbf3)}
&=\frac{e}{\sqrt2}\otimes
\begin{pmatrix}z_1&0\\0&z_2\\0&0\end{pmatrix},\quad
Z\big|_{(\overline{\mathbf6},\mathbf3)}
=\frac{\overline e}{\sqrt2}\otimes
\begin{pmatrix}z_1&0\\0&z_2\\0&0\end{pmatrix},\quad
Z\big|_{\rm other\ blocks}=0.
\end{aligned}
\label{eq:merged-Sp-SU6-singlets}
\end{equation}
The three rows have $J^3$ weights $1,0,-1$.
Since $T_I$ is anti-Hermitian, the $K$-generator bilinears cancel:
\begin{equation}
 \frac12\left(e^\dagger T_Ie+
 \overline e^{\,\dagger}\overline{T_I}\,\overline e\right)=0.
 \label{eq:merged-Sp-SU6-external-check}
\end{equation}
The diagonal equations are
\begin{equation}
 \sqrt2\,\overline z_1z_2=1,\quad
 \frac{1-2|z_1|^2}{1-|z_1|^2}=-\frac1{1-|z_2|^2},\quad
 |z_1|^2,|z_2|^2<1.
 \label{eq:merged-Sp-SU6-constraint}
\end{equation}
Choosing $z_1,z_2>0$, we obtain
\begin{equation}
 z_1^2=\frac{3+\sqrt3}{6},\quad
 z_2^2=\frac{3-\sqrt3}{2}.
 \label{eq:merged-Sp-SU6-vev}
\end{equation}
%
The model is ghost-free for every finite $\phi$. The connected unbroken group is $SU(5)$, and the remaining massless hypermultiplet spectrum is
\begin{equation}
 \begin{aligned}
\rho'_H={}18\cdot\mathbf1\oplus20\cdot\mathbf5\oplus12\cdot\mathbf{10}\oplus2\cdot\mathbf{15},\quad n'_H=268.
 \end{aligned}
\end{equation}

\subsubsection*{\texorpdfstring{\fbox{$SU(6)\times\mathrm{Spin}(12)\times Sp(1)_{R+}$ model}}{SU(6) x Spin(12) x Sp(1)R+ model}}

This model with two additional simple gauge factors has the spectrum
\begin{equation}
\begin{aligned}
\rho_H={}&(\mathbf{21},\mathbf1,\mathbf1)\oplus(\mathbf{21},\mathbf1,\mathbf2)
\oplus(\mathbf{15},\mathbf1,\mathbf1)\oplus(\mathbf{15},\mathbf1,\mathbf2)\\
&\oplus2(\mathbf1,\mathbf{12},\mathbf1)\oplus(\mathbf1,\mathbf{12},\mathbf2)
\oplus2(\mathbf{20},\mathbf1,\mathbf1)\oplus2(\mathbf{20},\mathbf1,\mathbf2)\\
&\oplus2(\mathbf6,\mathbf1,\mathbf1)\oplus2(\mathbf6,\mathbf1,\mathbf2)
\oplus2(\mathbf6,\mathbf1,\mathbf3),\quad n_H=348,
\end{aligned}
\label{eq:merged-Sp-SU6D6-example}
\end{equation}
with $b_1=(4,2)$, $b_2=(-2,1)$ and $b_{R+}=(3,11/2)$.
We choose
$N=SU(5)\times\mathrm{Spin}(11)$. The $SU(6)$ branchings in
\eqref{eq:merged-Sp-SU6-branching} apply, with the additional branchings
\begin{align}
 (\mathbf{15},\mathbf1)\longrightarrow
 (\mathbf{10},\mathbf1)\oplus(\mathbf5,\mathbf1),\quad
 (\mathbf1,\mathbf{12})\longrightarrow
 (\mathbf1,\mathbf{11})\oplus(\mathbf1,\mathbf1).
 \label{eq:merged-Sp-SU6D6-branching}
\end{align}
The $N$-singlets have diagonal hyperino representations
$\mathbf1(\times 5)\oplus\mathbf2(\times 4)\oplus\mathbf3(\times 2)$, giving
$n_s=19$.

For the two full triplet hypers, choose complex conjugate representation blocks
in $V_H$ as in the preceding example. Let $e$ be the real unit vector
fixed by $SU(5)$ in $\mathbf6$, and $u$ the real unit vector fixed by
$\mathrm{Spin}(11)$ in $\mathbf{12}$. In addition to the paired
triplet blocks, turn on one of the two $(\mathbf1,\mathbf{12},\mathbf1)$
blocks:
\begin{equation}
\begin{aligned}
Z\big|_{(\mathbf6,\mathbf1,\mathbf3)}
&=\frac{e}{\sqrt2}\otimes
\begin{pmatrix}z_1&0\\0&z_2\\0&0\end{pmatrix},\quad
Z\big|_{(\overline{\mathbf6},\mathbf1,\mathbf3)}
=\frac{\overline e}{\sqrt2}\otimes
\begin{pmatrix}z_1&0\\0&z_2\\0&0\end{pmatrix},\\
Z\big|_{(\mathbf1,\mathbf{12},\mathbf1)}
&=z_3\,u(1,0),\quad Z\big|_{\rm other\ blocks}=0,
\end{aligned}
\label{eq:merged-Sp-SU6D6-singlets}
\end{equation}
where $u$ is a fixed real unit vector in $\mathbf{12}$, $u^Tu=1$. The three triplet rows have $J^3$ weights $1,0,-1$. The $SU(6)$ moment maps cancel by \eqref{eq:merged-Sp-SU6-external-check}. The additional block has $Q_H^r=0$, and its $\mathrm{Spin}(12)$ bilinears vanish because $u^TT_Xu=0$ for real antisymmetric generators. The remaining equations are
\begin{equation}
\sqrt2\,\overline z_1z_2=1,\quad
\frac{1-2|z_1|^2}{1-|z_1|^2-|z_3|^2}
=-\frac1{1-|z_2|^2},\quad
|z_1|^2+|z_3|^2<1,\quad |z_2|^2<1.
\label{eq:merged-Sp-SU6D6-constraint}
\end{equation}
One solution with $z_1,z_2,z_3>0$ is
\begin{equation}
 z_1^2=\frac23,\quad z_2^2=\frac34,\quad z_3^2=\frac14.
 \label{eq:merged-Sp-SU6D6-vev}
\end{equation}
Here $Z^\dagger Z=\operatorname{diag}(11/12,3/4)$, and the matrices
$B(\sigma_r-Z^\dagger Q_H^rZ)B$ are zero for $r=1,2$ and
$-4\mathbf1_2$ for $r=3$. Thus all three diagonal moment maps vanish. The common positive kinetic domain is $e^{2\phi}<1/2$. 
The connected unbroken group at this vacuum is
$SU(5)\times\mathrm{Spin}(11)$, and the remaining massless hypermultiplet spectrum is
\begin{equation}
 \begin{aligned}
\rho'_H={}15(\mathbf1,\mathbf1)\oplus16(\mathbf5,\mathbf1)\oplus15(\mathbf{10},\mathbf1)\oplus3(\mathbf{15},\mathbf1)\oplus3(\mathbf1,\mathbf{11}),\quad n'_H=323.
 \end{aligned}
\end{equation}

\section{Charge bound and R-symmetry breaking}
\label{sec:charge-bound-wolf-general-unified45}

In this section, we shall show that the existence of a Minkowski$_6$ solution in $U(1)_{R+}$ gauged theories requires the maximum absolute charge $q_{\rm max}$ carried by the hyperinos to obey $q_{\rm max} \ge 2$, and in the $Sp(1)_{R+}$ gauged theories the maximum dimension $d_{\rm max}$ of any $Sp(1)_{R+}$ representation in the hyperino spectrum must obey the condition $d_{\rm max} \ge 3$. We shall also show that the $U(1)_{R+}$ and $Sp(1)_{R+}$ symmetries are necessarily broken in the Minkowski solutions.

\subsection{Charge bound}
\label{sec:charge-bound}

On a connected noncompact Wolf space, the Lie algebra
decomposes as
\begin{equation}
 \mathfrak g_H
 =\mathfrak h\oplus\mathfrak{sp}(1)_R\oplus\mathfrak p.
 \label{eq:general-wolf-decomposition-unified45}
\end{equation}
Here $\mathfrak p$ is the $4n_H$-dimensional real scalar
tangent representation. Let $T_{\widehat{I}}$ be a fixed generator of a compact gauged subgroup of $H\times Sp(1)_R$. We choose a compact Cartan subalgebra containing $T_{\widehat{I}}$, with basis $H_i$. Next, we define 
\begin{equation}
\kappa_{ij}=-\widehat{\tr}(H_iH_j),\quad
\lambda_{\widehat{I}i}=-2\widehat{\tr}(T_{\widehat{I}}H_i).
\label{eq:Cartan-components-unified45}
\end{equation}
where $\widehat{\tr}(T_{\widehat{I}}T_{\widehat{J}})$ is the invariant bilinear form on $\mathfrak g_H$
normalized so that its restriction to \(\mathfrak{sp}(1)_R\) equals the trace in the two-dimensional fundamental representation.\footnote{Here $\widehat{\tr}$ is the trace in the defining (fundamental) representation for $SU(n_H,2)$ and $Sp(n_H,1)$, and half the defining vector trace for $SO(n_H,4)$, whose vector contains two $Sp(1)_R$ doublets. For exceptional groups, the invariant form is similarly normalized to restrict to the $Sp(1)_R$ doublet trace.}
Cartan indices are $i,j=1,\ldots, \operatorname{rank}(\mathfrak h\oplus\mathfrak{sp}(1)_R)$. $\kappa_{ij}$ is positive definite on the compact Cartan, $\kappa^{ij}$ denotes its inverse, and $\lambda_{\widehat{I}i}:=\kappa_{ij}\lambda^j_{\widehat{I}}$. A root $\alpha$ is specified by real numbers $\alpha_i$ and a nonzero root vector $E_\alpha$ such that
\begin{equation}
 [H_i,E_\alpha]=-i\alpha_iE_\alpha,
 \quad (\alpha,\lambda_{\widehat{I}}):=\alpha_i\kappa^{ij}\lambda_{\widehat{I}j}.
 \label{eq:root-components-unified45}
\end{equation}
A noncompact root vector $E_\alpha\in\mathfrak p_{\mathbb C}$ has components $(E_\alpha)^{aA}$.

$T_{\widehat{I}}$ can be expressed as a linear combination of the Cartan generators $H_i$: $T_{\widehat{I}}=\frac12 \lambda^i_{\widehat{I}}H_i$. After a suitable isotropy rotation at each scalar point, Paradan's lemma~\cite[Section~2.5, Lemma~2.12]{Paradan:2015} gives the following projection onto any compact generator $T_{\widehat{J}}\in\mathfrak h\oplus\mathfrak{sp}(1)_R$:
\begin{equation}
-2\widehat{\tr}(L^{-1}T_{\widehat{I}}LT_{\widehat{J}})
=-2\widehat{\tr}(T_{\widehat{I}}T_{\widehat{J}})
-\sum_{\substack{\alpha\ {\rm noncompact}\\
(\alpha,\lambda_{\widehat{I}})>0}}
c_\alpha(L,T_{\widehat{I}})\alpha^i\widehat{\tr}(H_iT_{\widehat{J}}),
\quad c_\alpha(L,T_{\widehat{I}})\ge0,
\label{eq:paradan}
\end{equation}
We use the complex generator basis 
\begin{align}
T_{AB}=-\frac{i}{2}T_R^r\s^r_{AB}, 
\label{TAB}
\end{align}
which gives $(T_{AB})^C{}_D=\tfrac14(\delta_A^C\epsilon_{BD}+\delta_B^C\epsilon_{AD})$. Projecting onto the $Sp(1)_R$ generators $T_{AB}$ gives
\begin{equation}
\begin{gathered}
 C_{\widehat{I}}(L)_{AB}
 =-2\widehat{\tr}(L^{-1}T_{\widehat{I}}LT_{AB})
 =\lambda_{\widehat{I}AB}
 +\sum_{\substack{\alpha\ {\rm noncompact}\\
                  (\alpha,\lambda_{\widehat{I}})>0}}
 c_\alpha(L,T_{\widehat{I}})\alpha_{AB}.
\end{gathered}
 \label{eq:paradan-AB-unified45}
\end{equation}
Here $\lambda_{\widehat{I}AB}:=-2\widehat{\tr}(T_{\widehat{I}}T_{AB})$ and $\alpha_{AB}:=-\alpha^i\widehat{\tr}(H_iT_{AB})$. The same coefficients $c_\alpha(L,T_{\widehat{I}})$ apply to every projection.

\paragraph{Diagonal gauging.}
Take $T_{\widehat{I}}$ to be our diagonal gauging generator $T_{R+}^r=T_H^r+T_R^r$, with $r=3$ for $U(1)_{R+}$ and $r=1,2,3$ for $Sp(1)_{R+}$. We fix $r$ throughout the following argument, without summing over it, and denote the corresponding covector by $\lambda^{(r)}$. The root vectors satisfy
\begin{equation}
 [2T_H^r,E_\alpha]=-iq_\alpha E_\alpha,\quad
 [2T_R^r,E_\alpha]=-ir_\alpha E_\alpha,
 \quad r_\alpha=\pm1.
 \label{eq:diagonal-weight-components-unified45}
\end{equation}
In particular, for $r=3$,
\begin{align}
[T_{R+},E_\a]=-i\a_+ E_\a ,\qquad \a_+:=\frac12(q_\a+r_\a).
\end{align}
Here $q_\alpha$ and $r_\alpha$ are the charges under $2T_H^r$ and $2T_R^r$, respectively, and $q_{\max}=\max_\alpha|q_\alpha|$; $r$ labels the fixed generator. For an $Sp(1)_H$ representation of dimension $d=2j+1$,
\begin{equation}
 q_\alpha=-2j,-2j+2,\ldots,2j,
 \quad q_{\max}=d_{\max}-1.
 \label{eq:unified-Wolf-Cartan-charges}
\end{equation}
For each fixed $r$, we may choose the $Sp(1)_R$ basis without loss of generality so that $T_R^r=-i\sigma^3/2$. Consequently,
\begin{equation}
\begin{aligned}
(\alpha,\lambda^{(r)})=q_\alpha+r_\alpha,
\quad \alpha_{12}=\frac{r_\alpha}{2}\lambda^{(r)}_{12},\quad
\lambda^{(r)}_{12}=-\frac i2.
\end{aligned}
\label{eq:diagonal-R-components-unified45}
\end{equation}
We used $\mathfrak h\perp\mathfrak{sp}(1)_R$ and the trace normalization above.
For $q_{\max}\leq1$, $(\alpha,\lambda^{(r)})>0$ forces $r_\alpha=+1$, since $r_\alpha=-1$ would require $q_\alpha>1$. The $12$ component of \eqref{eq:paradan-AB-unified45}, with $c_\alpha^{(r)}(L):=c_\alpha(L,T_{R+}^r)$, gives 
\begin{equation}
\frac{(C_+^r)_{12}}{\lambda^{(r)}_{12}}=1+\frac12
\sum_{\substack{\alpha\ {\rm noncompact}\\(\alpha,\lambda^{(r)})>0}}c_\alpha^{(r)}(L)
\geq1.
\label{eq:positive-R-component-unified45}
\end{equation}
Thus $C_+^r\neq0$ at every finite scalar point, independently of the isotropy frame and without imposing a scalar ansatz. The Minkowski conditions of Section~\ref{sec:vacuum-conditions} therefore exclude finite vacua for $q_{\max}\leq1$ in the domain $f_z>0$. In particular, for $Sp(1)_{R+}$,
\begin{equation}
 C_+^r=0\quad(r=1,2,3)
 \quad\Longrightarrow\quad d_{\max}\geq3.
 \label{eq:unified-Sp-Wolf-bound}
\end{equation}

For these charges, \eqref{eq:positive-R-component-unified45} also bounds the invariant norm of each diagonal moment map below by its value at the origin. Since the remaining moment maps contribute nonnegative terms, the potential satisfies
\begin{align}
 V(L,\phi)\ge V(0,\phi)
\end{align}
at fixed $\phi$ in the domain $f_z>0$. Thus the origin is a minimum in the hyperscalar directions for $q_{\max}\leq1$ in the $U(1)_{R+}$ case and $d_{\max}\leq2$ in the $Sp(1)_{R+}$ case; flat directions are not excluded.

For $q_{\max}>1$, the same ratio instead reads
\begin{equation}
\frac{(C_+^r)_{12}}{\lambda^{(r)}_{12}}=1+\frac12\sum_{r_\alpha=+1}c_\alpha^{(r)}(L)
-\frac12\sum_{r_\alpha=-1}c_\alpha^{(r)}(L).
\label{eq:signed-R-component-unified45}
\end{equation}
Both sums run over noncompact roots satisfying $(\alpha,\lambda^{(r)})>0$. Roots with $q_\alpha>1$ and $r_\alpha=-1$ can contribute to the negative sum, so the positive lower bound in \eqref{eq:positive-R-component-unified45} no longer follows. Thus $q_{\max}>1$ is necessary for a finite Minkowski vacuum.

\subsection{\texorpdfstring{Solving $C_+=0$ for $U(1)_{R+}$}{Solving C+ = 0 for U(1)R+}}
\label{sec:u1-moment-map-zero}

For $U(1)_{R+}$, set $r=3$ and suppress this label.
We construct a finite zero of $C_+$ whenever $q_{\max}>1$, in particular for the $q_{\max}=2$ spectra.
Since the doublet factor $\mathbf2_R$ supplies both R-charges $\pm1$ for each hyperino weight, we can choose a noncompact root $\alpha_*$ whose hyperino and R-symmetry charges, $q_*$ and $r_*$, satisfy
\begin{equation}
 |q_*|>1,\quad r_*=-\operatorname{sgn}(q_*),\quad
 \alpha_{*+}=\frac{q_*+r_*}{2}.
 \label{eq:existence-root-unified45}
\end{equation}
Define the squared root length\footnote{In our normalization, noncompact roots have $\alpha_*^2=2$ for $SU(n_H,2)$, $SO(n_H,4)$ with even $n_H$, and the $E_{6(2)}$, $E_{7(-5)}$, $E_{8(-24)}$ Wolf spaces; $\alpha_*^2=1$ for $Sp(n_H,1)$ and $SO(1,4)$; $\alpha_*^2=1$ or $2$ for $SO(n_H,4)$ with odd $n_H\ge3$ and $F_{4(4)}$; and $\alpha_*^2=\tfrac23$ or $2$ for $G_{2(2)}$. When two lengths occur, the relevant value depends on the root satisfying $|q_*|>1$ for the chosen gauging.} and compact coroot by
\begin{equation}
 \alpha_*^2=(\alpha_*)_i\kappa^{ij}(\alpha_*)_j>0,\quad
 K_*=\frac{2\kappa^{ij}(\alpha_*)_j}{\alpha_*^2}H_i.
 \label{eq:coroot-normalization-unified45}
\end{equation}
Choose $E_{*\pm}:=E_{\pm\alpha_*}$ to be complex conjugates with respect to the real Lie algebra $\mathfrak g_H$, normalized so that $[E_{*+},E_{*-}]=iK_*$. The combinations $X_*=E_{*+}+E_{*-}$ and $Y_*=-i(E_{*+}-E_{*-})$ are real elements of $\mathfrak p$ and satisfy the $\mathfrak{sl}(2,\mathbb R)$ relations
\begin{equation}
\displaystyle
[K_*,X_*]=2Y_*,\quad
[K_*,Y_*]=-2X_*,\quad
[X_*,Y_*]=-2K_*.
\label{eq:root-real-triple-unified45}
\end{equation}
Choosing the group element $L(s)=\exp(sX_*)$, with $s\in\mathbb R$, 
we get 
\begin{equation}
L(s)^{-1}T_{R+}L(s)=T_{R+}
+\frac{q_*+r_*}{4}\left[(\cosh 2s-1)K_*+\sinh 2s\,Y_*\right].
\label{eq:exact-root-path-unified45}
\end{equation}
The $Y_*$ term is trace-orthogonal to $T_{AB}$ defined in \eqref{TAB}. Substituting \eqref{eq:coroot-normalization-unified45} and using \eqref{eq:paradan-AB-unified45} gives
\begin{align}
 -2\widehat{\tr}(K_*T_{AB})
 =-\frac{4}{\alpha_*^2}\kappa^{ij}(\alpha_*)_j
       \widehat{\tr}(H_iT_{AB})
 =\frac{4}{\alpha_*^2}(\alpha_*)_{AB}.
\end{align}
In our chosen compact Cartan, the $Sp(1)_R$ projections of $K_*$ and $T_{R+}$ are both along $T_R$. Hence $(\alpha_*)_{AB}$ is proportional to $\lambda_{AB}$; the $12$ component in \eqref{eq:diagonal-R-components-unified45} fixes the coefficient:
\begin{align}
 (\alpha_*)_{AB}
 =\frac{(\alpha_*)_{12}}{\lambda_{12}}\lambda_{AB}
 =\frac{r_*}{2}\lambda_{AB}.
\end{align}
Combining these identities, we obtain
\begin{equation}
 -2\widehat{\tr}(K_* T_{AB})
 =\frac{2r_*\lambda_{AB}}{\alpha_*^2}.
 \label{eq:coroot-R-projection-unified45}
\end{equation}
Projecting \eqref{eq:exact-root-path-unified45} on $Sp(1)_R$ and using $r_*q_*=-|q_*|$ therefore yields
\begin{equation}
\displaystyle
(C_+)_{AB}\big|_{L(s)}
=\left[1-\frac{|q_*|-1}{\alpha_*^2}\sinh^2s\right]\lambda_{AB}.
\label{eq:exact-diagonal-zero-path-unified45}
\end{equation}
All $AB$ components of $C_+$ vanish at the same finite point $L(s_*)$, where
\begin{equation}
\sinh^2s_* = \frac{\alpha_*^2}{|q_*|-1}.
\label{eq:finite-diagonal-zero-unified45}
\end{equation}
Together with \eqref{eq:positive-R-component-unified45}, this proves
that $q_{\max}>1$ is necessary and sufficient for a finite zero of
$C_+$ on every admissible Wolf space.

The same path also shows that the abelian de Sitter point \eqref{eq:origin-dS-stationary}, when it exists, has a tachyonic hyperscalar direction for $q_{\max}>1$. The remaining $K$ moment maps start at order $s^2$, so they contribute only at order $s^4$ to the potential. Expanding \eqref{eq:exact-diagonal-zero-path-unified45} therefore gives
\begin{align}
 V(L(s),\phi_0)=V_0\left[1-\frac{2(|q_*|-1)}{\alpha_*^2}s^2\right]+O(s^4),
\end{align}
whose quadratic term is negative since $|q_*|>1$.

\subsection{\texorpdfstring{Solving of $C^r_+=0$ for $Sp(1)_{R+}$}{Solving of C(r)+ = 0 for Sp(1)R+}}
\label{sec:sp1-moment-map-zeros}

We consider the three classical Wolf-space families\footnote{The five exceptional Wolf spaces have quaternionic dimensions $n_H=2,7,10,16,28$. None can be considered here, since $n_H=n_V+244\ge 244$ for the anomaly free models we are considering here. The subscript $0$ in $SO_0(n_H,4)$ denotes the connected component containing the identity.}:
\begin{equation}
 \frac{SU(n_H,2)}{S(U(n_H)\times U(2))},
 \quad
 \frac{Sp(n_H,1)}{Sp(n_H)\times Sp(1)_R},
 \quad
 \frac{SO_0(n_H,4)}{SO(n_H)\times SO(4)}.
 \label{eq:proposed-common-Wolf-spaces}
\end{equation}
For the following $Sp(1)_{R+}$ construction, we assume that the gauging is compatible with the chosen manifold and every representation block in \eqref{sp:eq:spec} is nontrivial under the additional connected semisimple group $K$. This restriction guarantees the representation multiplicities used below: two equivalent copies for the common unitary/symplectic construction, and the multiplicities needed for the orthogonal construction. Under these assumptions, $d_{\max}\geq3$ is sufficient for a common zero of all three diagonal moment maps on each of these spaces.

\subsection*{The case of $SU(n_H,2)$ and $Sp(n_H,1)$ Wolf spaces}
We use the same nonzero scalar block for the $SU(n_H,2)$ and $Sp(n_H,1)$ Wolf spaces in \eqref{eq:proposed-common-Wolf-spaces}. The orthogonal space is treated separately below. Let $V_j$ denote the spin-$j$ representation, where $j=(d-1)/2$. On the unitary space we work in $V_H$, and on the symplectic space in the hyperino representation $\mathcal R_H$. Since $K$ commutes with $Sp(1)_H$, each irreducible block has the form $R\otimes V_j$. Our matter restriction makes $R$ nontrivial; connected semisimple $K$ then implies $\dim R\geq2$. Thus each occupied $V_j$ has at least two copies.

Choose two equivalent copies in conjugate bases, forming a subspace $V_j\oplus\overline{V_j}$ on which
\begin{equation}
 Q_H^r\big|_{V_j\oplus\overline{V_j}}
 =2\begin{pmatrix}J^r_{(d)}&0\\0&-\overline{J^r_{(d)}}\end{pmatrix},
 \quad r=1,2,3,\quad  \text{where}\ \overline{J^r_{(d)}}=(J^r_{(d)})^T.
 \label{eq:proposed-common-paired-generators}
\end{equation}
Note that $\overline{J^r_{(d)}}$ denotes the complex conjugate of $J^r_{(d)}$ which in turn is Hermitian. On the symplectic space, let $\Theta$ denote the quaternionic structure on $\mathcal R_H$ with $\Theta^2=-1$ that commutes with the $Sp(1)_H$ group action.\footnote{On the underlying real representation space, multiplication by $i$, $\Theta$, and $i\Theta$ are the three complex structures satisfying the quaternionic relations. Thus one antilinear map $\Theta$ suffices to express the reality condition in a complex frame.} Choose the two copies so that $\Theta$ exchanges them. This is possible because real spin representations $V_j$ have even multiplicity, while two copies of a pseudoreal $V_j$ can be recombined into such a pair. In adapted bases, $\Theta(u,v)=(-\overline v,\overline u)$.

To construct a common zero of the three diagonal moment maps, we choose the following scalar configuration with real amplitudes $a,b>0$:
\begin{equation}
 Z=\begin{pmatrix}
 a|j,j\rangle&b|j,j-1\rangle\\
 -b\overline{|j,j-1\rangle}&a\overline{|j,j\rangle}
 \end{pmatrix},\quad a,b>0,
 \qquad Z=0\ \text{on the remaining rows}.
 \label{eq:proposed-common-scalar-profile}
\end{equation}
Including the zero rows, $Z$ is $n_H\times2$ on the unitary space and $2n_H\times2$ on the symplectic space; $Z^\dagger Q_H^rZ$ and $Z^\dagger Z$ are therefore $2\times2$. On the symplectic space, the displayed columns satisfy $Z=(z,\Theta z)$: the first column $z\in\mathbb C^{2n_H}$ determines the second, with $\Theta(\Theta z)=-z$. This is the quaternionic reality condition; for a general scalar matrix $Z$, it leaves $4n_H$ independent real components.

Direct multiplication gives
\begin{align}
 Z^\dagger Q_H^rZ&=2ab\sqrt{d-1}\,\sigma_r,
 &&r=1,2,\nonumber\\
 Z^\dagger Q_H^3Z&=\big[(d-1)a^2-(d-3)b^2\big]\sigma_3,
 & Z^\dagger Z&=(a^2+b^2)\mathbf1_2.
 \label{eq:proposed-common-compressions}
\end{align}
For this profile, both spaces give the following moment maps, provided $t:=a^2+b^2<1$:
\begin{equation}
 C_+^r=-\frac{i}{2(1-t)}(\sigma_r-Z^\dagger Q_H^rZ).
 \label{eq:proposed-common-moment-map}
\end{equation}
On the unitary space, substituting \eqref{eq:proposed-common-compressions} into \eqref{eq:SU2R-projection-QH-general_v2} gives this expression. On the symplectic space, the same block representative, subject to $Z=(z,\Theta z)$, gives the same expression. Thus $C_+^r=0$ for all three generators is equivalent to $Z^\dagger Q_H^rZ=\sigma_r$, which reduces to
\begin{equation}
 2ab\sqrt{d-1}=1,\quad
 (d-1)a^2-(d-3)b^2=1.
 \label{eq:proposed-common-amplitude-equations}
\end{equation}
For every $d\geq3$, these equations have the solution
\begin{equation}
 a^2=\frac{1+\sqrt{d-2}}{2(d-1)},\quad
 b^2=\frac{1}{2(1+\sqrt{d-2})}.
 \label{eq:proposed-common-amplitudes}
\end{equation}
The solution satisfies
\begin{equation}
 1-t=\frac{(d-2)^{3/2}}{(d-1)(1+\sqrt{d-2})}>0.
 \label{eq:proposed-common-ball-bound}
\end{equation}
These amplitudes therefore solve $C_+^r=0$, $r=1,2,3$, on both spaces.

\subsection*{The case of orthogonal Wolf space}

The isotropy algebra is $\mathfrak{so}(n_H)\oplus\mathfrak{sp}(1)_L\oplus\mathfrak{sp}(1)_R$. The $Sp(1)_H$ generators may have a zero or nonzero $Sp(1)_L$ component, denoted by $\varepsilon=0,1$, respectively. In the latter case, choose the $Sp(1)_L$ components to be $S_L^r$. Let $A_r\in\mathfrak{so}(n_H)$ denote the $SO(n_H)$ components of the three embedded $Sp(1)_H$ generators. These are real matrices satisfying $A_r^T=-A_r$ and $[A_r,A_s]=\epsilon_{rs}{}^tA_t$. Integer spin $\ell$ gives a real block of dimension $2\ell+1$; half-integer spin uses the underlying real space of dimension $2(2\ell+1)$.\footnote{For half-integer $\ell$, the spin representation is pseudoreal and has no invariant real form of dimension $2\ell+1$. Using the real and imaginary parts of its complex components as independent coordinates gives real dimension $2(2\ell+1)$. For integer $\ell$, an invariant real form of dimension $2\ell+1$ exists, so doubling is unnecessary.}

We use the following real $4\times4$ generators throughout the orthogonal-space construction:
\begin{equation}\label{eq:proposed-orthogonal-real-bases}
 \begin{aligned}
 S_L^1&=\tfrac12(i\sigma^2)\otimes\sigma^3,&
 S_R^1&=\tfrac12\sigma^1\otimes(-i\sigma^2),\\
 S_L^2&=\tfrac12(i\sigma^2)\otimes\sigma^1,&
 S_R^2&=\tfrac12(-i\sigma^2)\otimes\mathbf1_2,\\
 S_L^3&=\tfrac12\mathbf1_2\otimes(-i\sigma^2),&
 S_R^3&=\tfrac12\sigma^3\otimes(-i\sigma^2).
 \end{aligned}
\end{equation}
These generators satisfy $(S_L^r)^2=(S_R^r)^2=-\mathbf1_4/4$, and obey $[S_L^r,S_L^s]=\epsilon_{rs}{}^tS_L^t$, $[S_R^r,S_R^s]=\epsilon_{rs}{}^tS_R^t$, and $[S_L^r,S_R^s]=0$. The diagonal $Sp(1)_{R+}$ generators are represented in the defining representation of $SO(n_H,4)$ by
\begin{equation}
 T^r_{R+}=\operatorname{diag}(A^r,S_R^r+\varepsilon S_L^r).
 \label{eq:proposed-orthogonal-generators}
\end{equation}
We require the $Sp(1)_R$ projection of the $SO(4)$ block of $L^{-1}T^r_{R+}L$ to vanish.

Let $\ell$ be the largest $Sp(1)_H$ spin carried by the defining $SO(n_H)$ representation, and let $j$ be the largest spin in the full hyperino representation. The hyperinos transform as $(\mathbf{n_H},\mathbf2_L)$ under $\mathfrak{so}(n_H)\oplus\mathfrak{sp}(1)_L$. The $\mathbf2_L$ factor contributes spin $\tfrac12$ when $\varepsilon=1$. Hence
\begin{equation}
 j=\ell+\frac{\varepsilon}{2},\quad d_{\max}=2j+1=2\ell+1+\varepsilon.
 \label{eq:proposed-orthogonal-largest-spin}
\end{equation}
For a real $n_H\times4$ coordinate $Z$ with $\mathbf1_4-Z^TZ>0$, we have
\begin{equation}
 L(Z)=\begin{pmatrix}A&ZB\\BZ^T&B\end{pmatrix},
 \quad A=(\mathbf1_{n_H}-ZZ^T)^{-1/2},
 \quad B=(\mathbf1_4-Z^T Z)^{-1/2}.
 \label{eq:orthogonal-L-representative}
\end{equation}
The $SO(4)$ block of $L^{-1}T^r_{R+}L$ is
\begin{equation}
 D_r=B(S_R^r+\varepsilon S_L^r-Z^TA_rZ)B.
 \label{eq:proposed-orthogonal-map}
\end{equation}
Since $\operatorname{tr}_4(S_R^sS_R^t)=-\delta_{st}$ and $\operatorname{tr}_4(S_L^sS_R^t)=0$, its R projection and the corresponding moment map are
\begin{equation}
 (D_r)_R=-\sum_{s=1}^3\operatorname{tr}_4(D_rS_R^s)S_R^s,
 \quad C_+^r=-\sum_{s=1}^3\operatorname{tr}_4(D_rS_R^s)T_R^s.
 \label{eq:proposed-orthogonal-R-projection}
\end{equation}
These represent the same $\mathfrak{sp}(1)_R$ element as a real $4\times4$ matrix and a complex $2\times2$ matrix, respectively. Thus $C_+^r=0$ exactly when $(D_r)_R=0$.


Next, we use the Hermitian spin matrices of \eqref{eq:Sp-spin-weights}--\eqref{eq:Sp-spin-ladder}, with $j=\ell$. For integer $\ell$, choose the real orthonormal basis
\begin{equation}
 \begin{gathered}
 b_m=\frac{|\ell,m\rangle+(-1)^{\ell-m}|\ell,-m\rangle}{\sqrt2},\quad
 c_m=\frac{|\ell,m\rangle-(-1)^{\ell-m}|\ell,-m\rangle}{i\sqrt2}
 \quad(m>0),\quad
 b_0=i^\ell|\ell,0\rangle.
 \end{gathered}
 \label{eq:proposed-orthogonal-integer-real-basis}
\end{equation}
For integer $\ell$, write $J_r$ in the basis $(b_\ell,c_\ell,\ldots,b_1,c_1,b_0)$. Then
\begin{equation}
 A_r=\begin{cases}
 -iJ_r,&\ell\text{ integer},\\[6pt]
 \begin{pmatrix}
 \operatorname{Re}(-iJ_r)&-\operatorname{Im}(-iJ_r)\\
 \operatorname{Im}(-iJ_r)&\operatorname{Re}(-iJ_r)
 \end{pmatrix},&\ell\text{ half-integer}.
 \end{cases}
 \label{eq:proposed-orthogonal-realification}
\end{equation}
For integer $\ell\geq2$, the first four coordinates correspond to $b_\ell,c_\ell,b_{\ell-1},c_{\ell-1}$. All columns below are expressed in the same real basis as $A_r$.


Choose $0<a,b<1$. We consider \(\ell\geq\frac32\), for which the chosen spin block supplies the four orthonormal real directions used in the construction below. In the real bases specified above, take the following scalar configuration on the chosen spin block:
\begin{equation}
 Z=\begin{cases}
 \bigl(a\,b_\ell,\ a\,c_\ell,\ b\,b_{\ell-1},\ b\,c_{\ell-1}\bigr),
 &\ell\text{ integer},\\[8pt]
 \begin{pmatrix}
 a e_1&0&b e_2&0\\
 0&-a e_1&0&-b e_2
 \end{pmatrix},&\ell\text{ half-integer}.
 \end{cases}
 \label{eq:proposed-orthogonal-profile}
\end{equation}
Here the integer-spin columns are the real vectors defined above, while $e_1=(1,0,\ldots,0)^T$ and $e_2=(0,1,0,\ldots,0)^T$ belong to $\mathbb R^{2\ell+1}$ and correspond to weights $\ell,\ell-1$ in the half-integer case of \eqref{eq:proposed-orthogonal-realification}. Adding zero rows in the remaining $SO(n_H)$ directions gives a real $n_H\times4$ matrix.

The weights $\ell,\ell-1$ and the ladder coefficient $\sqrt{2\ell}$ give
\begin{equation}
 \begin{aligned}
 Z^TZ&=\operatorname{diag}(a^2\mathbf1_2,b^2\mathbf1_2),\\
 Z^TA_rZ&=\sqrt{2\ell}\,ab\,S_R^r\quad(r=1,2),\\
 Z^TA_3Z&=\bigl[\ell a^2-(\ell-1)b^2\bigr]S_R^3
          +\bigl[\ell a^2+(\ell-1)b^2\bigr]S_L^3.
 \end{aligned}
 \label{eq:proposed-orthogonal-compressions}
\end{equation}
Thus $Z^TZ<\mathbf1_4$ for $0<a,b<1$. Substituting into \eqref{eq:proposed-orthogonal-map} and taking the R projection gives
\begin{equation}
 \begin{aligned}
 (D_r)_R&=\frac{1-\sqrt{2\ell}\,ab}{\sqrt{(1-a^2)(1-b^2)}}S_R^r
 \quad(r=1,2),\\
 (D_3)_R&=\left[1-\frac{2\ell-1-\varepsilon}{2}
 \left(\frac{a^2}{1-a^2}-\frac{b^2}{1-b^2}\right)\right]S_R^3.
 \end{aligned}
 \label{eq:proposed-orthogonal-ab-projections}
\end{equation}
Consequently, all three diagonal moment maps vanish when
\begin{equation}
 ab=\frac1{\sqrt{2\ell}},\quad
 \frac{a^2}{1-a^2}-\frac{b^2}{1-b^2}
 =\frac2{2\ell-1-\varepsilon}.
 \label{eq:proposed-orthogonal-zero-equations}
\end{equation}
For $\ell\geq\tfrac32$ and $\varepsilon=0,1$, the right-hand side of the second equation is positive. Set $b=1/(\sqrt{2\ell}\,a)$ and vary $a$ over $((2\ell)^{-1/4},1)$. Then $0<b<a<1$, and the left-hand side of the second equation is continuous, approaching zero at the lower endpoint and $+\infty$ as $a\to1$. Hence these equations have a solution with $Z^TZ<\mathbf1_4$, at which $C_+^r=0$ for all three generators.

So far we have considered the case $\ell\ge3/2$.
For $\ell<\tfrac32$, the bound $d_{\max}>2$ leaves three cases:  $(\varepsilon,\ell)=(0,1),(1,\tfrac12),(1,1)$. Define the orthogonal matrix
\begin{equation}
 O=\frac1{\sqrt2}\begin{pmatrix}
 0&0&1&1\\1&-1&0&0\\1&1&0&0\\0&0&-1&1
 \end{pmatrix},\qquad O^TO=\mathbf1_4.
 \label{eq:proposed-orthogonal-coordinate-change}
\end{equation}
It converts these coordinates to the one-plus-three basis:
$O(S_R^r+S_L^r)O^T=\operatorname{diag}(0,G_r)$, where
$(G_r)_{st}=-\epsilon_{rst}$, $r,s,t=1,2,3$, and $\epsilon_{123}=1$.

\noindent\textbf{Case 1: $\varepsilon=0$, $\ell=1$ ($d_{\max}=3$).}
Use one real triplet with $A_r=G_r$. Substitution into \eqref{eq:proposed-orthogonal-map} gives
\begin{equation}
 Z=\left(0,\frac{\sqrt3}{2}\mathbf1_3\right)O,\quad D_r=-2S_L^r.
 \label{eq:proposed-orthogonal-low-triplet}
\end{equation}
\noindent\textbf{Case 2: $\varepsilon=1$, $\ell=\tfrac12$ ($d_{\max}=3$).}
Use one real four-dimensional block with $A_r=S_R^r$. Then
\begin{equation}
Z=O^T\operatorname{diag}\left(\frac3{\sqrt{10}},\frac{\mathbf1_3}{\sqrt2}\right)O,\quad D_r=3S_L^r.
 \label{eq:proposed-orthogonal-low-doublet}
\end{equation}
\medskip
\noindent\textbf{Case 3: $\varepsilon=1$, $\ell=1$ ($d_{\max}=4$).}
Here $K$ acts on the real triplet multiplicities, since it commutes with $Sp(1)_L$. There are at least three triplets: otherwise connected semisimple $K$ acts trivially, contrary to our matter restriction. Write three triplets as a real $3\times3$ matrix $M$, with the matrix units taken as an orthonormal basis and $A_rM=G_rM$. For $y\in\mathbb R^4$, write $Oy=(x_0,x^T)^T$, with $x=(x_1,x_2,x_3)^T$. The choice
\begin{equation}
 Zy=\frac{x_0}{2}\mathbf1_3-\frac1{\sqrt3}\sum_{r=1}^3x_rG_r
 \label{eq:proposed-orthogonal-three-triplets}
\end{equation}
gives
\begin{equation}
Z^TZ=O^T\operatorname{diag}\left(\frac34,\frac23\mathbf1_3\right)O,
\quad D_r=4S_L^r.
\label{eq:proposed-orthogonal-three-triplet-zero}
\end{equation}
All other rows of $Z$ are zero. In the three cases, the largest eigenvalues of $Z^TZ$ are $3/4$, $9/10$ and $3/4$, respectively, so $Z^TZ<\mathbf1_4$. Each $D_r$ is proportional to $S_L^r$ and hence trace-orthogonal to every $S_R^s$; \eqref{eq:proposed-orthogonal-R-projection} therefore gives $C_+^r=0$ for all three generators.

Under the stated assumptions, on the three classical Wolf-space families, the conditions
\begin{equation}
 \boxed{U(1)_{R+}:\quad q_{\max}>1,\qquad Sp(1)_{R+}:\quad d_{\max}>2}
 \label{eq:proposed-common-diagonal-criterion}
\end{equation}
are necessary and sufficient for the diagonal moment maps to vanish at a finite scalar point. For $Sp(1)_{R+}$, all three maps vanish at the same point. A Minkowski vacuum additionally requires all $K$ moment maps to vanish at that point, with $f_z>0$.

\subsection{R-symmetry breaking}
\label{sec:r-symmetry-breaking}

Every Minkowski vacuum breaks the continuous diagonal R-symmetry completely, even
allowing mixing with the other gauge factors.
Consider an arbitrary gauge generator
\begin{equation}
 \begin{aligned}
 \e^XT_X&=\epsilon^rT_{R+}^r+\epsilon^I T_I\ ,\\
 T_{R+}^r&=T_H^r+T_R^r\ ,
 \end{aligned}
 \label{eq:R-breaking-generator}
\end{equation}
where $T_I$ are the generators of the additional group factors acting on the
hyperscalars. Only $r=3$ is retained for $U(1)_{R+}$, whereas
$r=1,2,3$ for $Sp(1)_{R+}$.
Considering the transformation of the coset scalar fields under the $X$ transformations with local gauge parameter $\epsilon^X$ given by $\delta \varphi^\alpha= \epsilon^X K^\alpha_X$, where $K^\alpha_X$ is the Killing vector, it is straightforward to derive the relation~\cite{Percacci:1998ag,Jong:2006ud}
\begin{equation}
 \epsilon^X K_X^\alpha Q_\alpha^{AB}= \epsilon^X C_X^{AB}-\epsilon^r (T^r_R)^{AB}\ ,
 \label{eq:R-breaking-coset-identity}
\end{equation}
Here $Q_\alpha^{AB}$ is the composite $Sp(1)_R$ connection and
$C_X^{AB}$ is the corresponding $\mathfrak{sp}(1)_R$-valued
moment map. If $\epsilon^X T_X$ is unbroken at the vacuum $\varphi_*$,
then $\epsilon^X K_X(\varphi_*)=0$. Moreover, the Minkowski conditions
set every gauged moment map to zero, so linearity gives
$\epsilon^X C_X(\varphi_*)=0$. Consequently,
\begin{equation}
 0=\epsilon^rT_R^r \quad\Longrightarrow\quad \epsilon^r=0\ ,
\end{equation}
by linear independence of the retained R-symmetry generators.
Thus every unbroken continuous gauge generator lies entirely
in the additional gauge factors, with no diagonal
R-symmetry component. Discrete remnants are not excluded.

\section{Conclusions}
\label{sec:conclusions}

We have constructed bounded catalogues of anomaly-free models with $U(1)_{R+}$ and $Sp(1)_{R+}$ diagonal gauging and one or two additional simple gauge factors. We require every hypermultiplet block to be nontrivial under at least one additional simple factor to keep the search space manageable. Further bounds on the search are stated in \eq{G1}, \eq{G1xG2} and \eq{eq:shared-diagonal-scan}. The displayed appendix spectra form selected subsets of the complete counts given in Section~\ref{sec:search-strategy}. 

We have shown that all the models we have constructed admit fully supersymmetric Minkowski vacua provided that the maximum $U(1)_{R+}$ hyperino charge $q_{\rm max}$ $\ge 2$ or the maximum  $Sp(1)_{R+}$ irrep dimension $d_{\rm max}$ $\ge 3$, respectively. They all break $U(1)_{R+}$ or $Sp(1)_{R+}$, respectively, while a subgroup of the additional gauge groups remains unbroken. In Section~\ref{sec:explicit-Minkowski} we construct six explicit examples on the unitary Wolf space, with the following connected unbroken gauge groups:
\begin{equation}
\begin{aligned}
 SU(7)\times U(1)_{R+}
 &\to SO(7),\\
 G_2\times U(1)_{R+}
 &\to SU(3),\\
 SU(7)\times\mathrm{Spin}(18)\times U(1)_{R+}
 &\to SU(7)\times\mathrm{Spin}(17),\\
 \mathrm{Spin}(12)\times E_6\times U(1)_{R+}
 &\to\mathrm{Spin}(11)\times E_6,\\
 SU(6)\times Sp(1)_{R+}
 &\to SU(5),\\
 SU(6)\times\mathrm{Spin}(12)\times Sp(1)_{R+}
 &\to SU(5)\times\mathrm{Spin}(11).
\end{aligned}
\end{equation}

\par 

For the abelian catalogue, the vacuum statement concerns the displayed signed representatives, rather than every assignment counted by $N_{\rm sign}$. These Minkowski vacua  preserve all eight supersymmetries, and the tensor dilaton remains free in the common positive kinetic domain. 

In the main scans with $q_{\max}\leq2$ and $d_{\max}\leq3$, including ordinary $U(1)_R$ and $Sp(1)_R$ gauging, the largest total gauge-group ranks found are $8$ with one additional simple factor and $26$ with two. Both totals include the rank-one factor $G_+$. It is an interesting open problem to determine if a bound on the rank of the gauge group exists upon consideration of a much wider class of diagonally gauged models. 

Compactifications and string solutions remain natural directions for
these theories. The abelian toy model of \cite{Guo:2025mlb} already
admits supersymmetric Mink$_4\times S^2$ and nonsupersymmetric
(A)dS$_4\times S^2$ solutions. String-worldsheet anomaly cancellation
could supply further consistency conditions. A higher-dimensional
origin of the diagonal models remains an open question.

Allowing hypermultiplets that are singlets under the additional gauge factors, we also find an anomaly-free $E_7\times E_8\times U(1)_{R+}$ model with a supersymmetric Minkowski vacuum in which $U(1)_{R+}$ is broken.\footnote{The full-hypermultiplet spectrum is
$10(\mathbf{56},\mathbf1)_1+46(\mathbf1,\mathbf1)_0+10(\mathbf1,\mathbf1)_1+2(\mathbf1,\mathbf1)_2+8(\mathbf1,\mathbf1)_3$,
and its anomaly vectors are
$a=(-2,-2)$, $b_{E_7}=(1,6)$, $b_{E_8}=(1,-6)$ and $b_{R+}=(12,12)$.} Its remaining massless hypermultiplet spectrum consists of ten full hypermultiplets in $(\mathbf{56},\mathbf1)$ and 65 singlets, as in the standard heterotic compactification on K3~\cite{Schwarz:1995-K3}. This agreement does not establish a string origin for the parent gauged model: its massive spectrum and interactions must also be reproduced. Moreover, the standard heterotic compactification on K3 does not gauge R-symmetry and therefore does not directly realize this parent model~\cite{Gueven:2003uw}. Whether a more general string construction can realize the diagonally gauged model remains an open question.

\subsection*{Code and catalogue availability}
\label{sec:code-availability}

The Mathematica notebook and supplementary catalogues~\cite{DGcode} are stored at
\begin{center}
\url{https://github.com/XuGuo-hep/diagonal-gauging}.
\end{center}
The notebook \texttt{DG\_search.nb} performs anomaly searches for both diagonal gaugings with one or two additional simple factors. Usage instructions are in the repository README.

The catalogues \texttt{q=2\_tables.pdf} and \texttt{d=3\_tables.pdf} list all spectra for the $q_{\max}=2$ and $d_{\max}=3$ sectors, respectively. Both cover one or two additional simple group factors. The former omits the one-factor $A_3$ catalogue because of its very large size, though the notebook supports that search. Minkowski-vacuum calculations are not included.

\subsection*{Acknowledgments}
The authors thank Katrin Becker, Guillaume Bossard, Monica Kang, Ryo Suzuki, Yuji Tachikawa, Qi You and Yi Zhang for useful discussions. We also acknowledge substantial use of ChatGPT and Claude in exploratory and editorial stages of this work, with all technical results checked independently. The work of Y. P. is supported by the National Natural Science Foundation of China (NSFC) under Grant No. 12575076 and No. 12247103. The work of E. S. is supported in part by the NSF grant PHY-2413006 and TUBITAK, and X.G. is supported in part by the George P. and Cynthia W. Mitchell Institute for Fundamental Physics and Astronomy. E.S. and H.S. would like to thank their respective institutions for hospitality during the course of this work.

\begin{appendix}
\section{Conventions}
\label{app:conventions}


We use the following indices to label the generators of the indicated groups:
\begin{equation}
G=
\overbrace{
  \overbrace{
    \underbrace{G_+}_{r,s}
    \times
    \underbrace{K}_{I,J}
  }^{X,Y}
  \times
  \underbrace{G_E}_{P,Q}
}^{\ell,m},
\qquad
\underbrace{H\times Sp(1)_R}_{\widehat I, \widehat J}.
\label{eq:app-gauge-group-indices}
\end{equation}
Our space-time signature is $(-+++++)$. The $SO(1,1)$ matrix $v(\phi)$ is defined by 
\be
v_\alpha^0 v_\beta^0 -v_\alpha^1 v_\beta^1= \eta_{\alpha\beta}\ ,\qquad \mbox{with} \quad \eta_{\alpha\beta}=\begin{pmatrix} 0&1 \\1 &0\end{pmatrix}
\ .
\ee
This defining relation is satisfied by 
\begin{equation}
(v^+,v^-,v_1^+,v_1^-)=\frac1{\sqrt2}(e^{-\phi},e^\phi,e^{-\phi},-e^\phi)\ ,
\end{equation}
where we have used the notation $v^\alpha=(v^+, v^-)$, $v^\a=\eta^{\a\b}v_\b^0$ and $v^\a_1=\eta^{\a\b}v_\b^1$.
Our Clifford and chirality conventions are the same as those spelled out in \cite{Bossard:2024ffp}.
We use $\epsilon_{12}=\epsilon^{12}=1$ and raise and lower indices according to $\Psi^A=\epsilon^{AB}\Psi_B$ and $\Psi_A=\Psi^B\epsilon_{BA}$, respectively. The symplectic Majorana--Weyl condition relates the doublet components by $(\Psi^1)^c=\Psi^2$ and $(\Psi^2)^c=-\Psi^1$, or equivalently $(\Psi^A)^c=-\Psi_A$ where $\Psi^c=\mathcal B^{-1}\Psi^*$ and $*$ is the complex conjugation, and $\mathcal B$ is the six-dimensional spinor conjugation matrix acting on Lorentz spinor indices, with $\mathcal B^*\mathcal B=-\mathbf1$. We introduce complex Weyl spinors by
\begin{equation}
 \epsilon=\epsilon^1,\quad \epsilon^c=\epsilon^2,\qquad
 \Psi=\Psi^1,\quad \Psi^c=\Psi^2,\quad (\Psi^c)^c=-\Psi,
 \label{eq:abelian-complex-spinors}
\end{equation}
where $\Psi$ denotes the gravitino, tensorino or a gaugino, and the same conjugation rule applies to $\epsilon$. 
The moment map $C$ denotes the matrix with entries $C_A{}^B$, where $A$ labels the row and $B$ the column. For any moment map, $C^{AB}=\epsilon^{AC}C_C{}^B$ is symmetric and obeys $C^{22}=(C^{11})^*$ and $(C^{12})^*=-C^{12}$.

\section{\texorpdfstring{Selected $U(1)_{R+}$ spectra}{Selected U(1)R+ spectra}}
\label{app:abelian-spectra}

The charge tuples are $(N_{+1},N_0,N_{-1})$ for $q_{\max}\leq1$ and
$(N_{+2},N_{+1},N_0,N_{-1},N_{-2})$ for $q_{\max}=2$.
In $mR_{(N_q)}$, $N_q$ counts full hypermultiplets and $m=\sum_qN_q$,
except that a separately displayed neutral pseudoreal block uses a unit
neutral entry and its prefactor gives the multiplicity in full-hypermultiplet units. Thus $\frac{k}{2}R$ denotes $k$ neutral half-hypermultiplets. Every charged hypermultiplet is full. One cubic-compatible signed representative is shown per charge-allocation spectrum in Tables~\ref{tab:u1-one-combined} and~\ref{tab:u1-two-combined}. 

For each row, $N_{\rm sign}$ counts signed multiplicity assignments to its
complex nonabelian product representations, at fixed absolute-charge
multiplicities and irrep labels, satisfying the mixed cubic-anomaly
conditions. For the \(U(1)_{R+}\) entries, each absolute-charge allocation with fixed anomaly vectors is counted once if a compatible sign assignment exists; different admissible sign assignments do not increase the spectrum count. Simultaneous reversal of all charges is identified; signs on
self-conjugate product representations are not counted, since $R_q$ and
$R_{-q}$ describe the same full hypermultiplet when $R\simeq\overline R$.
No further quotient by diagram automorphisms or exchange of identical
factors is taken in $N_{\rm sign}$, so this column does not count physically
inequivalent theories. A fully neutral row has $N_{\rm sign}=1$. Daggers mark spectra requiring $I_{1,1}$; unmarked spectra admit $U$. Single rules separate nonabelian spectra or anomaly vectors.

\subsection{One additional simple factor}
\label{app:u1-one-factor}
\begingroup
\footnotesize
\setlength{\tabcolsep}{0pt}
\setlength{\LTleft}{0pt}\setlength{\LTright}{0pt}
\renewcommand{\arraystretch}{1.12}
\color{black}\hypersetup{linkcolor=blue,citecolor=blue,urlcolor=black}
\captionsetup{hypcap=false,position=bottom}
\setlength{\LTleft}{0pt plus 1fill}\setlength{\LTright}{0pt plus 1fill}
\setlength{\LTcapwidth}{\textwidth}
\renewcommand{\theHtable}{catalogue.\arabic{table}}
\begin{longtable}{@{}>{\centering\arraybackslash}p{14pt}@{\hspace{3pt}}p{25pt}@{\hspace{3pt}}>{\centering\arraybackslash}p{21pt}@{\hspace{3pt}}>{\raggedright\arraybackslash}p{\dimexpr\textwidth-145pt\relax}@{\hspace{3pt}}>{\centering\arraybackslash}p{33pt}@{\hspace{3pt}}>{\centering\arraybackslash}p{36pt}@{}}
\toprule
No. & $G_1$ & $N_{\rm sign}$ & Split spectrum & $b_1$ & $b_{R+}$\\
\midrule
\endfirsthead
\multicolumn{6}{l}{\small\textbf{Table~\ref{tab:u1-one-combined} (continued)}}\\[3pt]
\toprule
No. & $G_1$ & $N_{\rm sign}$ & Split spectrum & $b_1$ & $b_{R+}$\\
\midrule
\endhead
\midrule
\multicolumn{6}{r}{Continued on next page}\\
\endfoot
\endlastfoot
\multicolumn{6}{c}{\rule{0pt}{14pt}\small\bfseries\boldmath$q_{\max}\leq1$}\\*[3pt]
\midrule
1 & $A_{5}$ & 2 & $18\cdot{\mathbf{6}}_{(7,8,3)}$ $+\allowbreak\,$ $2\cdot{\mathbf{15}}_{(0,0,2)}$ $+\allowbreak\,$ $5\cdot{\mathbf{20}}_{(5,0,0)}$ $+\allowbreak\,$ $2\cdot{\mathbf{21}}_{(1,0,1)}$ & $(2,3)$ & $(16,2)$\\[2pt]
\midrule
2 & $A_{5}$ & 2 & $18\cdot{\mathbf{6}}_{(13,4,1)}$ $+\allowbreak\,$ $2\cdot{\mathbf{15}}_{(0,1,1)}$ $+\allowbreak\,$ $2\cdot{\mathbf{20}}_{(2,0,0)}$ $+\allowbreak\,$ $3\cdot{\mathbf{20}}_{(0,1,0)}$ $+\allowbreak\,$ $2\cdot{\mathbf{21}}_{(0,1,1)}$ & $(2,3)$ & $(2,10)$\\[2pt]
\midrule
3 & $A_{5}^{\dagger}$ & 2 & $18\cdot{\mathbf{6}}_{(14,2,2)}$ $+\allowbreak\,$ $2\cdot{\mathbf{15}}_{(0,1,1)}$ $+\allowbreak\,$ $2\cdot{\mathbf{20}}_{(2,0,0)}$ $+\allowbreak\,$ $3\cdot{\mathbf{20}}_{(0,1,0)}$ $+\allowbreak\,$ $2\cdot{\mathbf{21}}_{(0,1,1)}$ & $(2,3)$ & $(2,11)$\\[2pt]
\midrule
4 & $A_{5}$ & 2 & $18\cdot{\mathbf{6}}_{(15,0,3)}$ $+\allowbreak\,$ $2\cdot{\mathbf{15}}_{(0,1,1)}$ $+\allowbreak\,$ $2\cdot{\mathbf{20}}_{(2,0,0)}$ $+\allowbreak\,$ $3\cdot{\mathbf{20}}_{(0,1,0)}$ $+\allowbreak\,$ $2\cdot{\mathbf{21}}_{(0,1,1)}$ & $(2,3)$ & $(2,12)$\\[2pt]
\midrule
5 & $A_{5}^{\dagger}$ & 1 & $18\cdot{\mathbf{6}}_{(8,10,0)}$ $+\allowbreak\,$ $2\cdot{\mathbf{15}}_{(1,1,0)}$ $+\allowbreak\,$ $2\cdot{\mathbf{20}}_{(2,0,0)}$ $+\allowbreak\,$ $3\cdot{\mathbf{20}}_{(0,1,0)}$ $+\allowbreak\,$ $2\cdot{\mathbf{21}}_{(0,1,1)}$ & $(2,3)$ & $(2,7)$\\[2pt]
\midrule
6 & $A_{5}$ & 1 & $18\cdot{\mathbf{6}}_{(9,8,1)}$ $+\allowbreak\,$ $2\cdot{\mathbf{15}}_{(1,1,0)}$ $+\allowbreak\,$ $2\cdot{\mathbf{20}}_{(2,0,0)}$ $+\allowbreak\,$ $3\cdot{\mathbf{20}}_{(0,1,0)}$ $+\allowbreak\,$ $2\cdot{\mathbf{21}}_{(0,1,1)}$ & $(2,3)$ & $(2,8)$\\[2pt]
\midrule
7 & $A_{5}^{\dagger}$ & 2 & $18\cdot{\mathbf{6}}_{(12,6,0)}$ $+\allowbreak\,$ $2\cdot{\mathbf{15}}_{(0,1,1)}$ $+\allowbreak\,$ $2\cdot{\mathbf{20}}_{(2,0,0)}$ $+\allowbreak\,$ $3\cdot{\mathbf{20}}_{(0,1,0)}$ $+\allowbreak\,$ $2\cdot{\mathbf{21}}_{(0,1,1)}$ & $(2,3)$ & $(2,9)$\\[2pt]
\midrule
8 & $A_{5}$ & 2 & $18\cdot{\mathbf{6}}_{(4,14,0)}$ $+\allowbreak\,$ $2\cdot{\mathbf{15}}_{(0,0,2)}$ $+\allowbreak\,$ $2\cdot{\mathbf{20}}_{(2,0,0)}$ $+\allowbreak\,$ $3\cdot{\mathbf{20}}_{(0,1,0)}$ $+\allowbreak\,$ $2\cdot{\mathbf{21}}_{(1,0,1)}$ & $(2,3)$ & $(8,2)$\\[2pt]
\midrule
9 & $A_{5}^{\dagger}$ & 1 & $20\cdot{\mathbf{6}}_{(2,18,0)}$ $+\allowbreak\,$ $7\cdot{\mathbf{15}}_{(3,0,4)}$ $+\allowbreak\,$ ${\mathbf{20}}_{(1,0,0)}$ $+\allowbreak\,$ ${\mathbf{35}}_{(1,0,0)}$ & $(\tfrac{3}{2},3)$ & $(11,2)$\\[2pt]
\midrule
10 & $A_{6}$ & 1 & $6\cdot{\mathbf{7}}_{(2,3,1)}$ $+\allowbreak\,$ $5\cdot{\mathbf{21}}_{(0,4,1)}$ $+\allowbreak\,$ ${\mathbf{28}}_{(0,1,0)}$ $+\allowbreak\,$ $2\cdot{\mathbf{35}}_{(1,1,0)}$ $+\allowbreak\,$ ${\mathbf{48}}_{(1,0,0)}$ & $(2,3)$ & $(2,6)$\\[2pt]
\midrule
11 & $A_{6}^{\dagger}$ & 1 & $6\cdot{\mathbf{7}}_{(2,3,1)}$ $+\allowbreak\,$ $5\cdot{\mathbf{21}}_{(2,0,3)}$ $+\allowbreak\,$ ${\mathbf{28}}_{(0,1,0)}$ $+\allowbreak\,$ $2\cdot{\mathbf{35}}_{(1,1,0)}$ $+\allowbreak\,$ ${\mathbf{48}}_{(0,1,0)}$ & $(2,3)$ & $(2,9)$\\[2pt]
12 & $A_{6}^{\dagger}$ & 2 & $6\cdot{\mathbf{7}}_{(4,2,0)}$ $+\allowbreak\,$ $5\cdot{\mathbf{21}}_{(0,0,5)}$ $+\allowbreak\,$ ${\mathbf{28}}_{(1,0,0)}$ $+\allowbreak\,$ $2\cdot{\mathbf{35}}_{(0,2,0)}$ $+\allowbreak\,$ ${\mathbf{48}}_{(0,1,0)}$ & $(2,3)$ & $(2,9)$\\[2pt]
\midrule
13 & $B_{6}$ & 1 & $3\cdot{\mathbf{13}}_{(1,2,0)}$ $+\allowbreak\,$ ${\mathbf{64}}_{(1,0,0)}$ $+\allowbreak\,$ ${\mathbf{64}}_{(0,1,0)}$ $+\allowbreak\,$ $2\cdot{\mathbf{78}}_{(1,1,0)}$ & $(2,3)$ & $(2,6)$\\[2pt]
\midrule
14 & $C_{4}$ & 1 & $4\cdot{\mathbf{8}}_{(4,0,0)}$ $+\allowbreak\,$ $3\cdot{\mathbf{27}}_{(3,0,0)}$ $+\allowbreak\,$ $2\cdot{\mathbf{36}}_{(0,2,0)}$ $+\allowbreak\,$ ${\mathbf{48}}_{(1,0,0)}$ $+\allowbreak\,$ ${\mathbf{48}}_{(0,1,0)}$ & $(2,3)$ & $(2,10)$\\[2pt]
\midrule
15 & $G_{2}$ & 1 & $4\cdot{\mathbf{7}}_{(2,2,0)}$ $+\allowbreak\,$ $3\cdot{\mathbf{77}}_{(1,2,0)}$ & $(10,12)$ & $(2,6)$\\[2pt]
\specialrule{0.6pt}{5pt}{2pt}
\multicolumn{6}{c}{\rule{0pt}{14pt}\small\bfseries\boldmath$q_{\max}=2$}\\*[3pt]
\midrule
1 & $A_{6}$ & 3 & $6\cdot{\mathbf{7}}_{(1,2,3,0,0)}$ $+\allowbreak\,$ $5\cdot{\mathbf{21}}_{(1,0,3,1,0)}$ $+\allowbreak\,$ ${\mathbf{28}}_{(0,0,0,1,0)}$ $+\allowbreak\,$ $2\cdot{\mathbf{35}}_{(0,2,0,0,0)}$ $+\allowbreak\,$ ${\mathbf{48}}_{(0,1,0,0,0)}$ & $(2,3)$ & $(16,6)$\\[2pt]
\midrule
2 & $A_{6}^{\dagger}$ & 9 & $6\cdot{\mathbf{7}}_{(4,2,0,0,0)}$ $+\allowbreak\,$ $5\cdot{\mathbf{21}}_{(0,3,2,0,0)}$ $+\allowbreak\,$ ${\mathbf{28}}_{(0,0,0,1,0)}$ $+\allowbreak\,$ $2\cdot{\mathbf{35}}_{(0,0,0,0,2)}$ $+\allowbreak\,$ ${\mathbf{48}}_{(0,0,1,0,0)}$ & $(2,3)$ & $(30,9)$\\[2pt]
3 & $A_{6}^{\dagger}$ & 2 & $6\cdot{\mathbf{7}}_{(5,1,0,0,0)}$ $+\allowbreak\,$ $5\cdot{\mathbf{21}}_{(0,3,2,0,0)}$ $+\allowbreak\,$ ${\mathbf{28}}_{(0,0,0,0,1)}$ $+\allowbreak\,$ $2\cdot{\mathbf{35}}_{(1,0,0,1,0)}$ $+\allowbreak\,$ ${\mathbf{48}}_{(0,0,1,0,0)}$ & $(2,3)$ & $(30,9)$\\[2pt]
\midrule
4 & $A_{6}^{\dagger}$ & 4 & $6\cdot{\mathbf{7}}_{(2,1,1,2,0)}$ $+\allowbreak\,$ $5\cdot{\mathbf{21}}_{(0,0,2,0,3)}$ $+\allowbreak\,$ ${\mathbf{28}}_{(0,1,0,0,0)}$ $+\allowbreak\,$ $2\cdot{\mathbf{35}}_{(1,0,1,0,0)}$ $+\allowbreak\,$ ${\mathbf{48}}_{(0,1,0,0,0)}$ & $(2,3)$ & $(34,9)$\\[2pt]
5 & $A_{6}^{\dagger}$ & 2 & $6\cdot{\mathbf{7}}_{(2,1,2,1,0)}$ $+\allowbreak\,$ $5\cdot{\mathbf{21}}_{(1,0,2,0,2)}$ $+\allowbreak\,$ ${\mathbf{28}}_{(0,0,1,0,0)}$ $+\allowbreak\,$ $2\cdot{\mathbf{35}}_{(1,0,0,1,0)}$ $+\allowbreak\,$ ${\mathbf{48}}_{(0,1,0,0,0)}$ & $(2,3)$ & $(34,9)$\\[2pt]
6 & $A_{6}^{\dagger}$ & 2 & $6\cdot{\mathbf{7}}_{(3,1,1,1,0)}$ $+\allowbreak\,$ $5\cdot{\mathbf{21}}_{(3,0,2,0,0)}$ $+\allowbreak\,$ ${\mathbf{28}}_{(0,0,0,0,1)}$ $+\allowbreak\,$ $2\cdot{\mathbf{35}}_{(0,0,1,1,0)}$ $+\allowbreak\,$ ${\mathbf{48}}_{(0,1,0,0,0)}$ & $(2,3)$ & $(34,9)$\\[2pt]
\midrule
7 & $A_{6}$ & 2 & $6\cdot{\mathbf{7}}_{(4,0,2,0,0)}$ $+\allowbreak\,$ $5\cdot{\mathbf{21}}_{(0,3,0,2,0)}$ $+\allowbreak\,$ ${\mathbf{28}}_{(0,0,0,1,0)}$ $+\allowbreak\,$ $2\cdot{\mathbf{35}}_{(0,0,2,0,0)}$ $+\allowbreak\,$ ${\mathbf{48}}_{(0,1,0,0,0)}$ & $(2,3)$ & $(6,16)$\\[2pt]
\midrule
8 & $A_{7}^{\dagger}$ & 1 & $4\cdot{\mathbf{28}}_{(2,0,0,0,2)}$ $+\allowbreak\,$ $2\cdot{\mathbf{63}}_{(2,0,0,0,0)}$ $+\allowbreak\,$ ${\mathbf{70}}_{(1,0,0,0,0)}$ & $(2,3)$ & $(90,9)$\\[2pt]
\midrule
9 & $C_{4}$ & 1 & $4\cdot{\mathbf{8}}_{(2,2,0,0,0)}$ $+\allowbreak\,$ $3\cdot{\mathbf{27}}_{(0,3,0,0,0)}$ $+\allowbreak\,$ $2\cdot{\mathbf{36}}_{(0,0,2,0,0)}$ $+\allowbreak\,$ ${\mathbf{48}}_{(1,0,0,0,0)}$ $+\allowbreak\,$ ${\mathbf{48}}_{(0,0,1,0,0)}$ & $(2,3)$ & $(18,10)$\\[2pt]
\midrule
10 & $C_{4}$ & 1 & ${\mathbf{8}}_{(1,0,0,0,0)}$ $+\allowbreak\,$ $3\cdot{\mathbf{8}}_{(0,0,1,0,0)}$ $+\allowbreak\,$ $3\cdot{\mathbf{27}}_{(0,3,0,0,0)}$ $+\allowbreak\,$ $2\cdot{\mathbf{36}}_{(0,2,0,0,0)}$ $+\allowbreak\,$ $2\cdot{\mathbf{48}}_{(0,0,1,0,0)}$ & $(2,3)$ & $(4,10)$\\[2pt]
\midrule
11 & $C_{5}$ & 1 & $6\cdot{\mathbf{10}}_{(2,4,0,0,0)}$ $+\allowbreak\,$ $2\cdot{\mathbf{10}}_{(0,0,1,0,0)}$ $+\allowbreak\,$ $5\cdot{\mathbf{44}}_{(0,3,2,0,0)}$ & $(1,2)$ & $(6,12)$\\[2pt]
\midrule
12 & $D_{4}$ & 1 & $6\cdot{\mathbf{28}}_{(1,2,3,0,0)}$ $+\allowbreak\,$ ${\mathbf{35}_{s}}_{(0,1,0,0,0)}$ $+\allowbreak\,$ ${\mathbf{35}_{c}}_{(0,1,0,0,0)}$ $+\allowbreak\,$ ${\mathbf{35}_{v}}_{(0,1,0,0,0)}$ & $(4,6)$ & $(16,6)$\\[2pt]
\midrule
13 & $G_{2}^{\dagger}$ & 1 & $10\cdot{\mathbf{7}}_{(3,1,6,0,0)}$ $+\allowbreak\,$ $7\cdot{\mathbf{27}}_{(1,4,2,0,0)}$ & $(\tfrac{9}{2},7)$ & $(18,8)$\\[2pt]
\midrule
14 & $G_{2}^{\dagger}$ & 1 & $19\cdot{\mathbf{7}}_{(0,17,2,0,0)}$ $+\allowbreak\,$ $9\cdot{\mathbf{14}}_{(2,2,5,0,0)}$ & $(3,\tfrac{11}{2})$ & $(16,6)$\\[2pt]
15 & $G_{2}^{\dagger}$ & 1 & $19\cdot{\mathbf{7}}_{(2,9,8,0,0)}$ $+\allowbreak\,$ $9\cdot{\mathbf{14}}_{(1,6,2,0,0)}$ & $(3,\tfrac{11}{2})$ & $(16,6)$\\[2pt]
\midrule
16 & $G_{2}^{\dagger}$ & 1 & $19\cdot{\mathbf{7}}_{(8,3,8,0,0)}$ $+\allowbreak\,$ $9\cdot{\mathbf{14}}_{(4,3,2,0,0)}$ & $(3,\tfrac{11}{2})$ & $(34,9)$\\[2pt]
\midrule
17 & $G_{2}^{\dagger}$ & 1 & $25\cdot{\mathbf{7}}_{(2,21,2,0,0)}$ $+\allowbreak\,$ $6\cdot{\mathbf{14}}_{(2,2,2,0,0)}$ & $(\tfrac{5}{2},5)$ & $(23,6)$\\[2pt]
\midrule
18 & $G_{2}^{\dagger}$ & 1 & $25\cdot{\mathbf{7}}_{(4,19,2,0,0)}$ $+\allowbreak\,$ $6\cdot{\mathbf{14}}_{(0,1,5,0,0)}$ & $(\tfrac{5}{2},5)$ & $(6,16)$\\[2pt]
\midrule
19 & $G_{2}^{\dagger}$ & 1 & $4\cdot{\mathbf{7}}_{(0,2,2,0,0)}$ $+\allowbreak\,$ $10\cdot{\mathbf{14}}_{(9,1,0,0,0)}$ $+\allowbreak\,$ ${\mathbf{27}}_{(0,1,0,0,0)}$ $+\allowbreak\,$ ${\mathbf{64}}_{(0,0,1,0,0)}$ & $(\tfrac{13}{2},7)$ & $(9,38)$\\[2pt]
\midrule
20 & $G_{2}^{\dagger}$ & 1 & $4\cdot{\mathbf{7}}_{(2,2,0,0,0)}$ $+\allowbreak\,$ $10\cdot{\mathbf{14}}_{(0,1,9,0,0)}$ $+\allowbreak\,$ ${\mathbf{27}}_{(0,1,0,0,0)}$ $+\allowbreak\,$ ${\mathbf{64}}_{(0,1,0,0,0)}$ & $(\tfrac{13}{2},7)$ & $(9,6)$\\[2pt]
\midrule
21 & $G_{2}^{\dagger}$ & 1 & $7\cdot{\mathbf{7}}_{(0,3,4,0,0)}$ $+\allowbreak\,$ $15\cdot{\mathbf{14}}_{(3,11,1,0,0)}$ & $(\tfrac{7}{2},7)$ & $(23,6)$\\*[2pt]
\noalign{\nobreak\vskip\aboverulesep\nobreak\hrule height\heavyrulewidth\nobreak\vskip\belowrulesep\nobreak}
\caption{Selected $G_1\times U(1)_{R+}$ spectra. Numbering restarts in each block. The charge tuples are $(N_{+1},N_0,N_{-1})$ for $q_{\max}\leq1$ and
$(N_{+2},N_{+1},N_0,N_{-1},N_{-2})$ for $q_{\max}=2$. The $q_{\max}\leq1$ block contains 15, omitting $A_3,B_3,C_3$; the $q_{\max}=2$ block contains 21, also omitting $A_5$. $N_{\rm sign}$ follows the appendix introduction. For $D_4$, $\mathbf{35}_v,\mathbf{35}_s,\mathbf{35}_c$ have Dynkin labels $(2,0,0,0),(0,0,0,2),(0,0,2,0)$. For $G_2$, $\mathbf{77}$ has Dynkin label $(3,0)$, in the convention $\mathbf7=(1,0)$.}\label{tab:u1-one-combined}\label{tab:one-lowcharge}\label{t4}\\
\end{longtable}
\endgroup

\begingroup
\renewcommand{\baselinestretch}{1}\normalsize
\captionsetup{hypcap=false}
\setlength{\parskip}{0pt}
\noindent\begin{minipage}{\textwidth}
\noindent\textbf{Larger abelian charges}\par\vspace{8pt}

\par\vspace{10pt}
\color{black}\hypersetup{linkcolor=blue,citecolor=blue,urlcolor=black}
\centering
\footnotesize
\setlength{\tabcolsep}{0pt}
\renewcommand{\arraystretch}{1.12}
\begin{tabular}{@{}>{\centering\arraybackslash}p{14pt}@{\hspace{3pt}}p{25pt}@{\hspace{3pt}}>{\centering\arraybackslash}p{21pt}@{\hspace{3pt}}>{\centering\arraybackslash}p{21pt}@{\hspace{3pt}}>{\raggedright\arraybackslash}p{\dimexpr\textwidth-169pt\relax}@{\hspace{3pt}}>{\centering\arraybackslash}p{33pt}@{\hspace{3pt}}>{\centering\arraybackslash}p{36pt}@{}}
\toprule
No. & $G_1$ & $q_{\max}$ & $N_{\rm sign}$ & Split spectrum & $b_1$ & $b_{R+}$\\
\midrule
1 & $A_{3}^{\dagger}$ & $3$ & 1 & $20\cdot\mathbf{6}_{(5,1,1,13,0,0,0)}$ $+\allowbreak\,$ $8\cdot\mathbf{15}_{(0,8,0,0,0,0,0)}$ $+\allowbreak\,$ $2\cdot\mathbf{10}_{(0,0,1,0,1,0,0)}$ & $(3,6)$ & $(53,14)$\\[3pt]
\midrule
2 & $A_{5}$ & $3$ & 1 & $19\cdot\mathbf{6}_{(3,0,1,11,0,2,2)}$ $+\allowbreak\,$ $\frac{1}{2}\mathbf{20}_{(0,0,0,1,0,0,0)}$ $+\allowbreak\,$ $5\cdot\mathbf{20}_{(0,4,1,0,0,0,0)}$ $+\allowbreak\,$ $\mathbf{35}_{(0,0,0,1,0,0,0)}$ $+\allowbreak\,$ $\mathbf{21}_{(0,0,0,1,0,0,0)}$ & $(2,3)$ & $(36,18)$\\[3pt]
\midrule
3 & $A_{6}$ & $3$ & 2 & $6\cdot\mathbf{7}_{(1,0,0,4,0,0,1)}$ $+\allowbreak\,$ $5\cdot\mathbf{21}_{(0,0,0,2,1,2,0)}$ $+\allowbreak\,$ $2\cdot\mathbf{35}_{(0,0,2,0,0,0,0)}$ $+\allowbreak\,$ $\mathbf{48}_{(0,0,1,0,0,0,0)}$ $+\allowbreak\,$ $\mathbf{28}_{(0,0,1,0,0,0,0)}$ & $(2,3)$ & $(20,16)$\\[3pt]
\midrule
4 & $B_{3}$ & $3$ & 1 & $23\cdot\mathbf{7}_{(2,6,5,10,0,0,0)}$ $+\allowbreak\,$ $3\cdot\mathbf{35}_{(0,0,3,0,0,0,0)}$ & $(4,4)$ & $(16,20)$\\[3pt]
\midrule
5 & $B_{6}$ & $7$ & 1 & $3\cdot\mathbf{13}_{(2,0,0,0,1,0,0,0,0,0,0,0,0,0,0)}$ $+\allowbreak\,$ $2\cdot\mathbf{78}_{(0,0,0,0,2,0,0,0,0,0,0,0,0,0,0)}$ $+\allowbreak\,$ $\frac{4}{2}\mathbf{64}_{(0,0,0,0,0,0,0,1,0,0,0,0,0,0,0)}$ & $(2,3)$ & $(132,96)$\\[3pt]
\midrule
6 & $C_{3}^{\dagger}$ & $3$ & 1 & $17\cdot\mathbf{14}_{(1,3,10,3,0,0,0)}$ $+\allowbreak\,$ $\frac{1}{2}\mathbf{56}_{(0,0,0,1,0,0,0)}$ & $(3,\tfrac{7}{2})$ & $(16,20)$\\[3pt]
\midrule
7 & $C_{4}$ & $3$ & 1 & $\frac{6}{2}\mathbf{8}_{(0,0,0,1,0,0,0)}$ $+\allowbreak\,$ $\mathbf{8}_{(0,1,0,0,0,0,0)}$ $+\allowbreak\,$ $3\cdot\mathbf{27}_{(3,0,0,0,0,0,0)}$ $+\allowbreak\,$ $2\cdot\mathbf{36}_{(0,0,2,0,0,0,0)}$ $+\allowbreak\,$ $\frac{4}{2}\mathbf{48}_{(0,0,0,1,0,0,0)}$ & $(2,3)$ & $(40,28)$\\[3pt]
\midrule
8 & $C_{5}$ & $3$ & 1 & $\frac{4}{2}\mathbf{10}_{(0,0,0,1,0,0,0)}$ $+\allowbreak\,$ $6\cdot\mathbf{10}_{(1,0,5,0,0,0,0)}$ $+\allowbreak\,$ $5\cdot\mathbf{44}_{(0,1,4,0,0,0,0)}$ & $(1,2)$ & $(28,10)$\\[3pt]
\midrule
9 & $D_{4}$ & $4$ & 1 & $6\cdot\mathbf{28}_{(1,0,3,2,0,0,0,0,0)}$ $+\allowbreak\,$ $\mathbf{35}_{(0,0,0,1,0,0,0,0,0)}$ $+\allowbreak\,$ $\mathbf{35}'_{(0,0,0,1,0,0,0,0,0)}$ $+\allowbreak\,$ $\mathbf{35}''_{(0,0,0,1,0,0,0,0,0)}$ & $(4,6)$ & $(48,30)$\\[3pt]
\midrule
10 & $G_{2}^{\dagger}$ & $3$ & 1 & $15\cdot\mathbf{14}_{(0,4,1,10,0,0,0)}$ $+\allowbreak\,$ $7\cdot\mathbf{7}_{(1,4,2,0,0,0,0)}$ & $(\tfrac{7}{2},7)$ & $(16,20)$\\[3pt]
\bottomrule
\end{tabular}
\captionof{table}{The charge tuples are $(N_{+q_{\max}},\ldots,N_0,\ldots,N_{-q_{\max}})$. One cubic-compatible signed representative is shown per row. Other conventions, including $N_{\rm sign}$ and neutral half-hypermultiplets, follow the appendix introduction. For $C_3$, $\mathbf{14},\mathbf{56}$ have Dynkin labels $(0,1,0),(3,0,0)$. For $D_4$, $(\mathbf{35},\mathbf{35}',\mathbf{35}'')=(\mathbf{35}_v,\mathbf{35}_c,\mathbf{35}_s)$.}
\label{tab:qmax4-representatives}
\end{minipage}
\endgroup

\clearpage
\subsection{Two additional simple factors}
\label{app:u1-two-factors}

\begingroup
\footnotesize
\setlength{\tabcolsep}{0pt}
\setlength{\LTleft}{0pt}\setlength{\LTright}{0pt}
\renewcommand{\arraystretch}{1.12}
\color{black}\hypersetup{linkcolor=blue,citecolor=blue,urlcolor=black}
\captionsetup{hypcap=false,position=bottom}
\setlength{\LTleft}{0pt plus 1fill}\setlength{\LTright}{0pt plus 1fill}
\setlength{\LTcapwidth}{\textwidth}
\renewcommand{\theHtable}{catalogue.\arabic{table}}
\begin{longtable}{@{}>{\centering\arraybackslash}p{14pt}@{\hspace{3pt}}p{52pt}@{\hspace{3pt}}>{\centering\arraybackslash}p{21pt}@{\hspace{3pt}}>{\raggedright\arraybackslash}p{\dimexpr\textwidth-207pt\relax}@{\hspace{3pt}}>{\centering\arraybackslash}p{32pt}@{\hspace{3pt}}>{\centering\arraybackslash}p{32pt}@{\hspace{3pt}}>{\centering\arraybackslash}p{38pt}@{}}
\toprule
No. & $G_1\times G_2$ & $N_{\rm sign}$ & Split spectrum & $b_1$ & $b_2$ & $b_{R+}$\\
\midrule
\endfirsthead
\multicolumn{7}{l}{\small\textbf{Table~\ref{tab:u1-two-combined} (continued)}}\\[3pt]
\toprule
No. & $G_1\times G_2$ & $N_{\rm sign}$ & Split spectrum & $b_1$ & $b_2$ & $b_{R+}$\\
\midrule
\endhead
\midrule
\multicolumn{7}{r}{Continued on next page}\\
\endfoot
\endlastfoot
\multicolumn{7}{c}{\rule{0pt}{14pt}\small\bfseries\boldmath$q_{\max}\leq1$}\\*[3pt]
\midrule
1 & $A_{6}\times D_{7}^{\textcolor{black}{\dagger}}$ & 1 & $6\cdot{(\mathbf{1},\mathbf{14})}_{(4,2,0)}$ $+\allowbreak\,$ $9\cdot{(\mathbf{7},\mathbf{1})}_{(2,5,2)}$ $+\allowbreak\,$ $9\cdot{(\mathbf{21},\mathbf{1})}_{(4,1,4)}$ $+\allowbreak\,$ ${(\mathbf{48},\mathbf{1})}_{(0,1,0)}$ & $(\tfrac{3}{2},3)$ & $(1,-2)$ & $(9,2)$\\[2pt]
\midrule
2 & $A_{6}\times D_{9}$ & 1 & $10\cdot{(\mathbf{1},\mathbf{18})}_{(9,1,0)}$ $+\allowbreak\,$ $20\cdot{(\mathbf{7},\mathbf{1})}_{(1,18,1)}$ $+\allowbreak\,$ $6\cdot{(\mathbf{21},\mathbf{1})}_{(3,0,3)}$ & $(1,2)$ & $(1,-2)$ & $(8,2)$\\[2pt]
\midrule
3 & $A_{7}\times B_{8}^{\textcolor{black}{\dagger}}$ & 1 & ${(\mathbf{1},\mathbf{152})}_{(1,0,0)}$ $+\allowbreak\,$ ${(\mathbf{1},\mathbf{256})}_{(1,0,0)}$ $+\allowbreak\,$ ${(\mathbf{36},\mathbf{1})}_{(0,1,0)}$ & $(-1,\tfrac{1}{2})$ & $(4,2)$ & $(2,17)$\\[2pt]
\midrule
4 & $A_{7}\times E_{6}^{\textcolor{black}{\dagger}}$ & 3 & $2\cdot{(\mathbf{1},\mathbf{27})}_{(2,0,0)}$ $+\allowbreak\,$ $9\cdot{(\mathbf{8},\mathbf{1})}_{(0,9,0)}$ $+\allowbreak\,$ $6\cdot{(\mathbf{28},\mathbf{1})}_{(1,1,4)}$ $+\allowbreak\,$ ${(\mathbf{36},\mathbf{1})}_{(1,0,0)}$ $+\allowbreak\,$ ${(\mathbf{56},\mathbf{1})}_{(0,1,0)}$ & $(\tfrac{3}{2},3)$ & $(1,-2)$ & $(7,2)$\\[2pt]
\midrule
5 & $A_{7}\times F_{4}^{\textcolor{black}{\dagger}}$ & 1 & ${(\mathbf{1},\mathbf{324})}_{(1,0,0)}$ $+\allowbreak\,$ ${(\mathbf{36},\mathbf{1})}_{(0,1,0)}$ & $(-1,\tfrac{1}{2})$ & $(8,4)$ & $(2,17)$\\[2pt]
\midrule
6 & $A_{9}\times D_{6}$ & 1 & $8\cdot{(\mathbf{1},\mathbf{12})}_{(7,1,0)}$ $+\allowbreak\,$ ${(\mathbf{1},\mathbf{32})}_{(0,1,0)}$ $+\allowbreak\,$ ${(\mathbf{1},\mathbf{32}')}_{(1,0,0)}$ $+\allowbreak\,$ $16\cdot{(\mathbf{10},\mathbf{1})}_{(3,10,3)}$ $+\allowbreak\,$ $2\cdot{(\mathbf{45},\mathbf{1})}_{(1,0,1)}$ & $(0,1)$ & $(0,1)$ & $(2,8)$\\[2pt]
\midrule
7 & $A_{9}\times D_{6}$ & 1 & $8\cdot{(\mathbf{1},\mathbf{12})}_{(7,1,0)}$ $+\allowbreak\,$ ${(\mathbf{1},\mathbf{32})}_{(1,0,0)}$ $+\allowbreak\,$ ${(\mathbf{1},\mathbf{32})}_{(0,1,0)}$ $+\allowbreak\,$ $16\cdot{(\mathbf{10},\mathbf{1})}_{(3,10,3)}$ $+\allowbreak\,$ $2\cdot{(\mathbf{45},\mathbf{1})}_{(1,0,1)}$ & $(0,1)$ & $(0,1)$ & $(2,8)$\\[2pt]
\midrule
8 & $A_{9}\times D_{6}$ & 1 & $8\cdot{(\mathbf{1},\mathbf{12})}_{(7,1,0)}$ $+\allowbreak\,$ ${(\mathbf{1},\mathbf{32})}_{(1,0,0)}$ $+\allowbreak\,$ $\tfrac{1}{2}\cdot{(\mathbf{1},\mathbf{32})}_{(0,1,0)}$ $+\allowbreak\,$ $\tfrac{1}{2}\cdot{(\mathbf{1},\mathbf{32}')}_{(0,1,0)}$ $+\allowbreak\,$ $16\cdot{(\mathbf{10},\mathbf{1})}_{(3,10,3)}$ $+\allowbreak\,$ $2\cdot{(\mathbf{45},\mathbf{1})}_{(1,0,1)}$ & $(0,1)$ & $(0,1)$ & $(2,8)$\\[2pt]
\midrule
\rowcolor{gray!15}[1.5pt][1.5pt]9 & $B_{8}\times C_{9}$ & 1 & $\tfrac{1}{2}\cdot{(\mathbf{1},\mathbf{798})}_{(0,1,0)}$ $+\allowbreak\,$ $\tfrac{1}{2}\cdot{(\mathbf{17},\mathbf{18})}_{(0,1,0)}$ & $(-2,1)$ & $(3,1)$ & $(-26,2)$\\[2pt]
\midrule
\rowcolor{gray!15}[1.5pt][1.5pt]10 & $C_{9}\times F_{4}^{\textcolor{black}{\dagger}}$ & 1 & $\tfrac{1}{2}\cdot{(\mathbf{18},\mathbf{52})}_{(0,1,0)}$ & $(-\tfrac{1}{2},1)$ & $(10,2)$ & $(-19,2)$\\[2pt]
\midrule
\rowcolor{gray!15}[1.5pt][1.5pt]11 & $C_{10}\times E_{7}^{\textcolor{black}{\dagger}}$ & 1 & $\tfrac{1}{2}\cdot{(\mathbf{1},\mathbf{56})}_{(0,1,0)}$ $+\allowbreak\,$ $\tfrac{1}{2}\cdot{(\mathbf{1120},\mathbf{1})}_{(0,1,0)}$ & $(1,\tfrac{7}{2})$ & $(1,-\tfrac{7}{2})$ & $(2,-29)$\\[2pt]
\midrule
\rowcolor{gray!15}[1.5pt][1.5pt]12 & $E_{6}\times E_{7}$ & 1 & $\tfrac{1}{2}\cdot{(\mathbf{1},\mathbf{912})}_{(0,1,0)}$ & $(-3,1)$ & $(9,3)$ & $(-18,2)$\\[2pt]
\midrule
\rowcolor{gray!15}[1.5pt][1.5pt]13 & $A_{8}\times A_{17}$ & 1 & $ {(\overline{\mathbf{36}},\mathbf{18})}_{(0,1,0)}$ & $(1,8)$ & $(1,-1)$ & $(2,-34)$\\[2pt]
\midrule
\rowcolor{gray!15}[1.5pt][1.5pt]14 & $A_{17}\times E_{8}$ & 1 & $ {(\mathbf{816},\mathbf{1})}_{(0,1,0)}$ & $(1,6)$ & $(1,-6)$ & $(2,-48)$\\[2pt]
\midrule
\rowcolor{gray!15}[1.5pt][1.5pt]15 & $D_{18}\times E_{7}$ & 1 & $\tfrac{1}{2}\cdot{(\mathbf{36},\mathbf{56})}_{(0,1,0)}$ & $(-2,1)$ & $(14,1)$ & $(-64,2)$\\[2pt]
\specialrule{0.6pt}{5pt}{2pt}
\multicolumn{7}{c}{\rule{0pt}{14pt}\small\bfseries\boldmath$q_{\max}=2$}\\*[3pt]
\midrule
1 & $A_{6}\times A_{10}^{\dagger}$ & 1 & $6\cdot{(\mathbf{1},\mathbf{11})}_{(3,1,2,0,0)}$ $+\allowbreak\,$ $3\cdot{(\mathbf{1},\mathbf{55})}_{(0,0,2,0,1)}$ $+\allowbreak\,$ $6\cdot{(\mathbf{7},\mathbf{1})}_{(2,0,3,0,1)}$ $+\allowbreak\,$ ${(\overline{\mathbf{7}},\mathbf{11})}_{(0,1,0,0,0)}$ $+\allowbreak\,$ $3\cdot{(\mathbf{21},\mathbf{1})}_{(1,1,1,0,0)}$ & $(\tfrac{1}{2},1)$ & $(\tfrac{1}{2},1)$ & $(28,12)$\\[2pt]
\midrule
2 & $A_{6}\times A_{10}^{\dagger}$ & 15 & $6\cdot{(\mathbf{1},\mathbf{11})}_{(0,1,0,2,3)}$ $+\allowbreak\,$ $3\cdot{(\mathbf{1},\mathbf{55})}_{(0,2,0,0,1)}$ $+\allowbreak\,$ $6\cdot{(\mathbf{7},\mathbf{1})}_{(0,2,0,2,2)}$ $+\allowbreak\,$ ${(\overline{\mathbf{7}},\mathbf{11})}_{(0,1,0,0,0)}$ $+\allowbreak\,$ $3\cdot{(\mathbf{21},\mathbf{1})}_{(2,1,0,0,0)}$ & $(\tfrac{1}{2},1)$ & $(\tfrac{1}{2},1)$ & $(50,8)$\\[2pt]
\midrule
3 & $A_{6}\times D_{7}^{\dagger}$ & 2 & $6\cdot{(\mathbf{1},\mathbf{14})}_{(1,2,3,0,0)}$ $+\allowbreak\,$ $9\cdot{(\mathbf{7},\mathbf{1})}_{(0,0,3,6,0)}$ $+\allowbreak\,$ $9\cdot{(\mathbf{21},\mathbf{1})}_{(0,4,3,2,0)}$ $+\allowbreak\,$ ${(\mathbf{48},\mathbf{1})}_{(0,1,0,0,0)}$ & $(\tfrac{3}{2},3)$ & $(1,-2)$ & $(9,6)$\\[2pt]
\midrule
4 & $A_{6}\times D_{9}$ & 4 & $10\cdot{(\mathbf{1},\mathbf{18})}_{(1,7,2,0,0)}$ $+\allowbreak\,$ $20\cdot{(\mathbf{7},\mathbf{1})}_{(0,2,12,4,2)}$ $+\allowbreak\,$ $6\cdot{(\mathbf{21},\mathbf{1})}_{(0,4,0,2,0)}$ & $(1,2)$ & $(1,-2)$ & $(10,10)$\\[2pt]
\midrule
5 & $A_{6}\times D_{9}^{\dagger}$ & 4 & $10\cdot{(\mathbf{1},\mathbf{18})}_{(3,0,7,0,0)}$ $+\allowbreak\,$ $20\cdot{(\mathbf{7},\mathbf{1})}_{(0,1,0,19,0)}$ $+\allowbreak\,$ $6\cdot{(\mathbf{21},\mathbf{1})}_{(0,6,0,0,0)}$ & $(1,2)$ & $(1,-2)$ & $(11,14)$\\[2pt]
\midrule
6 & $A_{6}\times D_{9}^{\dagger}$ & 14 & $10\cdot{(\mathbf{1},\mathbf{18})}_{(3,2,5,0,0)}$ $+\allowbreak\,$ $20\cdot{(\mathbf{7},\mathbf{1})}_{(0,5,3,7,5)}$ $+\allowbreak\,$ $6\cdot{(\mathbf{21},\mathbf{1})}_{(1,2,3,0,0)}$ & $(1,2)$ & $(1,-2)$ & $(13,22)$\\[2pt]
7 & $A_{6}\times D_{9}^{\dagger}$ & 2 & $10\cdot{(\mathbf{1},\mathbf{18})}_{(3,2,5,0,0)}$ $+\allowbreak\,$ $20\cdot{(\mathbf{7},\mathbf{1})}_{(1,0,12,0,7)}$ $+\allowbreak\,$ $6\cdot{(\mathbf{21},\mathbf{1})}_{(0,5,0,1,0)}$ & $(1,2)$ & $(1,-2)$ & $(13,22)$\\[2pt]
\midrule
8 & $A_{6}\times D_{9}^{\dagger}$ & 1 & $10\cdot{(\mathbf{1},\mathbf{18})}_{(0,10,0,0,0)}$ $+\allowbreak\,$ $20\cdot{(\mathbf{7},\mathbf{1})}_{(1,0,18,0,1)}$ $+\allowbreak\,$ $6\cdot{(\mathbf{21},\mathbf{1})}_{(0,3,0,3,0)}$ & $(1,2)$ & $(1,-2)$ & $(9,6)$\\[2pt]
\midrule
9 & $A_{6}\times D_{9}$ & 3 & $3\cdot{(\mathbf{1},\mathbf{18})}_{(1,0,2,0,0)}$ $+\allowbreak\,$ $16\cdot{(\mathbf{7},\mathbf{1})}_{(0,0,6,8,2)}$ $+\allowbreak\,$ ${(\mathbf{7},\mathbf{18})}_{(0,1,0,0,0)}$ $+\allowbreak\,$ $4\cdot{(\mathbf{21},\mathbf{1})}_{(0,0,2,2,0)}$ $+\allowbreak\,$ $2\cdot{(\mathbf{35},\mathbf{1})}_{(0,1,0,1,0)}$ & $(1,4)$ & $(1,-2)$ & $(10,10)$\\[2pt]
\midrule
10 & $A_{6}\times D_{9}$ & 5 & $3\cdot{(\mathbf{1},\mathbf{18})}_{(0,2,1,0,0)}$ $+\allowbreak\,$ $16\cdot{(\mathbf{7},\mathbf{1})}_{(0,5,2,1,8)}$ $+\allowbreak\,$ ${(\mathbf{7},\mathbf{18})}_{(0,0,1,0,0)}$ $+\allowbreak\,$ $4\cdot{(\mathbf{21},\mathbf{1})}_{(3,0,0,0,1)}$ $+\allowbreak\,$ $2\cdot{(\mathbf{35},\mathbf{1})}_{(0,0,2,0,0)}$ & $(1,4)$ & $(1,-2)$ & $(22,16)$\\[2pt]
11 & $A_{6}\times D_{9}$ & 4 & $3\cdot{(\mathbf{1},\mathbf{18})}_{(0,2,1,0,0)}$ $+\allowbreak\,$ $16\cdot{(\mathbf{7},\mathbf{1})}_{(2,0,4,0,10)}$ $+\allowbreak\,$ ${(\mathbf{7},\mathbf{18})}_{(0,0,1,0,0)}$ $+\allowbreak\,$ $4\cdot{(\mathbf{21},\mathbf{1})}_{(1,2,1,0,0)}$ $+\allowbreak\,$ $2\cdot{(\mathbf{35},\mathbf{1})}_{(1,0,1,0,0)}$ & $(1,4)$ & $(1,-2)$ & $(22,16)$\\[2pt]
\midrule
12 & $A_{6}\times E_{7}^{\dagger}$ & 5 & $2\cdot{(\mathbf{1},\mathbf{56})}_{(0,2,0,0,0)}$ $+\allowbreak\,$ $19\cdot{(\mathbf{7},\mathbf{1})}_{(0,4,9,4,2)}$ $+\allowbreak\,$ $3\cdot{(\mathbf{21},\mathbf{1})}_{(0,0,3,0,0)}$ $+\allowbreak\,$ $2\cdot{(\mathbf{35},\mathbf{1})}_{(0,2,0,0,0)}$ $+\allowbreak\,$ ${(\mathbf{48},\mathbf{1})}_{(0,1,0,0,0)}$ & $(\tfrac{3}{2},3)$ & $(1,-2)$ & $(9,6)$\\[2pt]
\midrule
13 & $A_{6}\times F_{4}^{\dagger}$ & 3 & $2\cdot{(\mathbf{1},\mathbf{26})}_{(0,1,1,0,0)}$ $+\allowbreak\,$ $6\cdot{(\mathbf{7},\mathbf{1})}_{(1,0,1,1,3)}$ $+\allowbreak\,$ $5\cdot{(\mathbf{21},\mathbf{1})}_{(1,1,2,0,1)}$ $+\allowbreak\,$ ${(\mathbf{28},\mathbf{1})}_{(0,0,1,0,0)}$ $+\allowbreak\,$ $2\cdot{(\mathbf{35},\mathbf{1})}_{(0,1,1,0,0)}$ $+\allowbreak\,$ ${(\mathbf{48},\mathbf{1})}_{(0,1,0,0,0)}$ & $(2,3)$ & $(1,-\tfrac{3}{2})$ & $(16,12)$\\[2pt]
14 & $A_{6}\times F_{4}^{\dagger}$ & 4 & $2\cdot{(\mathbf{1},\mathbf{26})}_{(0,1,1,0,0)}$ $+\allowbreak\,$ $6\cdot{(\mathbf{7},\mathbf{1})}_{(1,2,0,0,3)}$ $+\allowbreak\,$ $5\cdot{(\mathbf{21},\mathbf{1})}_{(0,1,2,0,2)}$ $+\allowbreak\,$ ${(\mathbf{28},\mathbf{1})}_{(0,1,0,0,0)}$ $+\allowbreak\,$ $2\cdot{(\mathbf{35},\mathbf{1})}_{(0,0,2,0,0)}$ $+\allowbreak\,$ ${(\mathbf{48},\mathbf{1})}_{(0,1,0,0,0)}$ & $(2,3)$ & $(1,-\tfrac{3}{2})$ & $(16,12)$\\[2pt]
\midrule
15 & $A_{7}\times A_{9}^{\dagger}$ & 2 & $6\cdot{(\mathbf{1},\mathbf{10})}_{(1,0,1,0,4)}$ $+\allowbreak\,$ $3\cdot{(\mathbf{1},\mathbf{45})}_{(0,2,0,1,0)}$ $+\allowbreak\,$ $6\cdot{(\mathbf{8},\mathbf{1})}_{(0,0,2,0,4)}$ $+\allowbreak\,$ ${(\overline{\mathbf{8}},\mathbf{10})}_{(0,0,1,0,0)}$ $+\allowbreak\,$ $3\cdot{(\mathbf{28},\mathbf{1})}_{(1,0,2,0,0)}$ & $(\tfrac{1}{2},1)$ & $(\tfrac{1}{2},1)$ & $(12,24)$\\[2pt]
\midrule
16 & $A_{7}\times D_{6}$ & 3 & $4\cdot{(\mathbf{1},\mathbf{12})}_{(2,0,2,0,0)}$ $+\allowbreak\,$ $32\cdot{(\mathbf{8},\mathbf{1})}_{(0,12,12,4,4)}$ $+\allowbreak\,$ ${(\mathbf{70},\mathbf{1})}_{(0,1,0,0,0)}$ & $(1,2)$ & $(1,-2)$ & $(10,16)$\\[2pt]
\midrule
17 & $A_{7}\times D_{7}^{\dagger}$ & 1 & $6\cdot{(\mathbf{1},\mathbf{14})}_{(1,2,3,0,0)}$ $+\allowbreak\,$ $9\cdot{(\mathbf{28},\mathbf{1})}_{(0,3,3,3,0)}$ $+\allowbreak\,$ ${(\mathbf{63},\mathbf{1})}_{(0,1,0,0,0)}$ & $(\tfrac{3}{2},3)$ & $(1,-2)$ & $(9,6)$\\[2pt]
\midrule
18 & $A_{8}\times A_{8}$ & 2 & $5\cdot{(\mathbf{1},\mathbf{9})}_{(0,0,0,1,4)}$ $+\allowbreak\,$ $4\cdot{(\mathbf{1},\mathbf{36})}_{(0,1,2,1,0)}$ $+\allowbreak\,$ $7\cdot{(\mathbf{9},\mathbf{1})}_{(0,1,4,2,0)}$ $+\allowbreak\,$ ${(\overline{\mathbf{9}},\mathbf{9})}_{(0,1,0,0,0)}$ $+\allowbreak\,$ $2\cdot{(\mathbf{36},\mathbf{1})}_{(0,2,0,0,0)}$ & $(0,1)$ & $(1,1)$ & $(8,14)$\\[2pt]
19 & $A_{8}\times A_{8}$ & 2 & $5\cdot{(\mathbf{1},\mathbf{9})}_{(0,1,1,1,2)}$ $+\allowbreak\,$ $4\cdot{(\mathbf{1},\mathbf{36})}_{(0,1,1,2,0)}$ $+\allowbreak\,$ $7\cdot{(\mathbf{9},\mathbf{1})}_{(2,1,3,1,0)}$ $+\allowbreak\,$ ${(\overline{\mathbf{9}},\mathbf{9})}_{(0,1,0,0,0)}$ $+\allowbreak\,$ $2\cdot{(\mathbf{36},\mathbf{1})}_{(0,1,1,0,0)}$ & $(0,1)$ & $(1,1)$ & $(8,14)$\\[2pt]
20 & $A_{8}\times A_{8}$ & 1 & $5\cdot{(\mathbf{1},\mathbf{9})}_{(0,2,2,1,0)}$ $+\allowbreak\,$ $4\cdot{(\mathbf{1},\mathbf{36})}_{(0,1,0,3,0)}$ $+\allowbreak\,$ $7\cdot{(\mathbf{9},\mathbf{1})}_{(4,1,2,0,0)}$ $+\allowbreak\,$ ${(\overline{\mathbf{9}},\mathbf{9})}_{(0,1,0,0,0)}$ $+\allowbreak\,$ $2\cdot{(\mathbf{36},\mathbf{1})}_{(0,0,2,0,0)}$ & $(0,1)$ & $(1,1)$ & $(8,14)$\\[2pt]
\midrule
21 & $A_{9}\times D_{6}$ & 1 & $8\cdot{(\mathbf{1},\mathbf{12})}_{(0,8,0,0,0)}$ $+\allowbreak\,$ ${(\mathbf{1},\mathbf{32})}_{(0,1,0,0,0)}$ $+\allowbreak\,$ ${(\mathbf{1},\mathbf{32}')}_{(0,1,0,0,0)}$ $+\allowbreak\,$ $16\cdot{(\mathbf{10},\mathbf{1})}_{(2,0,12,0,2)}$ $+\allowbreak\,$ $2\cdot{(\mathbf{45},\mathbf{1})}_{(0,1,0,1,0)}$ & $(0,1)$ & $(0,1)$ & $(12,10)$\\[2pt]
\midrule
22 & $A_{9}\times D_{6}$ & 5 & $8\cdot{(\mathbf{1},\mathbf{12})}_{(0,4,4,0,0)}$ $+\allowbreak\,$ ${(\mathbf{1},\mathbf{32})}_{(0,0,1,0,0)}$ $+\allowbreak\,$ ${(\mathbf{1},\mathbf{32}')}_{(1,0,0,0,0)}$ $+\allowbreak\,$ $16\cdot{(\mathbf{10},\mathbf{1})}_{(0,2,4,6,4)}$ $+\allowbreak\,$ $2\cdot{(\mathbf{45},\mathbf{1})}_{(0,2,0,0,0)}$ & $(0,1)$ & $(0,1)$ & $(20,10)$\\[2pt]
\midrule
23 & $A_{9}\times D_{6}$ & 2 & $8\cdot{(\mathbf{1},\mathbf{12})}_{(4,0,4,0,0)}$ $+\allowbreak\,$ ${(\mathbf{1},\mathbf{32})}_{(0,0,1,0,0)}$ $+\allowbreak\,$ ${(\mathbf{1},\mathbf{32}')}_{(0,1,0,0,0)}$ $+\allowbreak\,$ $16\cdot{(\mathbf{10},\mathbf{1})}_{(0,0,10,0,6)}$ $+\allowbreak\,$ $2\cdot{(\mathbf{45},\mathbf{1})}_{(0,2,0,0,0)}$ & $(0,1)$ & $(0,1)$ & $(20,14)$\\[2pt]
\midrule
24 & $A_{9}\times D_{6}$ & 1 & $8\cdot{(\mathbf{1},\mathbf{12})}_{(1,1,6,0,0)}$ $+\allowbreak\,$ ${(\mathbf{1},\mathbf{32})}_{(0,1,0,0,0)}$ $+\allowbreak\,$ ${(\mathbf{1},\mathbf{32}')}_{(0,1,0,0,0)}$ $+\allowbreak\,$ $16\cdot{(\mathbf{10},\mathbf{1})}_{(0,5,6,5,0)}$ $+\allowbreak\,$ $2\cdot{(\mathbf{45},\mathbf{1})}_{(0,1,0,1,0)}$ & $(0,1)$ & $(0,1)$ & $(6,8)$\\[2pt]
\midrule
25 & $A_{9}\times D_{6}$ & 2 & $8\cdot{(\mathbf{1},\mathbf{12})}_{(2,2,4,0,0)}$ $+\allowbreak\,$ ${(\mathbf{1},\mathbf{32})}_{(0,0,1,0,0)}$ $+\allowbreak\,$ ${(\mathbf{1},\mathbf{32}')}_{(0,1,0,0,0)}$ $+\allowbreak\,$ $16\cdot{(\mathbf{10},\mathbf{1})}_{(0,0,4,12,0)}$ $+\allowbreak\,$ $2\cdot{(\mathbf{45},\mathbf{1})}_{(0,2,0,0,0)}$ & $(0,1)$ & $(0,1)$ & $(8,10)$\\[2pt]
\midrule
26 & $A_{9}\times D_{6}$ & 1 & $8\cdot{(\mathbf{1},\mathbf{12})}_{(0,8,0,0,0)}$ $+\allowbreak\,$ $2\cdot{(\mathbf{1},\mathbf{32})}_{(0,2,0,0,0)}$ $+\allowbreak\,$ $16\cdot{(\mathbf{10},\mathbf{1})}_{(2,0,12,0,2)}$ $+\allowbreak\,$ $2\cdot{(\mathbf{45},\mathbf{1})}_{(0,1,0,1,0)}$ & $(0,1)$ & $(0,1)$ & $(12,10)$\\[2pt]
\midrule
27 & $A_{9}\times D_{6}$ & 5 & $8\cdot{(\mathbf{1},\mathbf{12})}_{(0,4,4,0,0)}$ $+\allowbreak\,$ ${(\mathbf{1},\mathbf{32})}_{(1,0,0,0,0)}$ $+\allowbreak\,$ ${(\mathbf{1},\mathbf{32})}_{(0,0,1,0,0)}$ $+\allowbreak\,$ $16\cdot{(\mathbf{10},\mathbf{1})}_{(0,2,4,6,4)}$ $+\allowbreak\,$ $2\cdot{(\mathbf{45},\mathbf{1})}_{(0,2,0,0,0)}$ & $(0,1)$ & $(0,1)$ & $(20,10)$\\[2pt]
\midrule
28 & $A_{9}\times D_{6}$ & 2 & $8\cdot{(\mathbf{1},\mathbf{12})}_{(4,0,4,0,0)}$ $+\allowbreak\,$ ${(\mathbf{1},\mathbf{32})}_{(0,1,0,0,0)}$ $+\allowbreak\,$ ${(\mathbf{1},\mathbf{32})}_{(0,0,1,0,0)}$ $+\allowbreak\,$ $16\cdot{(\mathbf{10},\mathbf{1})}_{(0,0,10,0,6)}$ $+\allowbreak\,$ $2\cdot{(\mathbf{45},\mathbf{1})}_{(0,2,0,0,0)}$ & $(0,1)$ & $(0,1)$ & $(20,14)$\\[2pt]
\midrule
29 & $A_{9}\times D_{6}$ & 1 & $8\cdot{(\mathbf{1},\mathbf{12})}_{(1,1,6,0,0)}$ $+\allowbreak\,$ $2\cdot{(\mathbf{1},\mathbf{32})}_{(0,2,0,0,0)}$ $+\allowbreak\,$ $16\cdot{(\mathbf{10},\mathbf{1})}_{(0,5,6,5,0)}$ $+\allowbreak\,$ $2\cdot{(\mathbf{45},\mathbf{1})}_{(0,1,0,1,0)}$ & $(0,1)$ & $(0,1)$ & $(6,8)$\\[2pt]
\midrule
30 & $A_{9}\times D_{6}$ & 2 & $8\cdot{(\mathbf{1},\mathbf{12})}_{(2,2,4,0,0)}$ $+\allowbreak\,$ ${(\mathbf{1},\mathbf{32})}_{(0,1,0,0,0)}$ $+\allowbreak\,$ ${(\mathbf{1},\mathbf{32})}_{(0,0,1,0,0)}$ $+\allowbreak\,$ $16\cdot{(\mathbf{10},\mathbf{1})}_{(0,0,4,12,0)}$ $+\allowbreak\,$ $2\cdot{(\mathbf{45},\mathbf{1})}_{(0,2,0,0,0)}$ & $(0,1)$ & $(0,1)$ & $(8,10)$\\[2pt]
\midrule
31 & $A_{9}\times D_{6}$ & 5 & $8\cdot{(\mathbf{1},\mathbf{12})}_{(0,4,4,0,0)}$ $+\allowbreak\,$ ${(\mathbf{1},\mathbf{32})}_{(1,0,0,0,0)}$ $+\allowbreak\,$ $\tfrac{1}{2}\cdot{(\mathbf{1},\mathbf{32})}_{(0,0,1,0,0)}$ $+\allowbreak\,$ $\tfrac{1}{2}\cdot{(\mathbf{1},\mathbf{32}')}_{(0,0,1,0,0)}$ $+\allowbreak\,$ $16\cdot{(\mathbf{10},\mathbf{1})}_{(0,2,4,6,4)}$ $+\allowbreak\,$ $2\cdot{(\mathbf{45},\mathbf{1})}_{(0,2,0,0,0)}$ & $(0,1)$ & $(0,1)$ & $(20,10)$\\[2pt]
\midrule
32 & $A_{9}\times D_{6}$ & 2 & $8\cdot{(\mathbf{1},\mathbf{12})}_{(4,0,4,0,0)}$ $+\allowbreak\,$ ${(\mathbf{1},\mathbf{32})}_{(0,1,0,0,0)}$ $+\allowbreak\,$ $\tfrac{1}{2}\cdot{(\mathbf{1},\mathbf{32})}_{(0,0,1,0,0)}$ $+\allowbreak\,$ $\tfrac{1}{2}\cdot{(\mathbf{1},\mathbf{32}')}_{(0,0,1,0,0)}$ $+\allowbreak\,$ $16\cdot{(\mathbf{10},\mathbf{1})}_{(0,0,10,0,6)}$ $+\allowbreak\,$ $2\cdot{(\mathbf{45},\mathbf{1})}_{(0,2,0,0,0)}$ & $(0,1)$ & $(0,1)$ & $(20,14)$\\[2pt]
\midrule
33 & $A_{9}\times D_{6}$ & 2 & $8\cdot{(\mathbf{1},\mathbf{12})}_{(2,2,4,0,0)}$ $+\allowbreak\,$ ${(\mathbf{1},\mathbf{32})}_{(0,1,0,0,0)}$ $+\allowbreak\,$ $\tfrac{1}{2}\cdot{(\mathbf{1},\mathbf{32})}_{(0,0,1,0,0)}$ $+\allowbreak\,$ $\tfrac{1}{2}\cdot{(\mathbf{1},\mathbf{32}')}_{(0,0,1,0,0)}$ $+\allowbreak\,$ $16\cdot{(\mathbf{10},\mathbf{1})}_{(0,0,4,12,0)}$ $+\allowbreak\,$ $2\cdot{(\mathbf{45},\mathbf{1})}_{(0,2,0,0,0)}$ & $(0,1)$ & $(0,1)$ & $(8,10)$\\[2pt]
\midrule
34 & $D_{6}\times E_{6}^{\dagger}$ & 2 & $3\cdot{(\mathbf{1},\mathbf{27})}_{(0,3,0,0,0)}$ $+\allowbreak\,$ $4\cdot{(\mathbf{12},\mathbf{1})}_{(1,0,3,0,0)}$ $+\allowbreak\,$ $2\cdot{(\mathbf{32},\mathbf{1})}_{(0,0,1,0,0)}$ $+\allowbreak\,$ ${(\mathbf{32}',\mathbf{1})}_{(0,1,0,0,0)}$ $+\allowbreak\,$ ${(\mathbf{32}',\mathbf{1})}_{(0,0,1,0,0)}$ $+\allowbreak\,$ $2\cdot{(\mathbf{66},\mathbf{1})}_{(0,2,0,0,0)}$ & $(2,3)$ & $(1,-\tfrac{3}{2})$ & $(8,6)$\\[2pt]
\midrule
35 & $D_{6}\times E_{6}^{\dagger}$ & 2 & $3\cdot{(\mathbf{1},\mathbf{27})}_{(0,3,0,0,0)}$ $+\allowbreak\,$ $4\cdot{(\mathbf{12},\mathbf{1})}_{(1,0,3,0,0)}$ $+\allowbreak\,$ $3\cdot{(\mathbf{32},\mathbf{1})}_{(0,0,1,0,0)}$ $+\allowbreak\,$ ${(\mathbf{32}',\mathbf{1})}_{(0,1,0,0,0)}$ $+\allowbreak\,$ $2\cdot{(\mathbf{66},\mathbf{1})}_{(0,2,0,0,0)}$ & $(2,3)$ & $(1,-\tfrac{3}{2})$ & $(8,6)$\\[2pt]
36 & $D_{6}\times E_{6}^{\dagger}$ & 2 & $3\cdot{(\mathbf{1},\mathbf{27})}_{(0,3,0,0,0)}$ $+\allowbreak\,$ $4\cdot{(\mathbf{12},\mathbf{1})}_{(1,0,3,0,0)}$ $+\allowbreak\,$ ${(\mathbf{32},\mathbf{1})}_{(0,1,0,0,0)}$ $+\allowbreak\,$ $2\cdot{(\mathbf{32},\mathbf{1})}_{(0,0,1,0,0)}$ $+\allowbreak\,$ ${(\mathbf{32}',\mathbf{1})}_{(0,0,1,0,0)}$ $+\allowbreak\,$ $2\cdot{(\mathbf{66},\mathbf{1})}_{(0,2,0,0,0)}$ & $(2,3)$ & $(1,-\tfrac{3}{2})$ & $(8,6)$\\[2pt]
\midrule
37 & $D_{6}\times E_{6}^{\dagger}$ & 2 & $3\cdot{(\mathbf{1},\mathbf{27})}_{(0,3,0,0,0)}$ $+\allowbreak\,$ $4\cdot{(\mathbf{12},\mathbf{1})}_{(1,0,3,0,0)}$ $+\allowbreak\,$ ${(\mathbf{32},\mathbf{1})}_{(0,1,0,0,0)}$ $+\allowbreak\,$ $3\cdot{(\mathbf{32},\mathbf{1})}_{(0,0,1,0,0)}$ $+\allowbreak\,$ $2\cdot{(\mathbf{66},\mathbf{1})}_{(0,2,0,0,0)}$ & $(2,3)$ & $(1,-\tfrac{3}{2})$ & $(8,6)$\\[2pt]
\midrule
38 & $D_{6}\times E_{6}^{\dagger}$ & 2 & $3\cdot{(\mathbf{1},\mathbf{27})}_{(0,3,0,0,0)}$ $+\allowbreak\,$ $4\cdot{(\mathbf{12},\mathbf{1})}_{(1,0,3,0,0)}$ $+\allowbreak\,$ $\tfrac{5}{2}\cdot{(\mathbf{32},\mathbf{1})}_{(0,0,1,0,0)}$ $+\allowbreak\,$ ${(\mathbf{32}',\mathbf{1})}_{(0,1,0,0,0)}$ $+\allowbreak\,$ $\tfrac{1}{2}\cdot{(\mathbf{32}',\mathbf{1})}_{(0,0,1,0,0)}$ $+\allowbreak\,$ $2\cdot{(\mathbf{66},\mathbf{1})}_{(0,2,0,0,0)}$ & $(2,3)$ & $(1,-\tfrac{3}{2})$ & $(8,6)$\\[2pt]
39 & $D_{6}\times E_{6}^{\dagger}$ & 2 & $3\cdot{(\mathbf{1},\mathbf{27})}_{(0,3,0,0,0)}$ $+\allowbreak\,$ $4\cdot{(\mathbf{12},\mathbf{1})}_{(1,0,3,0,0)}$ $+\allowbreak\,$ ${(\mathbf{32},\mathbf{1})}_{(0,1,0,0,0)}$ $+\allowbreak\,$ $\tfrac{3}{2}\cdot{(\mathbf{32},\mathbf{1})}_{(0,0,1,0,0)}$ $+\allowbreak\,$ $\tfrac{3}{2}\cdot{(\mathbf{32}',\mathbf{1})}_{(0,0,1,0,0)}$ $+\allowbreak\,$ $2\cdot{(\mathbf{66},\mathbf{1})}_{(0,2,0,0,0)}$ & $(2,3)$ & $(1,-\tfrac{3}{2})$ & $(8,6)$\\[2pt]
\midrule
40 & $D_{6}\times E_{6}^{\dagger}$ & 2 & $3\cdot{(\mathbf{1},\mathbf{27})}_{(0,3,0,0,0)}$ $+\allowbreak\,$ $4\cdot{(\mathbf{12},\mathbf{1})}_{(1,0,3,0,0)}$ $+\allowbreak\,$ ${(\mathbf{32},\mathbf{1})}_{(0,1,0,0,0)}$ $+\allowbreak\,$ $\tfrac{5}{2}\cdot{(\mathbf{32},\mathbf{1})}_{(0,0,1,0,0)}$ $+\allowbreak\,$ $\tfrac{1}{2}\cdot{(\mathbf{32}',\mathbf{1})}_{(0,0,1,0,0)}$ $+\allowbreak\,$ $2\cdot{(\mathbf{66},\mathbf{1})}_{(0,2,0,0,0)}$ & $(2,3)$ & $(1,-\tfrac{3}{2})$ & $(8,6)$\\[2pt]
\midrule
41 & $D_{7}\times E_{7}^{\dagger}$ & 1 & $3\cdot{(\mathbf{1},\mathbf{56})}_{(0,3,0,0,0)}$ $+\allowbreak\,$ $\tfrac{3}{2}\cdot{(\mathbf{1},\mathbf{56})}_{(0,0,1,0,0)}$ $+\allowbreak\,$ ${(\mathbf{1},\mathbf{133})}_{(0,1,0,0,0)}$ $+\allowbreak\,$ $6\cdot{(\mathbf{14},\mathbf{1})}_{(1,2,3,0,0)}$ & $(-2,1)$ & $(3,\tfrac{3}{2})$ & $(6,9)$\\[2pt]
\midrule
42 & $A_{13}\times D_{9}^{\dagger}$ & 1 & $10\cdot{(\mathbf{1},\mathbf{18})}_{(3,0,7,0,0)}$ $+\allowbreak\,$ $10\cdot{(\mathbf{14},\mathbf{1})}_{(0,0,0,10,0)}$ $+\allowbreak\,$ $3\cdot{(\mathbf{91},\mathbf{1})}_{(0,2,0,1,0)}$ & $(\tfrac{1}{2},1)$ & $(1,-2)$ & $(11,14)$\\*[2pt]
\noalign{\nobreak\vskip\aboverulesep\nobreak\hrule height\heavyrulewidth\nobreak\vskip\belowrulesep\nobreak}
\caption{Selected $G_1\times G_2\times U(1)_{R+}$ spectra with additional simple gauge factors of rank $6$--$30$ or equal to $F_4$, omitting $D_6\times D_7$. Numbering restarts in each block. The $q_{\max}\leq1$ block has 15 spectra, including seven neutral spectra shaded gray; the $q_{\max}=2$ block has 42, including two $A_6\times F_4$ spectra. Among the shaded models, the ones with $A_8\times A_{17}$ and $D_{18}\times E_7$ are new. The conventions, including $N_{\rm sign}$, are given in the appendix introduction.
The $D_6$ spinors $\mathbf{32},\mathbf{32}^{\prime}$ have Dynkin labels $(0,0,0,0,0,1),(0,0,0,0,1,0)$. 
}
\label{tab:u1-two-combined}\label{tab:two-lowcharge}\label{t7}\\
\end{longtable}
\endgroup

\section{\texorpdfstring{Selected $Sp(1)_{R+}$ spectra}{Selected Sp(1)R+ spectra}}
\label{sp:app:spectra}

The final entry of each irrep is its $Sp(1)_{R+}$ dimension, and a
coefficient $m/2$ counts $m$ half-hypermultiplets; bars and $s,c$
distinguish conjugate and spinor irreps.
As in Appendix~\ref{app:abelian-spectra}, daggers mark spectra requiring
$I_{1,1}$; unmarked spectra admit $U$.
A superscript $\mathrm U$ marks compatibility with the unitary Wolf space
of Section~\ref{sec:unitary-embeddings}, independently of vacuum
existence; anomaly vectors use $a=(-2,-2)$ and the coordinates and
lattices in \eqref{eq:lightcone} and
\eqref{eq:lattices}.

\subsection{One additional simple factor}
\label{app:sp1-one-factor}

\begingroup\renewcommand{\baselinestretch}{1.15}\footnotesize
\setlength{\tabcolsep}{0pt}\renewcommand{\arraystretch}{1.12}
\captionsetup{hypcap=false,position=bottom}
\setlength{\LTleft}{0pt plus 1fill}\setlength{\LTright}{0pt plus 1fill}\setlength{\LTcapwidth}{\textwidth}
\color{black}\hypersetup{linkcolor=blue,citecolor=blue,urlcolor=black}
\begin{longtable}{@{}>{\centering\arraybackslash}p{14pt}@{\hspace{3pt}}p{27pt}@{\hspace{3pt}}>{\raggedright\arraybackslash}p{\dimexpr\textwidth-123pt\relax}@{\hspace{3pt}}>{\centering\arraybackslash}p{33pt}@{\hspace{3pt}}>{\centering\arraybackslash}p{37pt}@{}}
\toprule No. & $G_1$ & $\rho_H$ & $b_1$ & $b_{R+}$\\\midrule
\endfirsthead
\multicolumn{5}{l}{\small\textbf{Table~\ref{sp:tab:sp1-one-combined} (continued)}}\\[3pt]
\toprule No. & $G_1$ & $\rho_H$ & $b_1$ & $b_{R+}$\\\midrule
\endhead
\midrule
\multicolumn{5}{r}{Continued on next page}\\
\endfoot
\endlastfoot
\multicolumn{5}{c}{\rule{0pt}{14pt}\small\bfseries\boldmath$d_{\max}\leq2$}\\*[3pt]
\midrule
1 & $C_{4}$ & $\frac{52}{2}\cdot (\mathbf{8},\mathbf{1})$ $+\allowbreak\,$ $\frac{1}{2}\cdot (\mathbf{27},\mathbf{2})$ $+\allowbreak\,$ $\frac{2}{2}\cdot (\mathbf{48},\mathbf{1})$ & $(2,1)$ & $(2,-2)$\\[2pt]
2 & $C_{4}$ & $\frac{28}{2}\cdot (\mathbf{8},\mathbf{1})$ $+\allowbreak\,$ $6\cdot (\mathbf{8},\mathbf{2})$ $+\allowbreak\,$ $\frac{1}{2}\cdot (\mathbf{27},\mathbf{2})$ $+\allowbreak\,$ $\frac{2}{2}\cdot (\mathbf{48},\mathbf{1})$ & $(2,1)$ & $(4,0)$\\[2pt]
3 & $C_{4}$ & $\frac{1}{2}\cdot (\mathbf{27},\mathbf{2})$ $+\allowbreak\,$ $\frac{4}{2}\cdot (\mathbf{48},\mathbf{1})$ $+\allowbreak\,$ $\frac{2}{2}\cdot (\mathbf{160},\mathbf{1})$ & $(4,3)$ & $(2,-2)$\\[2pt]
4 & $C_{4}$ & $\frac{1}{2}\cdot (\mathbf{27},\mathbf{2})$ $+\allowbreak\,$ $(\mathbf{48},\mathbf{2})$ $+\allowbreak\,$ $\frac{2}{2}\cdot (\mathbf{160},\mathbf{1})$ & $(4,3)$ & $(4,0)$\\[2pt]
\specialrule{0.6pt}{5pt}{2pt}
\multicolumn{5}{c}{\rule{0pt}{14pt}\small\bfseries\boldmath$d_{\max}=3$}\\*[3pt]
\midrule
1 & $A_{5}^{\dagger}$ & $5\cdot (\mathbf{6},\mathbf{1})$ $+\allowbreak\,$ $8\cdot (\mathbf{6},\mathbf{2})$ $+\allowbreak\,$ $5\cdot (\mathbf{15},\mathbf{1})$ $+\allowbreak\,$ $(\mathbf{15},\mathbf{2})$ $+\allowbreak\,$ $\frac{1}{2}\cdot (\mathbf{20},\mathbf{3})$ $+\allowbreak\,$ $(\mathbf{21},\mathbf{1})$ & $(3,\tfrac{3}{2})$ & $(3,\tfrac{9}{2})$\\[2pt]
2 & $A_{5}^{\dagger}$ & $2\cdot (\mathbf{6},\mathbf{1})$ $+\allowbreak\,$ $5\cdot (\mathbf{6},\mathbf{2})$ $+\allowbreak\,$ $3\cdot (\mathbf{6},\mathbf{3})$ $+\allowbreak\,$ $5\cdot (\mathbf{15},\mathbf{1})$ $+\allowbreak\,$ $(\mathbf{15},\mathbf{2})$ $+\allowbreak\,$ $\frac{1}{2}\cdot (\mathbf{20},\mathbf{3})$ $+\allowbreak\,$ $(\mathbf{21},\mathbf{1})$ & $(3,\tfrac{3}{2})$ & $(6,6)$\\[2pt]
3 & $A_{5}^{\dagger}$ & $8\cdot (\mathbf{6},\mathbf{1})$ $+\allowbreak\,$ $5\cdot (\mathbf{6},\mathbf{2})$ $+\allowbreak\,$ $(\mathbf{6},\mathbf{3})$ $+\allowbreak\,$ $(\mathbf{15},\mathbf{1})$ $+\allowbreak\,$ $3\cdot (\mathbf{15},\mathbf{2})$ $+\allowbreak\,$ $\frac{1}{2}\cdot (\mathbf{20},\mathbf{3})$ $+\allowbreak\,$ $(\mathbf{21},\mathbf{1})$ & $(3,\tfrac{3}{2})$ & $(3,\tfrac{15}{2})$\\[2pt]
4 & $A_{5}^{\mathrm U}$ & $11\cdot (\mathbf{6},\mathbf{1})$ $+\allowbreak\,$ $3\cdot (\mathbf{6},\mathbf{3})$ $+\allowbreak\,$ $\frac{4}{2}\cdot (\mathbf{20},\mathbf{1})$ $+\allowbreak\,$ $2\cdot (\mathbf{20},\mathbf{2})$ $+\allowbreak\,$ $(\mathbf{21},\mathbf{2})$ & $(2,3)$ & $(6,4)$\\[2pt]
5 & $A_{5}$ & $3\cdot (\mathbf{6},\mathbf{1})$ $+\allowbreak\,$ $4\cdot (\mathbf{6},\mathbf{2})$ $+\allowbreak\,$ $3\cdot (\mathbf{6},\mathbf{3})$ $+\allowbreak\,$ $\frac{9}{2}\cdot (\mathbf{20},\mathbf{1})$ $+\allowbreak\,$ $\frac{1}{2}\cdot (\mathbf{20},\mathbf{3})$ $+\allowbreak\,$ $(\mathbf{21},\mathbf{2})$ & $(2,3)$ & $(6,6)$\\[2pt]
6 & $A_{5}^{\dagger}$ & $6\cdot (\mathbf{6},\mathbf{1})$ $+\allowbreak\,$ $8\cdot (\mathbf{6},\mathbf{2})$ $+\allowbreak\,$ $3\cdot (\mathbf{15},\mathbf{1})$ $+\allowbreak\,$ $(\mathbf{15},\mathbf{2})$ $+\allowbreak\,$ $\frac{1}{2}\cdot (\mathbf{20},\mathbf{1})$ $+\allowbreak\,$ $\frac{1}{2}\cdot (\mathbf{20},\mathbf{3})$ $+\allowbreak\,$ $(\mathbf{35},\mathbf{1})$ & $(3,\tfrac{3}{2})$ & $(3,\tfrac{9}{2})$\\[2pt]
7 & $A_{5}^{\dagger}$ & $3\cdot (\mathbf{6},\mathbf{1})$ $+\allowbreak\,$ $5\cdot (\mathbf{6},\mathbf{2})$ $+\allowbreak\,$ $3\cdot (\mathbf{6},\mathbf{3})$ $+\allowbreak\,$ $3\cdot (\mathbf{15},\mathbf{1})$ $+\allowbreak\,$ $(\mathbf{15},\mathbf{2})$ $+\allowbreak\,$ $\frac{1}{2}\cdot (\mathbf{20},\mathbf{1})$ $+\allowbreak\,$ $\frac{1}{2}\cdot (\mathbf{20},\mathbf{3})$ $+\allowbreak\,$ $(\mathbf{35},\mathbf{1})$ & $(3,\tfrac{3}{2})$ & $(6,6)$\\[2pt]
\midrule
8 & $C_{4}$ & $\frac{14}{2}\cdot (\mathbf{8},\mathbf{1})$ $+\allowbreak\,$ $8\cdot (\mathbf{8},\mathbf{2})$ $+\allowbreak\,$ $\frac{2}{2}\cdot (\mathbf{8},\mathbf{3})$ $+\allowbreak\,$ $\frac{1}{2}\cdot (\mathbf{27},\mathbf{2})$ $+\allowbreak\,$ $\frac{2}{2}\cdot (\mathbf{48},\mathbf{1})$ & $(2,1)$ & $(6,2)$\\[2pt]
9 & $C_{4}$ & $\frac{10}{2}\cdot (\mathbf{8},\mathbf{1})$ $+\allowbreak\,$ $6\cdot (\mathbf{8},\mathbf{2})$ $+\allowbreak\,$ $\frac{6}{2}\cdot (\mathbf{8},\mathbf{3})$ $+\allowbreak\,$ $\frac{1}{2}\cdot (\mathbf{27},\mathbf{2})$ $+\allowbreak\,$ $\frac{2}{2}\cdot (\mathbf{48},\mathbf{1})$ & $(2,1)$ & $(8,4)$\\[2pt]
10 & $C_{4}$ & $\frac{16}{2}\cdot (\mathbf{8},\mathbf{1})$ $+\allowbreak\,$ $\frac{12}{2}\cdot (\mathbf{8},\mathbf{3})$ $+\allowbreak\,$ $\frac{1}{2}\cdot (\mathbf{27},\mathbf{2})$ $+\allowbreak\,$ $\frac{2}{2}\cdot (\mathbf{48},\mathbf{1})$ & $(2,1)$ & $(10,6)$\\[2pt]
\midrule
11 & $C_{6}^{\dagger}$ & $(\mathbf{65},\mathbf{1})$ $+\allowbreak\,$ $\frac{1}{2}\cdot (\mathbf{65},\mathbf{2})$ $+\allowbreak\,$ $(\mathbf{65},\mathbf{3})$ & $(2,1)$ & $(5,\tfrac{33}{2})$\\*[2pt]
\noalign{\nobreak\vskip\aboverulesep\nobreak\hrule height\heavyrulewidth\nobreak\vskip\belowrulesep\nobreak}
\caption{Fifteen selected $G_1\times Sp(1)_{R+}$ spectra, omitting rank-three groups. Numbering restarts in each block. The $d_{\max}\leq2$ and $d_{\max}=3$ blocks contain four and eleven spectra, respectively.}\label{sp:tab:sp1-one-combined}\label{sp:tab:sp1-one-low}\label{sp:tab:sp1-one-high}\\
\end{longtable}\endgroup

\subsection{Two additional simple factors}
\label{app:sp1-two-factors}
\begingroup\renewcommand{\baselinestretch}{1.15}\footnotesize
\setlength{\tabcolsep}{0pt}\renewcommand{\arraystretch}{1.12}
\captionsetup{hypcap=false,position=bottom}
\setlength{\LTleft}{0pt plus 1fill}\setlength{\LTright}{0pt plus 1fill}\setlength{\LTcapwidth}{\textwidth}
\color{black}\hypersetup{linkcolor=blue,citecolor=blue,urlcolor=black}
\begin{longtable}{@{}>{\centering\arraybackslash}p{14pt}@{\hspace{3pt}}p{69pt}@{\hspace{3pt}}>{\raggedright\arraybackslash}p{\dimexpr\textwidth-200pt\relax}@{\hspace{3pt}}>{\centering\arraybackslash}p{33pt}@{\hspace{3pt}}>{\centering\arraybackslash}p{33pt}@{\hspace{3pt}}>{\centering\arraybackslash}p{36pt}@{}}
\toprule No. & $G_1\times G_2$ & $\rho_H$ & $b_1$ & $b_2$ & $b_{R+}$\\\midrule
\endfirsthead
\multicolumn{6}{l}{\small\textbf{Table~\ref{sp:tab:sp1-two-combined} (continued)}}\\[3pt]
\toprule No. & $G_1\times G_2$ & $\rho_H$ & $b_1$ & $b_2$ & $b_{R+}$\\\midrule
\endhead
\midrule
\multicolumn{6}{r}{Continued on next page}\\
\endfoot
\endlastfoot
\multicolumn{6}{c}{\rule{0pt}{14pt}\small\bfseries\boldmath$d_{\max}\leq2$}\\*[3pt]
\midrule
1 & $A_{10}\times D_{7}^{\mathrm U}$ & $2\cdot (\mathbf1,\mathbf{14},\mathbf{1})$ $+\allowbreak\,$ $\frac{4}{2}\cdot (\mathbf1,\mathbf{14},\mathbf{2})$ $+\allowbreak\,$ $2\cdot (\mathbf{11},\mathbf1,\mathbf{1})$ $+\allowbreak\,$ $(\mathbf{11},\mathbf1,\mathbf{2})$ $+\allowbreak\,$ $2\cdot (\mathbf{55},\mathbf1,\mathbf{1})$ $+\allowbreak\,$ $2\cdot (\mathbf{55},\mathbf1,\mathbf{2})$ & $(2,1)$ & $(-2,1)$ & $(0,4)$\\[2pt]
\midrule
2 & $B_{6}\times E_{7}$ & $(\mathbf1,\mathbf{56},\mathbf{2})$ $+\allowbreak\,$ $\frac{8}{2}\cdot (\mathbf{64},\mathbf1,\mathbf{1})$ $+\allowbreak\,$ $\frac{1}{2}\cdot (\mathbf{90},\mathbf1,\mathbf{2})$ & $(4,2)$ & $(-2,1)$ & $(-2,2)$\\[2pt]
\midrule
3 & $B_{7}\times C_{6}$ & $2\cdot (\mathbf1,\mathbf{65},\mathbf{1})$ $+\allowbreak\,$ $\frac{3}{2}\cdot (\mathbf1,\mathbf{65},\mathbf{2})$ $+\allowbreak\,$ $2\cdot (\mathbf{15},\mathbf1,\mathbf{1})$ $+\allowbreak\,$ $\frac{5}{2}\cdot (\mathbf{15},\mathbf1,\mathbf{2})$ & $(-2,1)$ & $(2,1)$ & $(0,4)$\\[2pt]
\midrule
4 & $B_{9}\times B_{13}$ & $\frac12\cdot (\mathbf{19},\mathbf{27},\mathbf2)$ $+\allowbreak\,$ $\frac12\cdot (\mathbf{512},\mathbf1,\mathbf1)$ & $(6,1)$ & $(-2,1)$ & $(-2,2)$\\[2pt]
\midrule
5 & $C_{8}\times D_{8}$ & $\frac{1}{2}\cdot (\mathbf1,{\mathbf{128}}_{c},\mathbf{2})$ $+\allowbreak\,$ $\frac{2}{2}\cdot (\mathbf{16},\mathbf{16},\mathbf{1})$ $+\allowbreak\,$ $\frac{1}{2}\cdot (\mathbf{119},\mathbf1,\mathbf{2})$ & $(0,1)$ & $(2,1)$ & $(-2,2)$\\[2pt]
\midrule
6 & $D_{8}\times E_{6}$ & $(\mathbf1,\mathbf{27},\mathbf{2})$ $+\allowbreak\,$ $2\cdot ({\mathbf{128}}_{c},\mathbf1,\mathbf{1})$ $+\allowbreak\,$ $\frac{1}{2}\cdot (\mathbf{135},\mathbf1,\mathbf{2})$ & $(4,2)$ & $(-2,1)$ & $(-2,2)$\\[2pt]
7 & $D_{8}\times E_{6}$ & $(\mathbf1,\mathbf{27},\mathbf{2})$ $+\allowbreak\,$ $({\mathbf{128}}_{s},\mathbf1,\mathbf{1})$ $+\allowbreak\,$ $({\mathbf{128}}_{c},\mathbf1,\mathbf{1})$ $+\allowbreak\,$ $\frac{1}{2}\cdot (\mathbf{135},\mathbf1,\mathbf{2})$ & $(4,2)$ & $(-2,1)$ & $(-2,2)$\\[2pt]
\midrule
8 & $D_{9}\times F_{4}$ & $\frac{1}{2}\cdot (\mathbf1,\mathbf{26},\mathbf{2})$ $+\allowbreak\,$ $\frac{1}{2}\cdot (\mathbf{170},\mathbf1,\mathbf{2})$ $+\allowbreak\,$ $({\mathbf{256}}_{s},\mathbf1,\mathbf{1})$ & $(4,2)$ & $(-2,1)$ & $(-2,2)$\\[2pt]
\midrule
9 & $E_{6}\times E_{7}^{\dagger}$ & $\frac{1}{2}\cdot (\mathbf1,\mathbf{56},\mathbf{1})$ $+\allowbreak\,$ $(\mathbf1,\mathbf{56},\mathbf{2})$ $+\allowbreak\,$ $2\cdot (\mathbf{27},\mathbf1,\mathbf{1})$ $+\allowbreak\,$ $2\cdot (\mathbf{27},\mathbf1,\mathbf{2})$ $+\allowbreak\,$ $(\mathbf{78},\mathbf1,\mathbf{1})$ $+\allowbreak\,$ $\frac{1}{2}\cdot (\mathbf{78},\mathbf1,\mathbf{2})$ & $(2,3)$ & $(1,-\tfrac{3}{2})$ & $(4,0)$\\[2pt]
\specialrule{0.6pt}{5pt}{2pt}
\multicolumn{6}{c}{\rule{0pt}{14pt}\small\bfseries\boldmath$d_{\max}=3$}\\*[3pt]
\noalign{\nobreak\vskip\aboverulesep\nobreak\hrule height\lightrulewidth\nobreak\vskip\belowrulesep\nobreak}
1 & $B_{6}\times F_{4}^{\dagger}$ & $4\cdot (\mathbf1,\mathbf{26},\mathbf{1})$ $+\allowbreak\,$ $\frac{2}{2}\cdot (\mathbf1,\mathbf{26},\mathbf{2})$ $+\allowbreak\,$ $\frac{3}{2}\cdot (\mathbf1,\mathbf{52},\mathbf{2})$ $+\allowbreak\,$ $(\mathbf{13},\mathbf1,\mathbf{1})$ $+\allowbreak\,$ $\frac{1}{2}\cdot (\mathbf{13},\mathbf1,\mathbf{2})$ $+\allowbreak\,$ $(\mathbf{13},\mathbf1,\mathbf{3})$ & $(-2,1)$ & $(4,2)$ & $(5,\tfrac{7}{2})$\\*[2pt]
\noalign{\nobreak\vskip\aboverulesep\nobreak\hrule height\lightrulewidth\nobreak\vskip\belowrulesep\nobreak}
2 & $C_{7}\times F_{4}$ & $(\mathbf1,\mathbf{26},\mathbf{1})$ $+\allowbreak\,$ $\frac{2}{2}\cdot (\mathbf1,\mathbf{26},\mathbf{2})$ $+\allowbreak\,$ $\frac{1}{2}\cdot (\mathbf{14},\mathbf1,\mathbf{1})$ $+\allowbreak\,$ $(\mathbf{14},\mathbf1,\mathbf{2})$ $+\allowbreak\,$ $\frac{1}{2}\cdot (\mathbf{14},\mathbf1,\mathbf{3})$ $+\allowbreak\,$ $(\mathbf{90},\mathbf1,\mathbf{1})$ $+\allowbreak\,$ $\frac{2}{2}\cdot (\mathbf{90},\mathbf1,\mathbf{2})$ & $(1,1)$ & $(-1,1)$ & $(2,5)$\\*[2pt]
\noalign{\nobreak\vskip\aboverulesep\nobreak\hrule height\heavyrulewidth\nobreak\vskip\belowrulesep\nobreak}
\caption{Eleven selected $G_1\times G_2\times Sp(1)_{R+}$ spectra, omitting spectra with an additional rank-five simple gauge factor while retaining $F_4$. Numbering restarts in each block. The $d_{\max}\leq2$ and $d_{\max}=3$ blocks contain nine and two spectra, respectively.}\label{sp:tab:sp1-two-combined}\label{sp:tab:sp1-two-low}\label{sp:tab:sp1-two-high}\\
\end{longtable}\endgroup

\section{\texorpdfstring{The $(K,N)$ pairs for all displayed models with $q_{\rm max}=2$ and $d_{\rm max}=3$ }{More Minkowski vacua for qmax=2 and dmax=3}}
\label{app:maximal-symmetry}

In Section~\ref{sec:explicit-Minkowski}, we gave a detailed description of the 4 Minkowski vacuum solutions of $U(1)_{R+}$ gauged models with $q_{\rm max}=2$, and 2 Minkowski vacuum solutions of $Sp(1)_{R+}$ gauged models with $d_{\rm max}=3$, in which the gauge group $K$ is broken down to $N$. In this appendix, we list $(K,N)$ pairs with the largest $\dim N$ found for all the displayed $q_{\rm max}=2$ and $d_{\rm max}=3$ models, respectively, in Tables~\ref{tab:Sp-maximal-combined} and~\ref{tab:Sp1-maximal-combined}. Both use the symplectic Wolf space $Sp(n_H,1)/(Sp(n_H)\times Sp(1)_R)$.

\begingroup
\setlength{\LTleft}{\fill}\setlength{\LTright}{\fill}
\setlength{\LTcapwidth}{\textwidth}
\captionsetup{hypcap=false,position=bottom}
\renewcommand{\baselinestretch}{1}
\small
\renewcommand{\arraystretch}{1}
\setlength{\LTpre}{6pt}
\begin{longtable}{@{}r@{\hspace{11.5pt}}l@{\hspace{11.5pt}}l@{\hspace{11.5pt}}l@{\hspace{24pt}}r@{\hspace{11.5pt}}l@{\hspace{11.5pt}}l@{\hspace{11.5pt}}l@{}}
\toprule
\multicolumn{4}{c}{One additional simple factor} & \multicolumn{4}{c}{Two additional simple factors}\\
\cmidrule(r{6pt}){1-4}\cmidrule(l{6pt}){5-8}
No(s). & $K$ & $N$ & $(n_2,n_1,n_0)$ & No(s). & $K$ & $N$ & $(n_2,n_1,n_0)$\\
\midrule
\endfirsthead
\multicolumn{8}{l}{\small\textbf{Table~\ref{tab:Sp-maximal-combined} (continued)}}\\[3pt]
\toprule
\multicolumn{4}{c}{One additional simple factor} & \multicolumn{4}{c}{Two additional simple factors}\\
\cmidrule(r{6pt}){1-4}\cmidrule(l{6pt}){5-8}
No(s). & $K$ & $N$ & $(n_2,n_1,n_0)$ & No(s). & $K$ & $N$ & $(n_2,n_1,n_0)$\\
\midrule
\endhead
\midrule
\multicolumn{8}{r}{Continued on next page}\\
\endfoot
\endlastfoot
1 & $A_{6}$ & $\mathrm{SU}(6)$ & $(1,4,3)$ & 1 & $A_{6}\!\times\! A_{10}$ & $\mathrm{SU}(6)\!\times\! \mathrm{SU}(11)$ & $(3,0,3)$\\
2,6 & $A_{6}$ & $\mathrm{SU}(6)$ & $(4,3,1)$ & 2 & $A_{6}\!\times\! A_{10}$ & $\mathrm{SU}(6)\!\times\! \mathrm{SU}(11)$ & $(2,4,0)$\\
3 & $A_{6}$ & $\mathrm{SU}(6)$ & $(6,1,1)$ & 3 & $A_{6}\!\times\! D_{7}$ & $\mathrm{SU}(7)\!\times\! \mathrm{Spin}(13)$ & $(1,2,3)$\\
4 & $A_{6}$ & $\mathrm{SU}(6)$ & $(2,5,1)$ & 4 & $A_{6}\!\times\! D_{9}$ & $\mathrm{SU}(6)\!\times\! \mathrm{Spin}(18)$ & $(2,6,12)$\\
5 & $A_{6}$ & $\mathrm{SU}(6)$ & $(2,3,3)$ & 5 & $A_{6}\!\times\! D_{9}$ & $\mathrm{SU}(7)\!\times\! \mathrm{Spin}(17)$ & $(3,0,7)$\\
7 & $A_{6}$ & $\mathrm{SU}(6)$ & $(4,2,2)$ & 6 & $A_{6}\!\times\! D_{9}$ & $\mathrm{SU}(6)\!\times\! \mathrm{Spin}(18)$ & $(5,12,3)$\\
8 & $A_{7}$ & $S(\mathrm{U}(7)\times\mathrm{U}(1))$ & $(2,0,0)$ & 7 & $A_{6}\!\times\! D_{9}$ & $\mathrm{SU}(6)\!\times\! \mathrm{Spin}(18)$ & $(8,0,12)$\\
9 & $C_{4}$ & $\mathrm{Sp}(3)$ & $(4,7,6)$ & 8 & $A_{6}\!\times\! D_{9}$ & $\mathrm{SU}(6)\!\times\! \mathrm{Spin}(18)$ & $(2,0,18)$\\
10 & $C_{4}$ & $\mathrm{Sp}(3)$ & $(2,9,6)$ & 9 & $A_{6}\!\times\! D_{9}$ & $\mathrm{SU}(7)\!\times\! \mathrm{Spin}(17)$ & $(1,0,2)$\\
11 & $C_{5}$ & $\mathrm{Sp}(4)$ & $(4,11,6)$ & 10 & $A_{6}\!\times\! D_{9}$ & $\mathrm{SU}(6)\!\times\! \mathrm{Spin}(18)$ & $(8,6,2)$\\
12 & $D_{4}$ & $\mathrm{SU}(4)\times\mathrm{U}(1)$ & $(1,3,3)$ & 11 & $A_{6}\!\times\! D_{9}$ & $\mathrm{SU}(6)\!\times\! \mathrm{Spin}(18)$ & $(12,0,4)$\\
13 & $G_{2}$ & $\mathrm{SU}(3)$ & $(4,5,8)$ & 12 & $A_{6}\!\times\! E_{7}$ & $\mathrm{SU}(6)\!\times\! E_{7}$ & $(2,9,9)$\\
14 & $G_{2}$ & $\mathrm{SU}(2)_{\rm long}\times\mathrm{U}(1)$ & $(2,19,7)$ & 13 & $A_{6}\!\times\! F_{4}$ & $\mathrm{SU}(6)\!\times\! F_{4}$ & $(4,2,2)$\\
14 & $G_{2}$ & $\mathrm{SU}(2)_{\rm short}\times\mathrm{U}(1)$ & $(2,2,5)$ & 14 & $A_{6}\!\times\! F_{4}$ & $\mathrm{SU}(6)\!\times\! F_{4}$ & $(4,4,0)$\\
15 & $G_{2}$ & $\mathrm{SU}(3)$ & $(2,9,8)$ & 15 & $A_{7}\!\times\! A_{9}$ & $\mathrm{SU}(8)\!\times\! \mathrm{SU}(9)$ & $(5,0,1)$\\
16 & $G_{2}$ & $\mathrm{SU}(3)$ & $(8,3,8)$ & 16 & $A_{7}\!\times\! D_{6}$ & $\mathrm{SU}(8)\!\times\! \mathrm{Spin}(11)$ & $(2,0,2)$\\
17 & $G_{2}$ & $\mathrm{SU}(3)$ & $(2,21,2)$ & 17 & $A_{7}\!\times\! D_{7}$ & $\mathrm{SU}(8)\!\times\! \mathrm{Spin}(13)$ & $(1,2,3)$\\
18 & $G_{2}$ & $\mathrm{SU}(3)$ & $(4,19,2)$ & 18 & $A_{8}\!\times\! A_{8}$ & $\mathrm{SU}(9)\!\times\! \mathrm{SU}(7)$ & $(8,4,2)$\\
19 & $G_{2}$ & $\mathrm{SU}(2)_{\rm long}\times\mathrm{U}(1)$ & $(9,5,4)$ & 19 & $A_{8}\!\times\! A_{8}$ & $\mathrm{SU}(8)\!\times\! \mathrm{SU}(9)$ & $(2,2,3)$\\
19 & $G_{2}$ & $\mathrm{SU}(2)_{\rm short}\times\mathrm{U}(1)$ & $(9,2,0)$ & 20 & $A_{8}\!\times\! A_{8}$ & $\mathrm{SU}(7)\!\times\! \mathrm{SU}(9)$ & $(8,2,6)$\\
20 & $G_{2}$ & $\mathrm{SU}(3)$ & $(2,3,0)$ & 21,26 & $A_{9}\!\times\! D_{6}$ & $\mathrm{SU}(9)\!\times\! \mathrm{Spin}(12)$ & $(4,0,12)$\\
21 & $G_{2}$ & $\mathrm{SU}(2)_{\rm long}\times\mathrm{U}(1)$ & $(3,14,5)$ & 22,27,31 & $A_{9}\!\times\! D_{6}$ & $\mathrm{SU}(9)\!\times\! \mathrm{Spin}(12)$ & $(4,8,4)$\\
21 & $G_{2}$ & $\mathrm{SU}(2)_{\rm short}\times\mathrm{U}(1)$ & $(3,11,1)$ & 23,28,32 & $A_{9}\!\times\! D_{6}$ & $\mathrm{SU}(10)\!\times\! \mathrm{Spin}(11)$ & $(4,0,4)$\\
 &  &  &  & 24,29 & $A_{9}\!\times\! D_{6}$ & $\mathrm{SU}(10)\!\times\! \mathrm{Spin}(11)$ & $(1,1,6)$\\
 &  &  &  & 25,30,33 & $A_{9}\!\times\! D_{6}$ & $\mathrm{SU}(10)\!\times\! \mathrm{Spin}(11)$ & $(2,2,4)$\\
 &  &  &  & 34--40 & $D_{6}\!\times\! E_{6}$ & $\mathrm{Spin}(11)\!\times\! E_{6}$ & $(1,0,3)$\\*
 &  &  &  & 41 & $D_{7}\!\times\! E_{7}$ & $\mathrm{Spin}(13)\!\times\! E_{7}$ & $(1,2,3)$\\*
 &  &  &  & 42 & $A_{13}\!\times\! D_{9}$ & $\mathrm{SU}(14)\!\times\! \mathrm{Spin}(17)$ & $(3,0,7)$\\*
\noalign{\nobreak\vskip\aboverulesep\nobreak\hrule height\heavyrulewidth\nobreak\vskip\belowrulesep\nobreak}
\caption{Minkowski vacua for selected $q_{\max}=2$ spectra. The numbers in the left and right panels refer to the numbered $q_{\max}=2$ blocks of Tables~\ref{tab:u1-one-combined} and~\ref{tab:u1-two-combined}, respectively. $K$ is the additional nonabelian gauge group and $N$ a connected unbroken group of the largest dimension found. The $n_j$ count $N$-singlet full-hypermultiplet units with original hyperino charge magnitude $j$, counting conjugate pairs once. These singlet counts are taken before imposing the vacuum equations and identifying gauge-equivalent configurations; they are not moduli dimensions.  The $G_2$ labels distinguish long- and short-root $SU(2)$ embeddings; the three triality-related $D_4$ embeddings have identical counts. The two-factor panel lists one unbroken group per displayed spectrum.}
\label{tab:Sp-maximal-combined}\label{tab:Sp-one-maximal}\label{tab:Sp-two-maximal}
\\
\end{longtable}
\endgroup

\begin{table}[H]
\centering
\renewcommand{\baselinestretch}{1}
\small
\renewcommand{\arraystretch}{1.05}
\begin{tabular}{@{}r@{\hspace{17pt}}l@{\hspace{17pt}}l@{\hspace{17pt}}l@{\hspace{24pt}}r@{\hspace{17pt}}l@{\hspace{17pt}}l@{\hspace{17pt}}l@{}}
\toprule
\multicolumn{4}{c}{One additional simple factor} & \multicolumn{4}{c}{Two additional simple factors}\\
\cmidrule(r{6pt}){1-4}\cmidrule(l{6pt}){5-8}
No(s). & $K$ & $N$ & $(m_3,m_2,m_1)$ & No(s). & $K$ & $N$ & $(m_3,m_2,m_1)$\\
\midrule
1 & $A_5$ & $\mathrm{SU}(3)\times\mathrm{SU}(2)$ & $(1,9,11)$ & 1 & $B_6\!\times\! F_4$ & $\mathrm{Spin}(12)\!\times\! F_4$ & $(1,\tfrac12,1)$\\
2,5 & $A_5$ & $\mathrm{SU}(5)$ & $(3,5,3)$ & 2 & $C_7\!\times\! F_4$ & $\mathrm{Sp}(6)\!\times\! F_4$ & $(1,3,2)$\\
3 & $A_5$ & $\mathrm{SU}(5)$ & $(1,5,9)$ &  &  &  & \\
4 & $A_5$ & $\mathrm{SU}(5)$ & $(3,1,11)$ &  &  &  & \\
6 & $A_5$ & $\mathrm{SU}(3)\times\mathrm{SU}(2)$ & $(1,9,12)$ &  &  &  & \\
7 & $A_5$ & $\mathrm{SU}(5)$ & $(3,5,4)$ &  &  &  & \\
8 & $C_4$ & $\mathrm{Sp}(3)$ & $(2,\tfrac{33}{2},14)$ &  &  &  & \\
9 & $C_4$ & $\mathrm{Sp}(3)$ & $(6,\tfrac{25}{2},10)$ &  &  &  & \\
10 & $C_4$ & $\mathrm{Sp}(3)$ & $(12,\tfrac12,16)$ &  &  &  & \\
11 & $C_6$ & $\mathrm{Sp}(5)\times\mathrm{Sp}(1)$ & $(1,\tfrac12,1)$ &  &  &  & \\
\bottomrule
\end{tabular}
\caption{Minkowski vacua for thirteen selected $d_{\max}=3$ spectra. The numbers in the left and right panels refer to the numbered $d_{\max}=3$ blocks of Tables~\ref{sp:tab:sp1-one-combined} and~\ref{sp:tab:sp1-two-combined}, respectively. $K$ is the additional nonabelian gauge group and $N$ a connected unbroken group of the largest dimension found. The $m_d$ count $N$-singlet diagonal $\mathbf d$ multiplets in full-hypermultiplet units, allowing half-integer $m_2$. These singlet counts are taken before imposing the vacuum equations and identifying gauge-equivalent configurations; they are not moduli dimensions. The diagonal $Sp(1)_{R+}$ is completely broken.}
\label{tab:Sp1-maximal-combined}
\end{table}

\end{appendix}


\begin{thebibliography}{10}

\bibitem{Nishino:1986dc}
H.~Nishino and E.~Sezgin, \emph{{The Complete $N=2$, $d=6$ Supergravity With Matter and {Yang-Mills} Couplings}}, \href{https://doi.org/10.1016/0550-3213(86)90218-X}{\emph{Nucl. Phys. B} {\bfseries 278} (1986) 353}.

\bibitem{Nishino:1997ff}
H.~Nishino and E.~Sezgin, \emph{{New couplings of six-dimensional supergravity}}, \href{https://doi.org/10.1016/S0550-3213(97)00357-X}{\emph{Nucl. Phys. B} {\bfseries 505} (1997) 497} [\href{https://arxiv.org/abs/hep-th/9703075}{{\ttfamily hep-th/9703075}}].

\bibitem{Riccioni:2001bg}
F.~Riccioni, \emph{{All couplings of minimal six-dimensional supergravity}}, \href{https://doi.org/10.1016/S0550-3213(01)00199-7}{\emph{Nucl. Phys. B} {\bfseries 605} (2001) 245} [\href{https://arxiv.org/abs/hep-th/0101074}{{\ttfamily hep-th/0101074}}].

\bibitem{Randjbar-Daemi:1985tdc}
S.~Randjbar-Daemi, A.~Salam, E.~Sezgin and J.~A. Strathdee, \emph{{An Anomaly Free Model in Six-Dimensions}}, \href{https://doi.org/10.1016/0370-2693(85)91653-3}{\emph{Phys. Lett. B} {\bfseries 151} (1985) 351}.

\bibitem{Avramis:2005qt}
S.~D. Avramis, A.~Kehagias and S.~Randjbar-Daemi, \emph{{A New anomaly-free gauged supergravity in six dimensions}}, \href{https://doi.org/10.1088/1126-6708/2005/05/057}{\emph{JHEP} {\bfseries 05} (2005) 057} [\href{https://arxiv.org/abs/hep-th/0504033}{{\ttfamily hep-th/0504033}}].

\bibitem{Avramis:2005hc}
S.~D. Avramis and A.~Kehagias, \emph{{A Systematic search for anomaly-free supergravities in six dimensions}}, \href{https://doi.org/10.1088/1126-6708/2005/10/052}{\emph{JHEP} {\bfseries 10} (2005) 052} [\href{https://arxiv.org/abs/hep-th/0508172}{{\ttfamily hep-th/0508172}}].

\bibitem{Becker:2023zyb}
K.~Becker, A.~Kehagias, E.~Sezgin, D.~Tennyson and A.~Violaris, \emph{{New anomaly free supergravities in six dimensions}}, \href{https://doi.org/10.1007/JHEP05(2024)144}{\emph{JHEP} {\bfseries 05} (2024) 144} [\href{https://arxiv.org/abs/2311.03337}{{\ttfamily 2311.03337}}].

\bibitem{Becker:2025xgy}
K.~Becker, E.~Sezgin, D.~Tennyson and Y.~Tachikawa, \emph{{Global anomalies in 6D gauged supergravities}}, \href{https://doi.org/10.1007/JHEP01(2026)092}{\emph{JHEP} {\bfseries 01} (2026) 092} [\href{https://arxiv.org/abs/2507.22127}{{\ttfamily 2507.22127}}].

\bibitem{Suzuki:2005vu}
R.~Suzuki and Y.~Tachikawa, \emph{{More anomaly-free models of six-dimensional gauged supergravity}}, \href{https://doi.org/10.1063/1.2209767}{\emph{J. Math. Phys.} {\bfseries 47} (2006) 062302} [\href{https://arxiv.org/abs/hep-th/0512019}{{\ttfamily hep-th/0512019}}].

\bibitem{Becker:2026etl}
K.~Becker and Q.~You, \emph{{Anomaly-free spectra, unimodular lattices and 6D R-symmetry gauged supergravity}}, \href{https://doi.org/10.1007/JHEP09(2026)186}{\emph{JHEP} {\bfseries 09} (2026) 186} [\href{https://arxiv.org/abs/2605.17643}{{\ttfamily 2605.17643}}].

\bibitem{Bagger:1983tt}
J.~Bagger and E.~Witten, \emph{{Matter Couplings in N=2 Supergravity}}, \href{https://doi.org/10.1016/0550-3213(83)90605-3}{\emph{Nucl. Phys. B} {\bfseries 222} (1983) 1}.

\bibitem{Sezgin:2023hkc}
E.~Sezgin, \emph{{Survey of supergravities}},  \href{https://arxiv.org/abs/2312.06754}{{\ttfamily 2312.06754}}.

\bibitem{Paradan:2015}
P.-E. Paradan, \emph{{Quantization commutes with reduction in the non-compact setting: the case of holomorphic discrete series}}, {\emph{J. Eur. Math. Soc.} {\bfseries 17} (2015) 955} [\href{https://arxiv.org/abs/1201.5451}{{\ttfamily 1201.5451}}].

\bibitem{Tsokur:1996su}
V.~A. Tsokur and Y.~M. Zinovev, \emph{{N=2 supergravity models based on the nonsymmetric quaternionic manifolds. 1. Symmetries and Lagrangians}},  \href{https://arxiv.org/abs/hep-th/9605134}{{\ttfamily hep-th/9605134}}.

\bibitem{Tsokur:1996vk}
V.~A. Tsokur and Y.~M. Zinovev, \emph{{N=2 supergravity models based on the nonsymmetric quaternionic manifolds. 2. Gauge interactions}},  \href{https://arxiv.org/abs/hep-th/9605143}{{\ttfamily hep-th/9605143}}.

\bibitem{Raghuram:2021wvj}
N.~Raghuram and A.~P. Turner, \emph{{Orders of Vanishing and U(1) Charges in F-theory}}, \href{https://doi.org/10.1007/JHEP03(2022)051}{\emph{JHEP} {\bfseries 03} (2022) 051} [\href{https://arxiv.org/abs/2110.10159}{{\ttfamily 2110.10159}}].

\bibitem{Alvarez-Gaume:1983ihn}
L.~Alvarez-Gaume and E.~Witten, \emph{{Gravitational Anomalies}}, \href{https://doi.org/10.1016/0550-3213(84)90066-X}{\emph{Nucl. Phys. B} {\bfseries 234} (1984) 269}.

\bibitem{Alvarez-Gaume:1984zlq}
L.~Alvarez-Gaume and P.~H. Ginsparg, \emph{{The Structure of Gauge and Gravitational Anomalies}}, \href{https://doi.org/10.1016/0003-4916(85)90087-9}{\emph{Annals Phys.} {\bfseries 161} (1985) 423}.

\bibitem{Bilal:2008qx}
A.~Bilal, \emph{{Lectures on Anomalies}},  \href{https://arxiv.org/abs/0802.0634}{{\ttfamily 0802.0634}}.

\bibitem{Taylor:2011wt}
W.~Taylor, \emph{{TASI Lectures on Supergravity and String Vacua in Various Dimensions}},  \href{https://arxiv.org/abs/1104.2051}{{\ttfamily 1104.2051}}.

\bibitem{Seiberg:2011dr}
N.~Seiberg and W.~Taylor, \emph{{Charge Lattices and Consistency of 6D Supergravity}}, \href{https://doi.org/10.1007/JHEP06(2011)001}{\emph{JHEP} {\bfseries 06} (2011) 001} [\href{https://arxiv.org/abs/1103.0019}{{\ttfamily 1103.0019}}].

\bibitem{Monnier:2017oqd}
S.~Monnier, G.~W. Moore and D.~S. Park, \emph{{Quantization of anomaly coefficients in 6D $\mathcal{N}=(1,0)$ supergravity}}, \href{https://doi.org/10.1007/JHEP02(2018)020}{\emph{JHEP} {\bfseries 02} (2018) 020} [\href{https://arxiv.org/abs/1711.04777}{{\ttfamily 1711.04777}}].

\bibitem{Lee:2020ewl}
Y.~Lee and Y.~Tachikawa, \emph{{Some comments on 6D global gauge anomalies}}, \href{https://doi.org/10.1093/ptep/ptab015}{\emph{PTEP} {\bfseries 2021} (2021) 08B103} [\href{https://arxiv.org/abs/2012.11622}{{\ttfamily 2012.11622}}].

\bibitem{Monnier:2018}
S.~Monnier and G.~W. Moore, \emph{{Remarks on the Green--Schwarz terms of six-dimensional supergravity theories}}, \href{https://doi.org/10.1007/s00220-019-03341-7}{\emph{Commun. Math. Phys.} {\bfseries 372} (2019) 963} [\href{https://arxiv.org/abs/1808.01334}{{\ttfamily 1808.01334}}].

\bibitem{Bershadsky:1997}
M.~Bershadsky and C.~Vafa, \emph{{Global anomalies and geometric engineering of critical theories in six dimensions}},  \href{https://arxiv.org/abs/hep-th/9703167}{{\ttfamily hep-th/9703167}}.

\bibitem{Bossard:2024ffp}
G.~Bossard, A.~Kleinschmidt and E.~Sezgin, \emph{{Higher derivative couplings with multi-tensor multiplets in 6D supergravity, action and anomalies}}, \href{https://doi.org/10.1007/JHEP03(2025)108}{\emph{JHEP} {\bfseries 03} (2025) 108} [\href{https://arxiv.org/abs/2412.05365}{{\ttfamily 2412.05365}}].

\bibitem{Park:2011wv}
D.~S. Park and W.~Taylor, \emph{{Constraints on 6D Supergravity Theories with Abelian Gauge Symmetry}}, \href{https://doi.org/10.1007/JHEP01(2012)141}{\emph{JHEP} {\bfseries 01} (2012) 141} [\href{https://arxiv.org/abs/1110.5916}{{\ttfamily 1110.5916}}].

\bibitem{Ozkan:2024euj}
M.~Ozkan, Y.~Pang and E.~Sezgin, \emph{{Higher derivative supergravities in diverse dimensions}}, \href{https://doi.org/10.1016/j.physrep.2024.07.002}{\emph{Phys. Rept.} {\bfseries 1086} (2024) 1} [\href{https://arxiv.org/abs/2401.08945}{{\ttfamily 2401.08945}}].

\bibitem{DGcode}
X.~Guo, Y.~Pang, H.~Samtleben and E.~Sezgin, ``{Diagonal gauging in six-dimensional supergravity: Mathematica code and supplementary catalogues}.'' GitHub repository, \url{https://github.com/XuGuo-hep/diagonal-gauging}, 2026.

\bibitem{Hamada:2025vga}
Y.~Hamada and G.~J. Loges, \emph{{A finite 6d supergravity landscape from anomalies}},  \href{https://arxiv.org/abs/2507.20949}{{\ttfamily 2507.20949}}.

\bibitem{Percacci:1998ag}
R.~Percacci and E.~Sezgin, \emph{{Properties of gauged sigma models}},  in \emph{{Richard Arnowitt Fest: A Symposium on Supersymmetry and Gravitation}}, pp.~255--278, 4, 1998, \href{https://arxiv.org/abs/hep-th/9810183}{{\ttfamily hep-th/9810183}}.

\bibitem{Jong:2006ud}
D.~C. Jong, \emph{{6D Dyonic String With Active Hyperscalars}},  \href{https://arxiv.org/abs/hep-th/0608034}{{\ttfamily hep-th/0608034}}.

\bibitem{Guo:2025mlb}
X.~Guo, Y.~Pang and E.~Sezgin, \emph{{4D de Sitter from 6D gauged supergravity with Green-Schwarz counterterm}}, \href{https://doi.org/10.1007/JHEP07(2026)006}{\emph{JHEP} {\bfseries 07} (2026) 006} [\href{https://arxiv.org/abs/2510.11794}{{\ttfamily 2510.11794}}].

\bibitem{Schwarz:1995-K3}
J.~H. Schwarz, \emph{{Anomaly-Free Supersymmetric Models in Six Dimensions}}, \href{https://doi.org/10.1016/0370-2693(95)01610-4}{\emph{Phys. Lett. B} {\bfseries 371} (1996) 223} [\href{https://arxiv.org/abs/hep-th/9512053}{{\ttfamily hep-th/9512053}}].

\bibitem{Gueven:2003uw}
R.~Gueven, J.~T. Liu, C.~N. Pope and E.~Sezgin, \emph{{Fine tuning and six-dimensional gauged N=(1,0) supergravity vacua}}, \href{https://doi.org/10.1088/0264-9381/21/4/019}{\emph{Class. Quant. Grav.} {\bfseries 21} (2004) 1001} [\href{https://arxiv.org/abs/hep-th/0306201}{{\ttfamily hep-th/0306201}}].

\end{thebibliography}

\providecommand{\href}[2]{#2}\begingroup\raggedright\endgroup

\end{document}